\documentclass[a4paper,10pt,preprint,5p]{elsarticle}
\usepackage{graphicx,graphics}
\usepackage{dcolumn}
\usepackage{bm}
\usepackage{lineno,hyperref}
\usepackage{amsmath}
\usepackage{cuted, ragged2e}
\usepackage{lipsum, array, booktabs, caption}
\usepackage{float}
\usepackage[english]{babel}
\usepackage{tabularx}
\usepackage{epstopdf}
\usepackage{multirow}
\usepackage{babel,blindtext}
\usepackage{subcaption}
\usepackage{amsthm}
\newtheorem{prop}{Proposition}

\journal{Chaos, Solitons \& Fractals}

\begin{document}

\begin{frontmatter}

\title{Degenerating the butterfly attractor in a plasma perturbation model using nonlinear controllers}

\author[mymainaddress]{Hayder Natiq}
\author[mymainaddress,mysecondaryaddress]{Santo Banerjee}
\cortext[mycorrespondingauthor]{Corresponding author}
\ead{santoban@gmail.com}
\author[myfourthaddress]{A.P. Misra}
\author[mymainaddress,mysecondaryaddress,mythirdaddress]{M.R.M. Said}

\address[mymainaddress]{Institute for Mathematical Research, Universiti Putra Malaysia, Serdang, Malaysia}
\address[mysecondaryaddress]{Malaysia-Italy Centre of Excellence for Mathematical Science, Universiti Putra Malaysia, Serdang, Malaysia}
\address[mythirdaddress]{Department of Mathematics, Universiti Putra Malaysia, Serdang, Malaysia}
\address[myfourthaddress]{Department of Mathematics, Siksha Bhavana, Visva-Bharati University, Santiniketan, 731 235, India}

\begin{abstract}
In this work, the dynamical behaviors of a low-dimensional model, which governs the interplay between a driver associated with pressure gradient and relaxation of instability due to magnetic field perturbations, are investigated. Besides that, two nonlinear controllers are constructed precisely to shift the equilibria of the plasma model apart from each other. Simulation results show that shifting the equilibria can change the spacing of chaotic attractors, and subsequently break the butterfly wings into one or two symmetric pair of coexisting chaotic attractors. Furthermore, stretching the equilibria of the system apart enough from each other gives rise to degenerate the butterfly wings into several periodic orbits. In addition, with appropriate initial conditions, the complex multistability behaviors including the coexistence of butterfly chaotic attractor with two point attractors, the coexistence of transient transition chaos with completely quasi-periodic behavior, and the coexistence of symmetric Hopf bifurcations are also observed.
\end{abstract}


\end{frontmatter}

\section{Introduction}
\label{section:section1}
Chaos as a complicated nonlinear phenomenon has attracted considerable interest in the past few decades, and it has been widely applied in various fields such as biomedical engineering \cite{mukherjee2015can,banerjee2016complexity}, secure communications \cite{banerjee2004optically,jeeva2012comment}, and cryptographic applications \cite{banerjee2011synchronization,banerjee2012multi,natiq2018new}.

Recently, with further investigations of chaos, it was unexpected to find that many chaotic systems have several possible final stable states (attractors) for a given
set of parameters. This nonlinear phenomenon is known as multistability behavior or coexisting attractors. More interestingly, multistability behaviors have been observed in numerous natural systems and usually plays a crucial role in their performance \cite{angeli2004detection,natiq2019dynamics}. The clear evidence of multistability behavior was first manifested experimentally in a Q-switched gas laser \cite{arecchi1982experimental}, since then chaotic models with multistability behaviors have been extensively reported in both continuous and discrete systems. Li and Sprott pointed out that the butterfly attractor of Lorenz system for certain values of the parameters can exhibit coexisting symmetric pair of chaotic attractors, or coexisting symmetric pair of limit cycles \cite{li2014multistability}. Li et al. \cite{li2017infinite} introduced a new method for constructing self-reproducing chaotic systems with extreme multistability behaviors, in which the coexisting attractors reside in the phase space along a specific coordinate axis. Natiq et al. \cite{natiq2018self} presented a simple chaotic system with trigonometric function term generating extreme multistability behaviors. Wang et al. \cite{wang2019dynamics} proposed a four-wing memristive chaotic system with various multistability behaviors. Besides that, the unusual multistability behavior of coexisting hyperchaotic with periodic attractors has been obtained in discrete hyperchaotic systems \cite{natiq2018designing,natiq2019can}. However, multistability behavior, as a new research direction in chaos theory, is still in its infancy. Therefore, the chaotic systems with multistability behaviors require further research.

Chaos control is an attractive theoretical subject and very important in studying chaotic dynamics and their applications, which can be distinguished into two categories: directing a chaotic system to a desired dynamical properties, which is known as chaotification or anti-control of chaos \cite{chen2003chaotification,yu2003chaotic}; developing control strategies to suppress the chaotic behavior when it is harmful \cite{christini1996using}. In \cite{chen1999yet}, the non-chaotic state of Lorenz system is derived to the chaotic state via a simple linear partial state-feedback controller, which has led to the discovery of Chen system. In \cite{chen2006generating}, the hyperchaotic L\"{u} attractor is constructed based on L\"{u} system by using a state feedback controller. In \cite{qi2006analysis}, a constant control is introduced on a 4D autonomous system to generate coexisting asymmetric double-wing chaotic attractors as well as  coexisting two single-wing chaotic attractors. In \cite{sharma2015controlling}, a linear augmentation control is used for stabilizing a chaotic system with hidden attractors to a fixed point state even when the original chaotic system has no fixed points. In  \cite{tahir2015novel}, a chaotic system with a butterfly hidden attractor is generated by using a state feedback controller. In \cite{zhang2018generating}, a simple state feedback controller is introduced on the 3D L\"{u} system to construct a 4D hidden chaotic attractor with either no equilibria or a line of equilibria. In \cite{yadav2018control}, a feedback for an initial duration of time is proposed for controlling chaotic systems with extreme multistability behaviors. 

Motivated by Ref. \cite{qi2006analysis} with using nonlinear controllers, in this paper, we address the degenerating and breaking the butterfly attractor of 3D plasma perturbation model. Of most interest is how to choose nonlinear controllers, in which these controllers can shift the equilibria of plasma perturbation model apart from each other. That means, the spacing of butterfly attractor are changed, and then the double-wing is degenerated into one or two symmetric pair of coexisting chaotic attractors, or one or two symmetric pair of coexisting period-2 limit cycles. Furthermore, the unusual and striking dynamic behavior of coexisting two point attractors with butterfly attractor, coexisting transient transition chaos with quasi-periodic behavior, and coexisting symmetric Hopf bifurcations are revealed numerically.

The rest of this paper is organized as follows: Section~\ref{section:section2} presents the 3D plasma perturbation model. Section~\ref{section:section3} introduces the proposed nonlinear controllers. In Section~\ref{section:section4}, we investigate the effect of shifting the equilibria of the system. In Section~\ref{section:section5}, the coexistence of various attractors are analyzed numerically. The conclusions are presented in Section~\ref{section:section6}.

\begin{figure}[t]
	\centering
	\begin{subfigure}[h]{0.4\textwidth}
		\includegraphics[width=8cm, height=5cm]{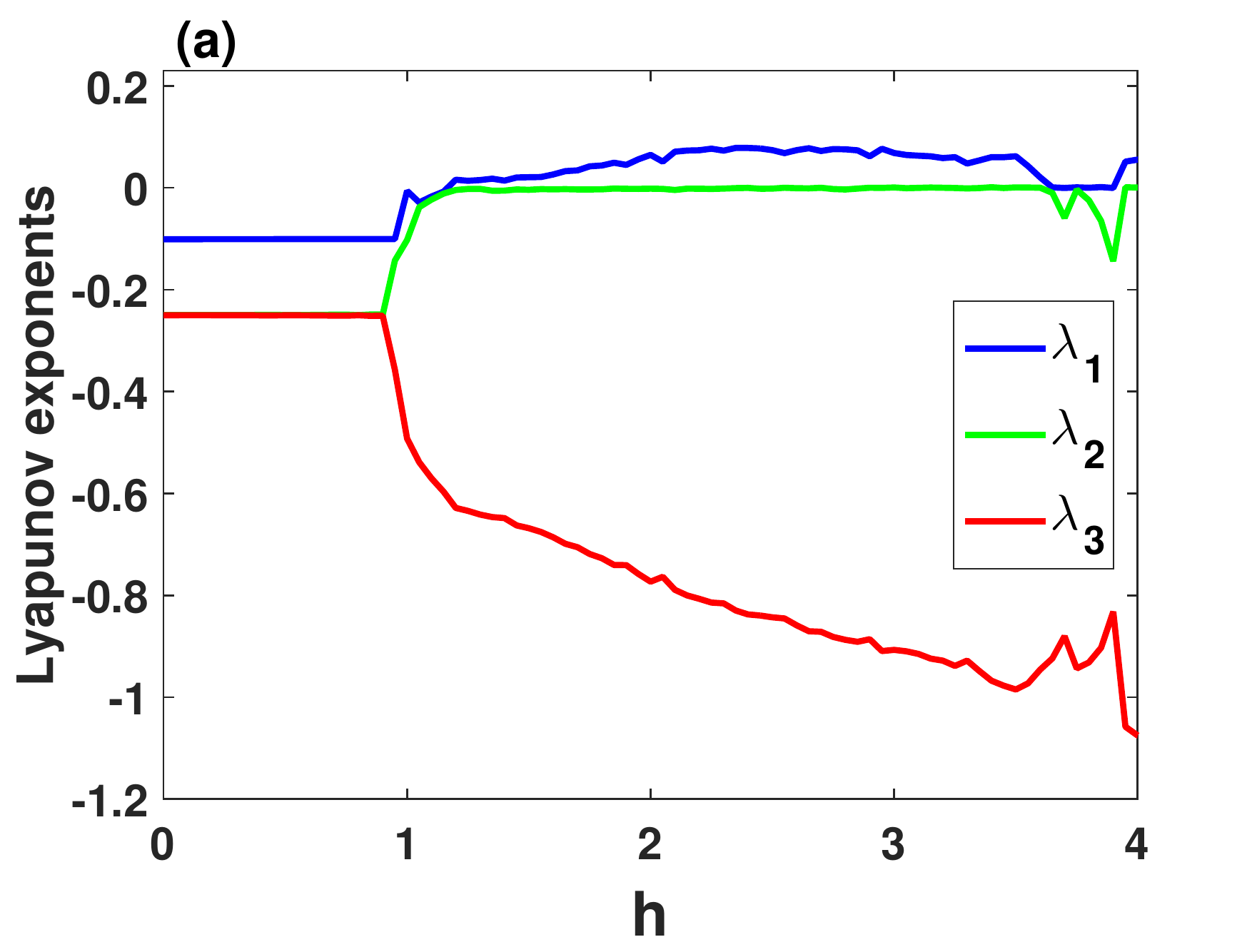}
	\end{subfigure}
	
	\begin{subfigure}[h]{0.4\textwidth}
		\includegraphics[width=8cm, height=5cm]{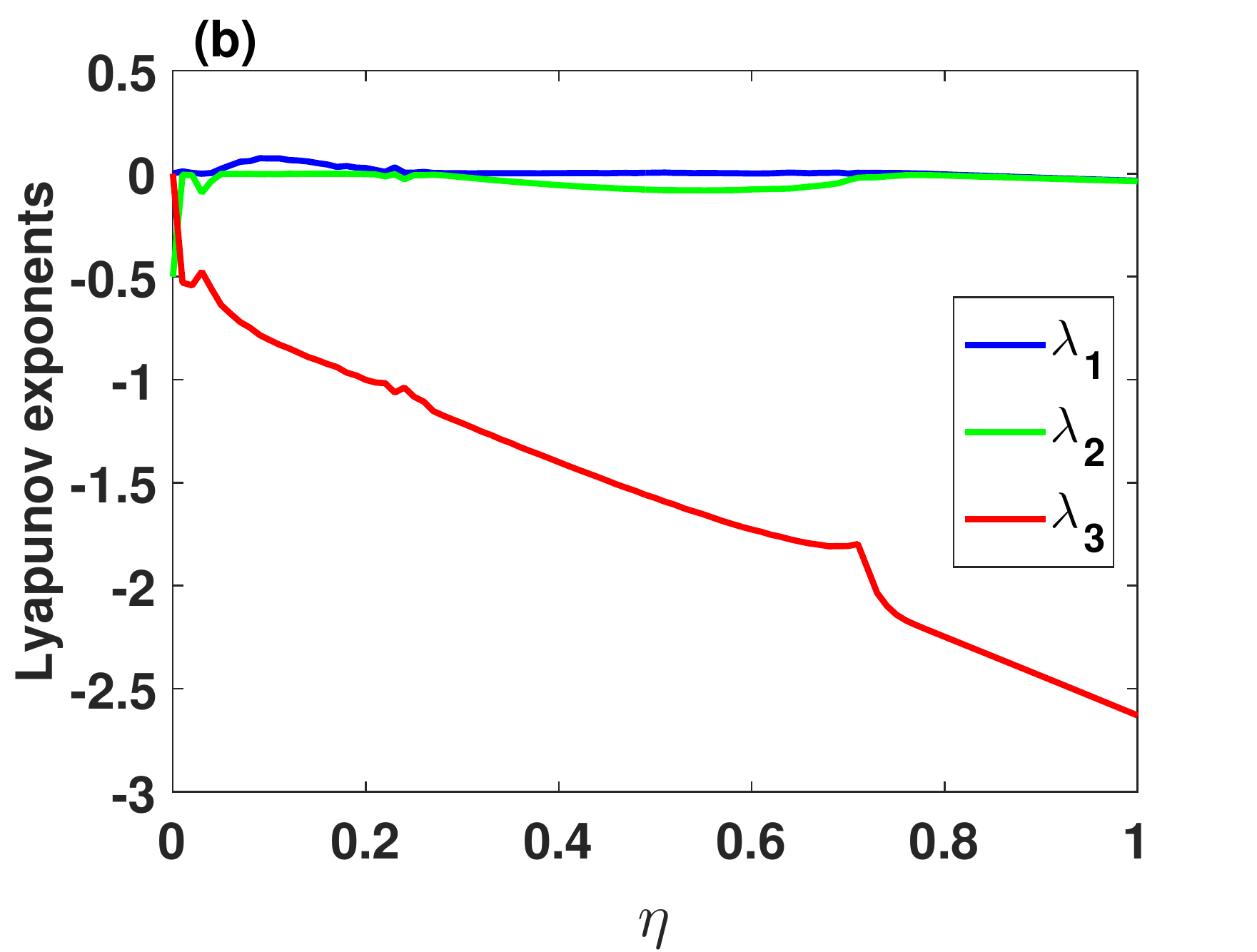}
	\end{subfigure}
	
	\begin{subfigure}[h]{0.4\textwidth}
		\includegraphics[width=8cm, height=5cm]{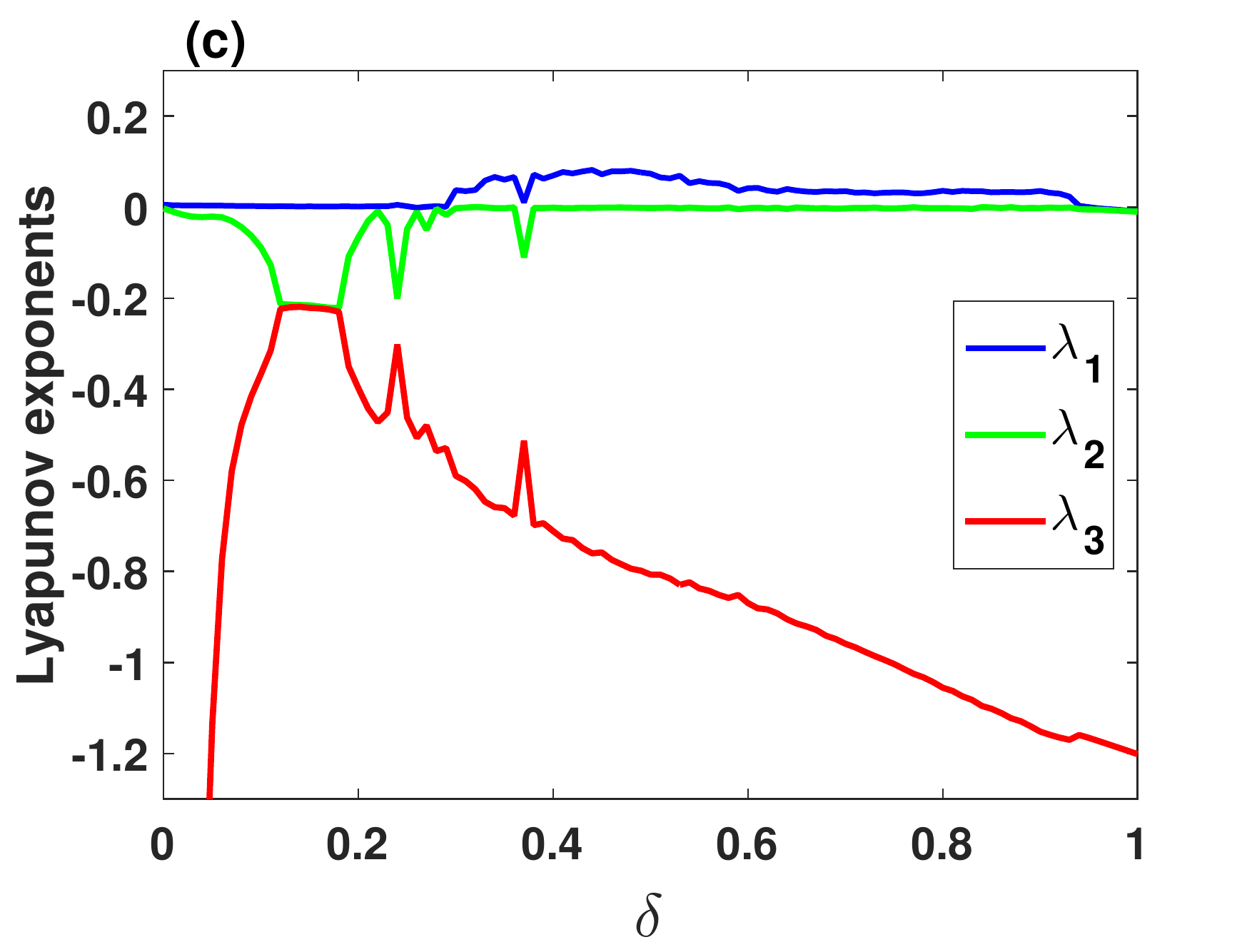}
	\end{subfigure}
	\centering
	\caption{The Lyapunov exponents of the system~(\ref{2}) for the initial conditions $ (1, 0.5, 1) $ versus parameter varying: (a) $ \delta=0.5 $, $ \eta=0.1 $ and $0 \leq h\leq 4$; (b) $ \delta=0.5 $, $ h=2.2 $ and $0 \leq \eta \leq 1$; (c) $ \eta=0.1 $, $ h=2.2 $ and $0 \leq \delta \leq 1$.}
	\label{fig:lya1}
\end{figure}

\begin{figure*}[t]
	\centering
	\begin{subfigure}[h]{0.32\textwidth}
		\includegraphics[width=6.3cm, height=6.5cm]{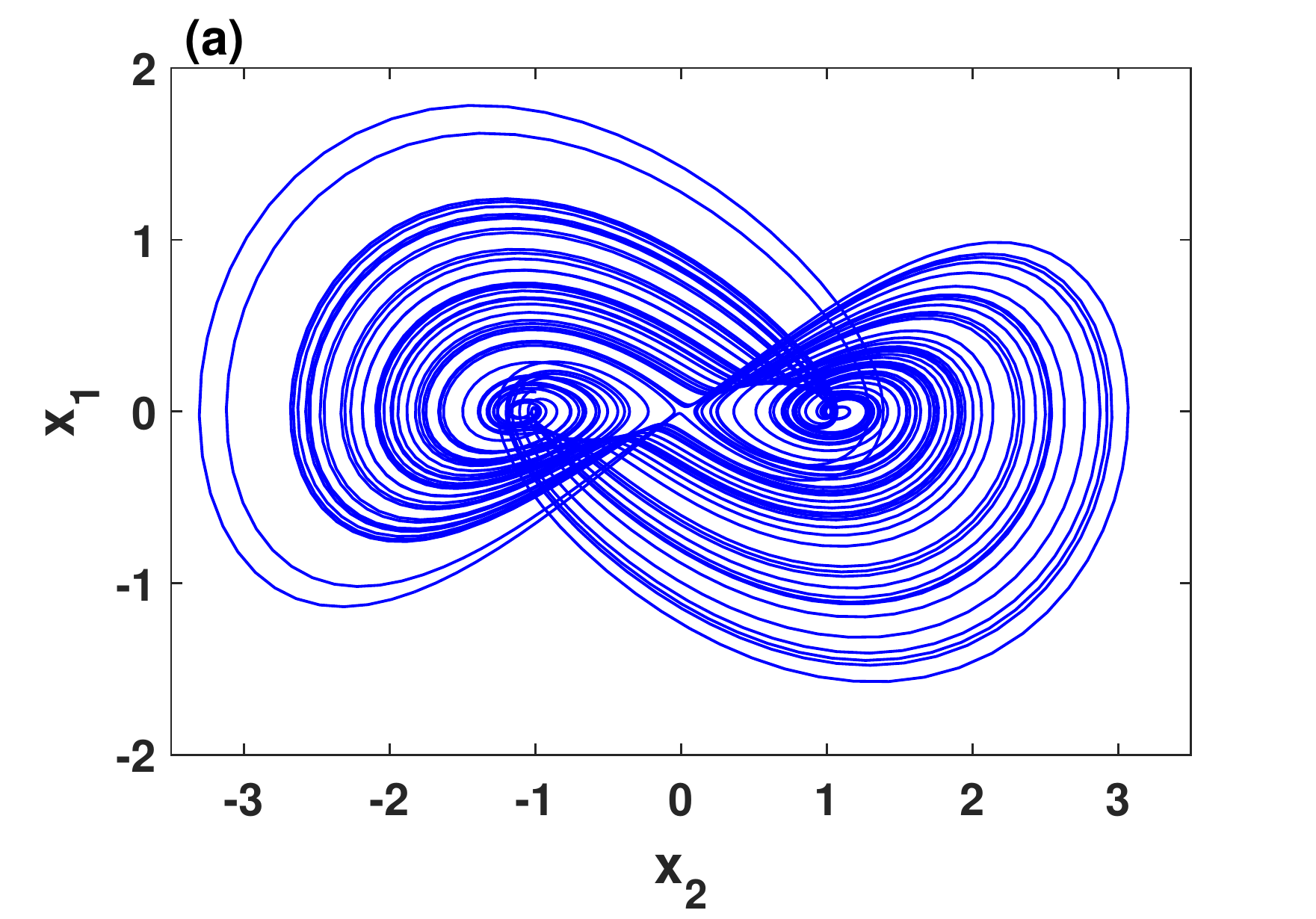}
	\end{subfigure}
	\begin{subfigure}[h]{0.32\textwidth}
		\includegraphics[width=6.3cm, height=6.5cm]{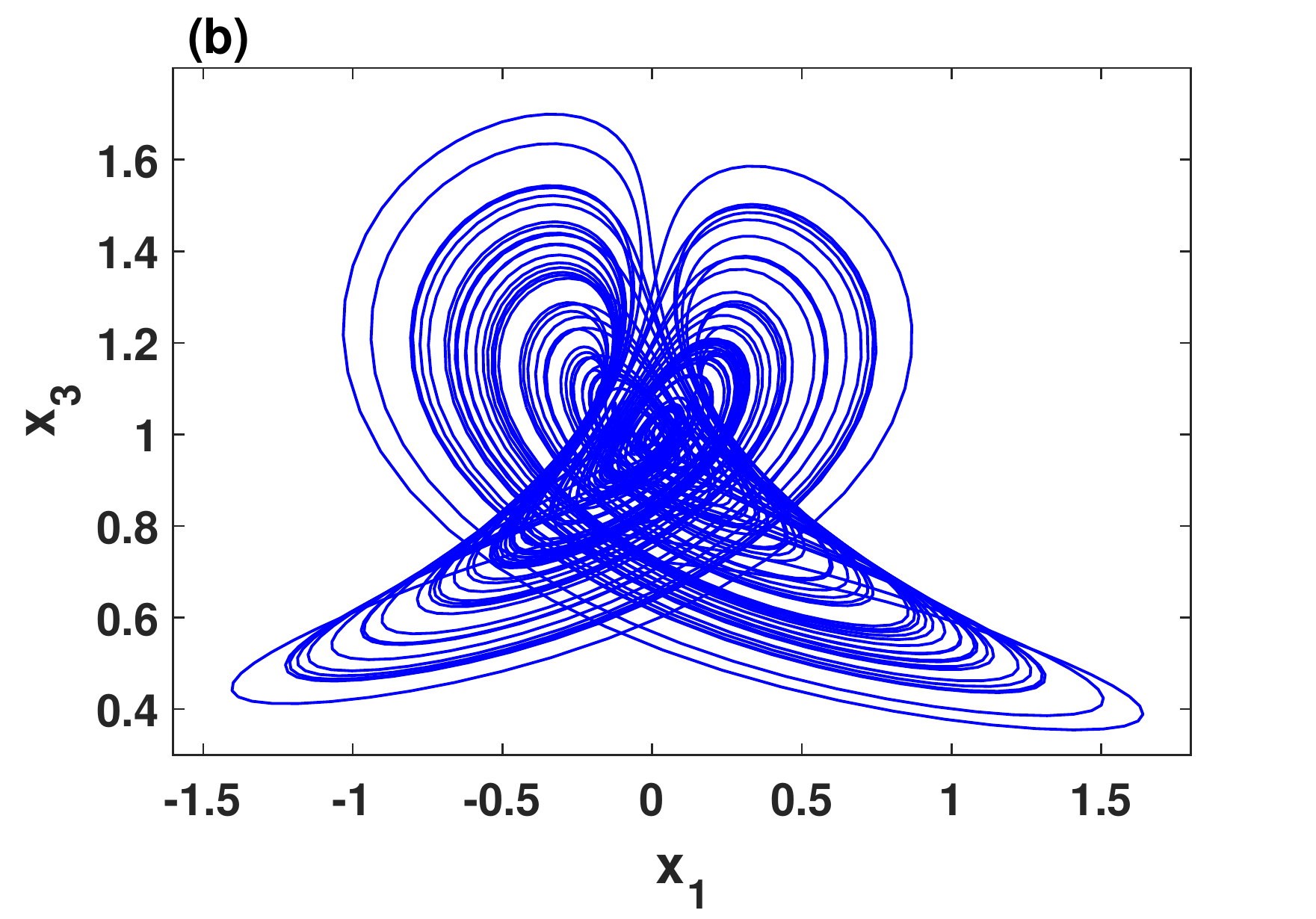}
	\end{subfigure}
	\begin{subfigure}[h]{0.32\textwidth}
		\includegraphics[width=6.3cm, height=6.5cm]{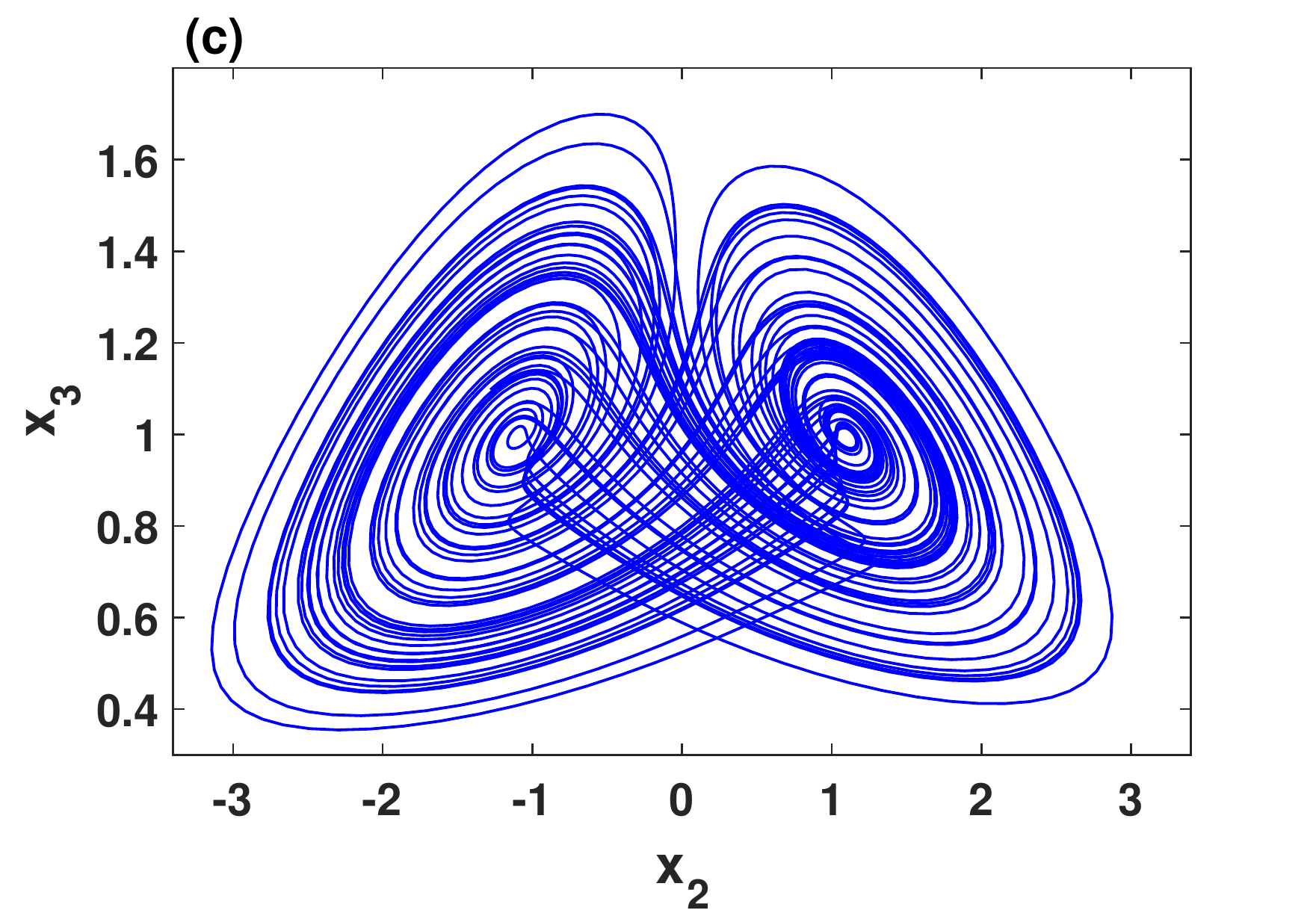}
	\end{subfigure}
	\caption{Butterfly attractor of the system~(\ref{2}) with the parameters $ h = 2.2 $ , $ \eta= 0.1 $, and $ \delta= 0.5 $ for the initial conditions $ (1, 0.5, 1) $.}
	\centering
	\label{fig:phase1}
\end{figure*}

\begin{figure}[!b]
	\centering 
	\includegraphics[width=7cm, height=5cm]{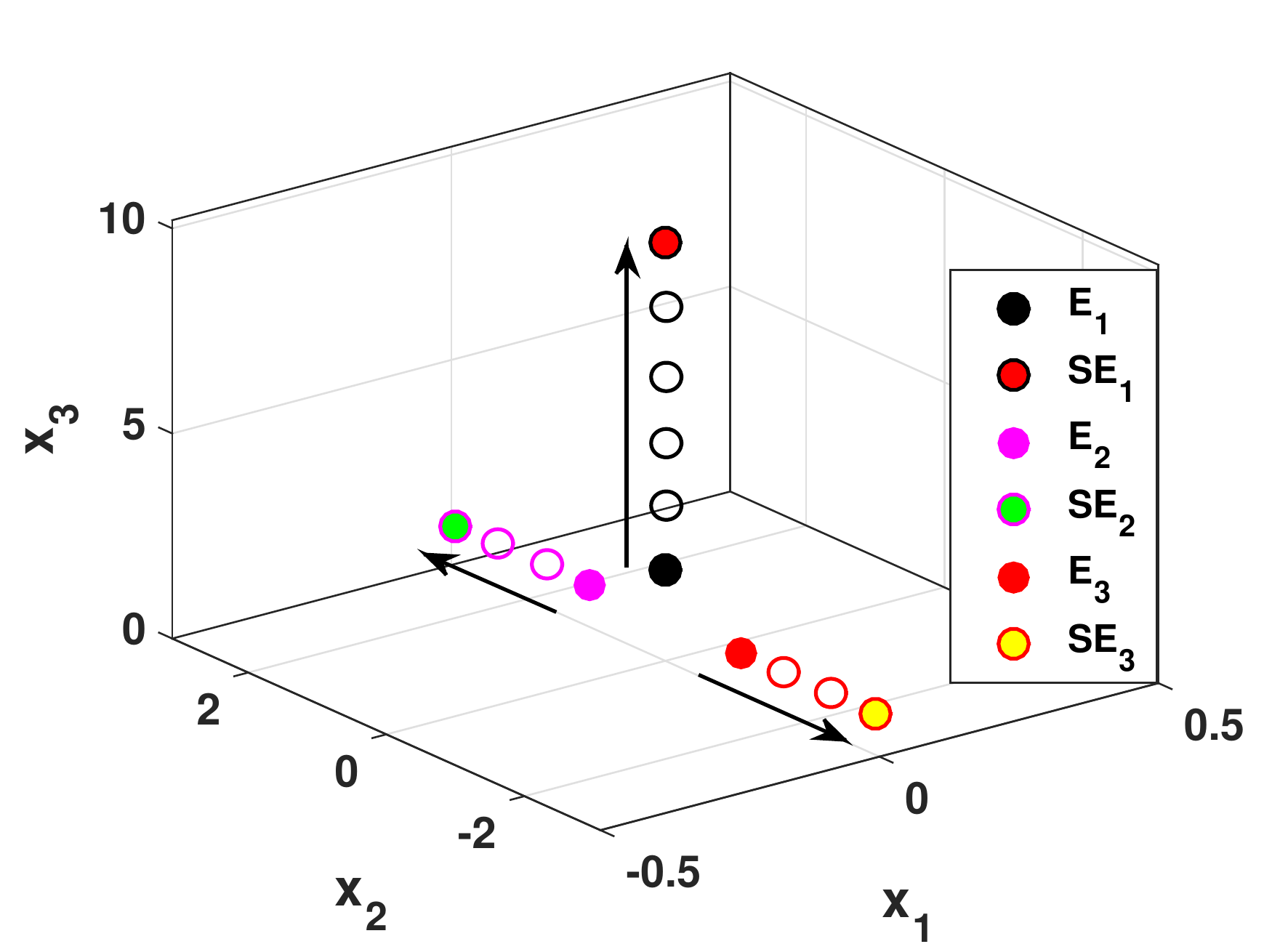}
	\caption{The shifted equilibria diagram: $ E_{1} $, $ E_{2} $, $ E_{3} $ are the equilibria of the system~(\ref{2}); and the corresponding shifted equilibria are $ SE_{1}  $, $ SE_{2} $, $ SE_{3} $, respectively.}
	\centering
	\label{fig:fixed}
\end{figure}

\begin{figure*}[t]
	\centering
	\begin{subfigure}[h]{0.31\textwidth}
		\includegraphics[width=6.1cm, height=3.83cm]{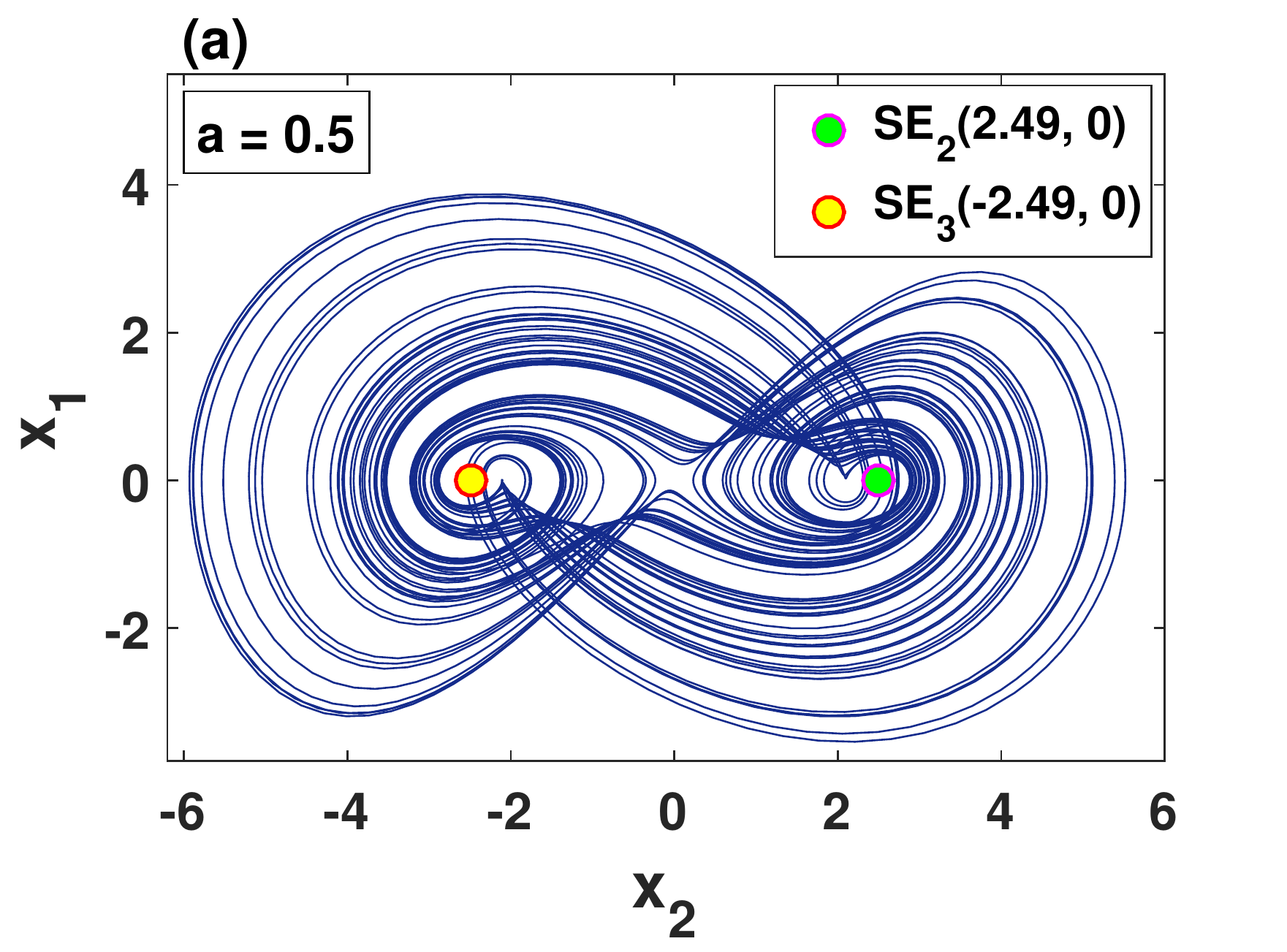}
	\end{subfigure}
	\centering
	\begin{subfigure}[h]{0.31\textwidth}
		\includegraphics[width=6.1cm, height=3.83cm]{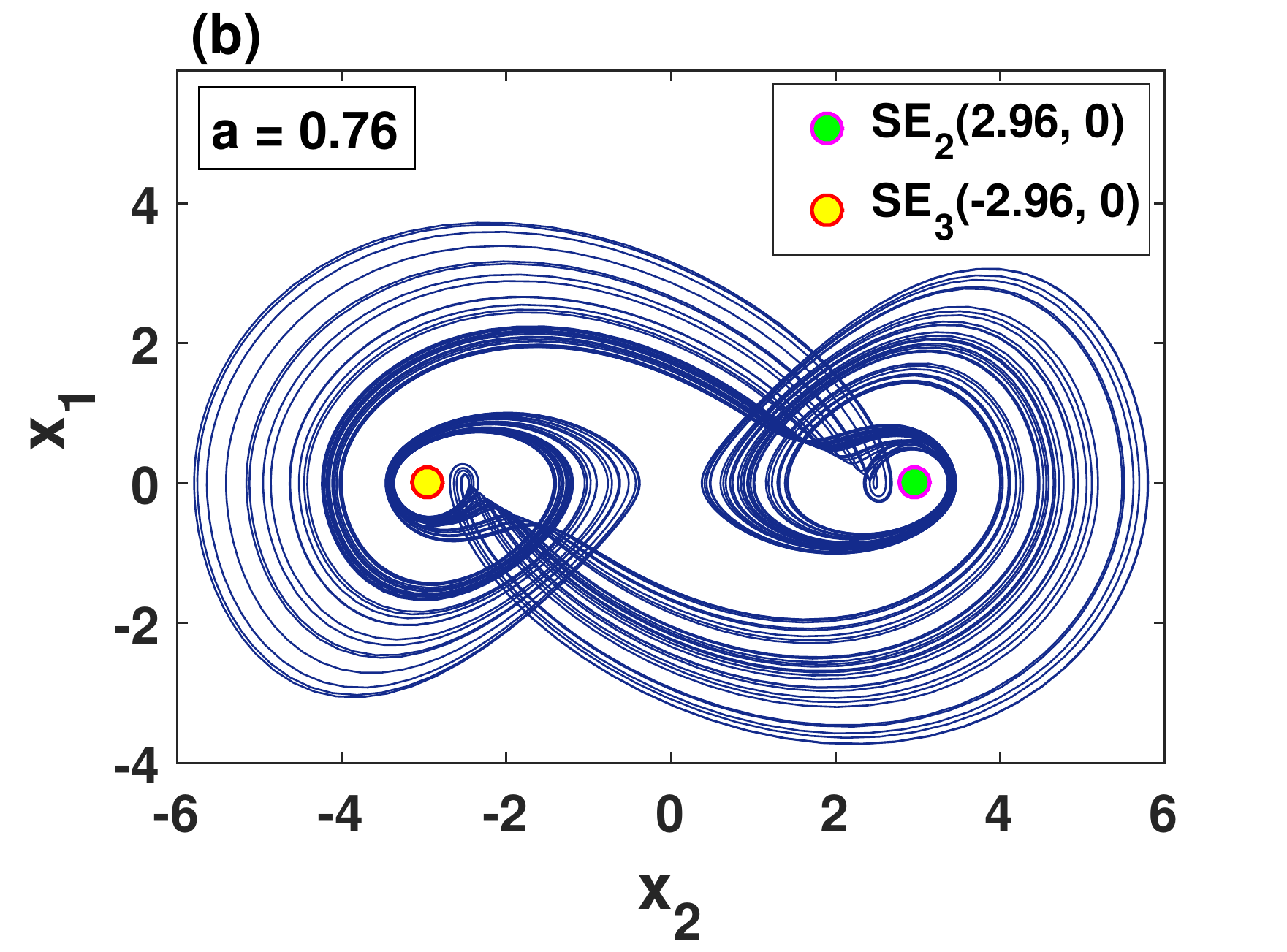}
	\end{subfigure}
	\centering
	\begin{subfigure}[h]{0.31\textwidth}
		\includegraphics[width=6.1cm, height=3.83cm]{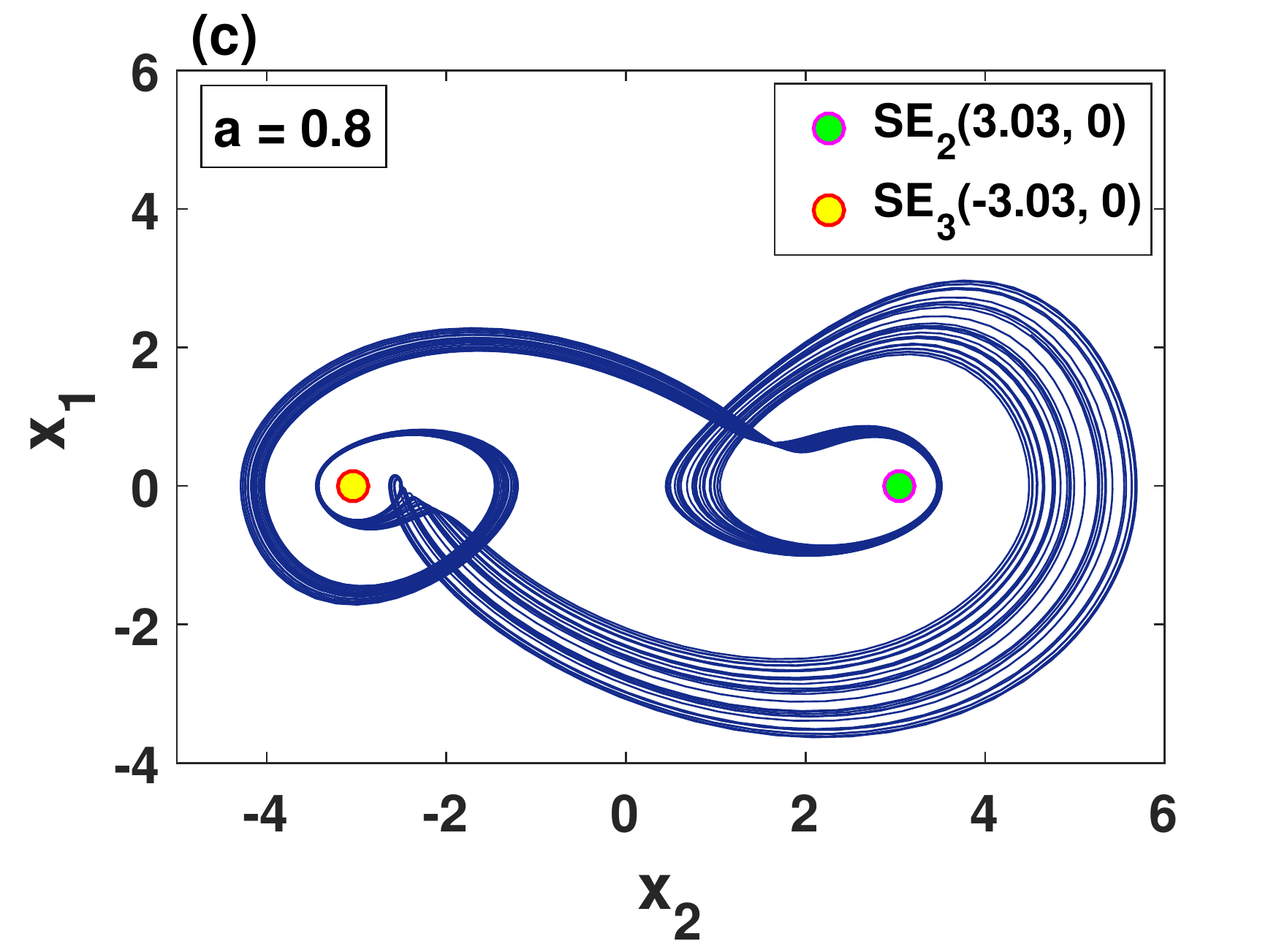}
	\end{subfigure}

	\begin{subfigure}[h]{0.31\textwidth}
		\includegraphics[width=6.1cm, height=3.83cm]{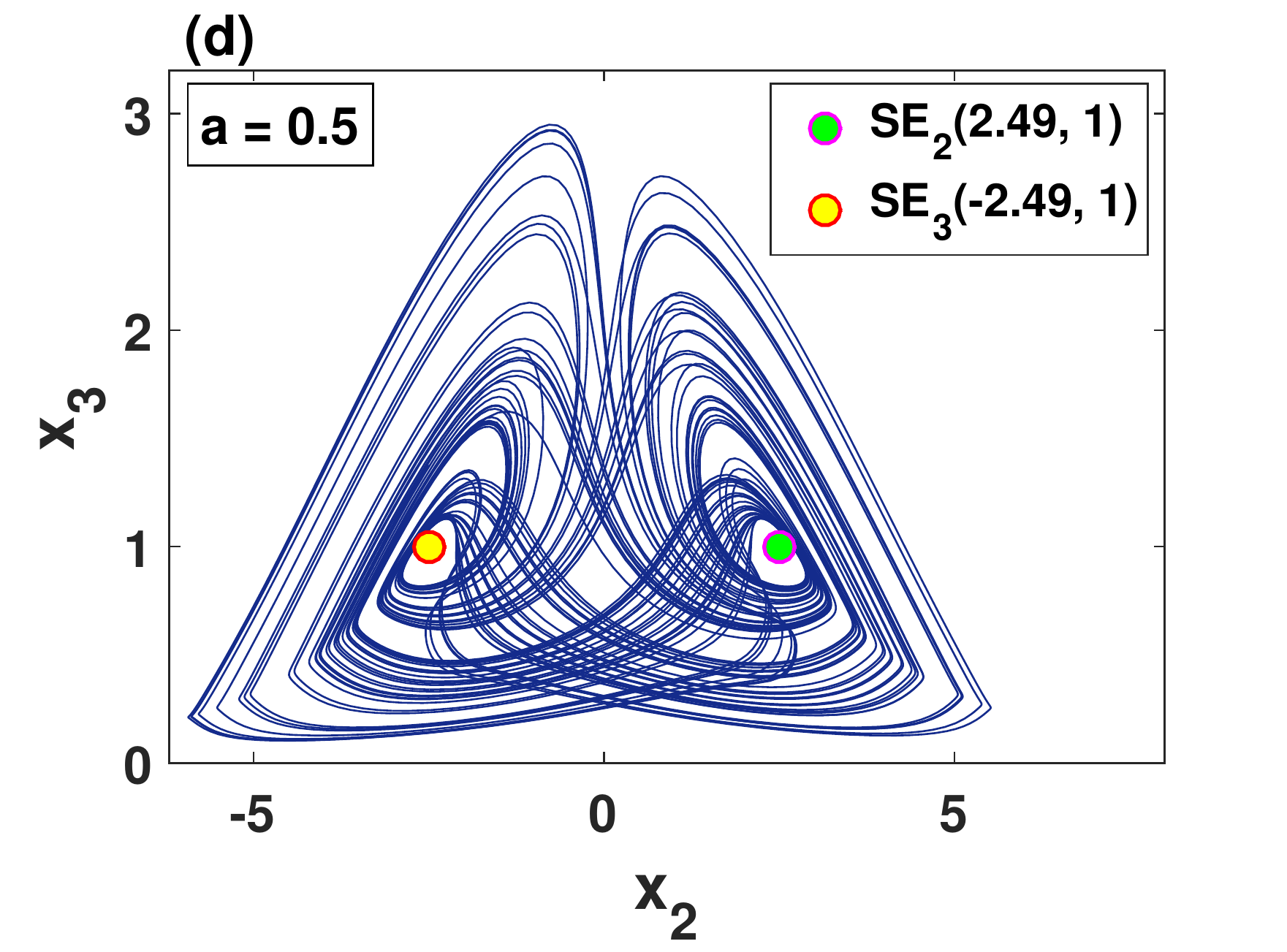}
	\end{subfigure}
	\centering
	\begin{subfigure}[h]{0.31\textwidth}
		\includegraphics[width=6.1cm, height=3.83cm]{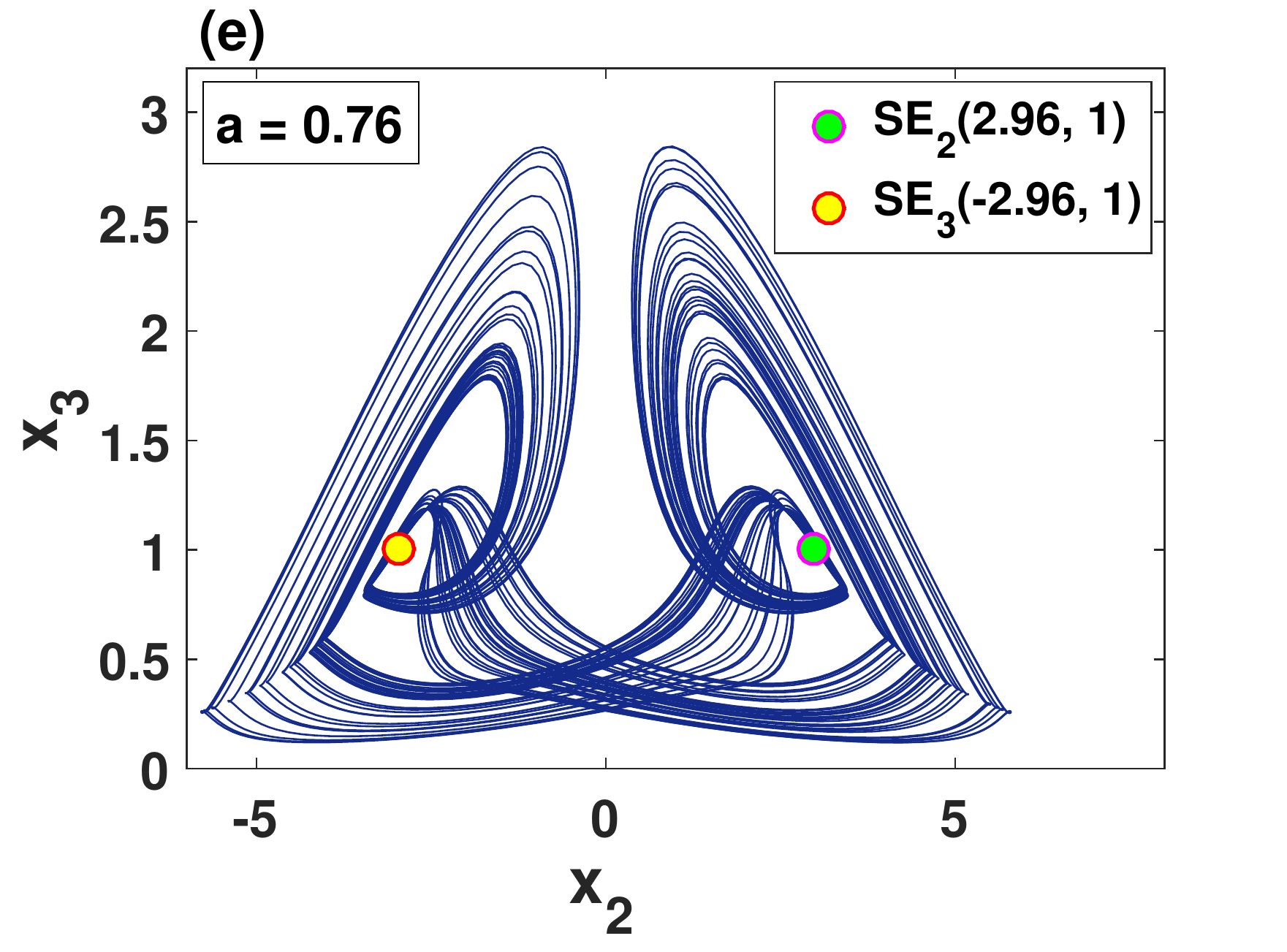}
	\end{subfigure}
	\centering
	\begin{subfigure}[h]{0.31\textwidth}
		\includegraphics[width=6.1cm, height=3.83cm]{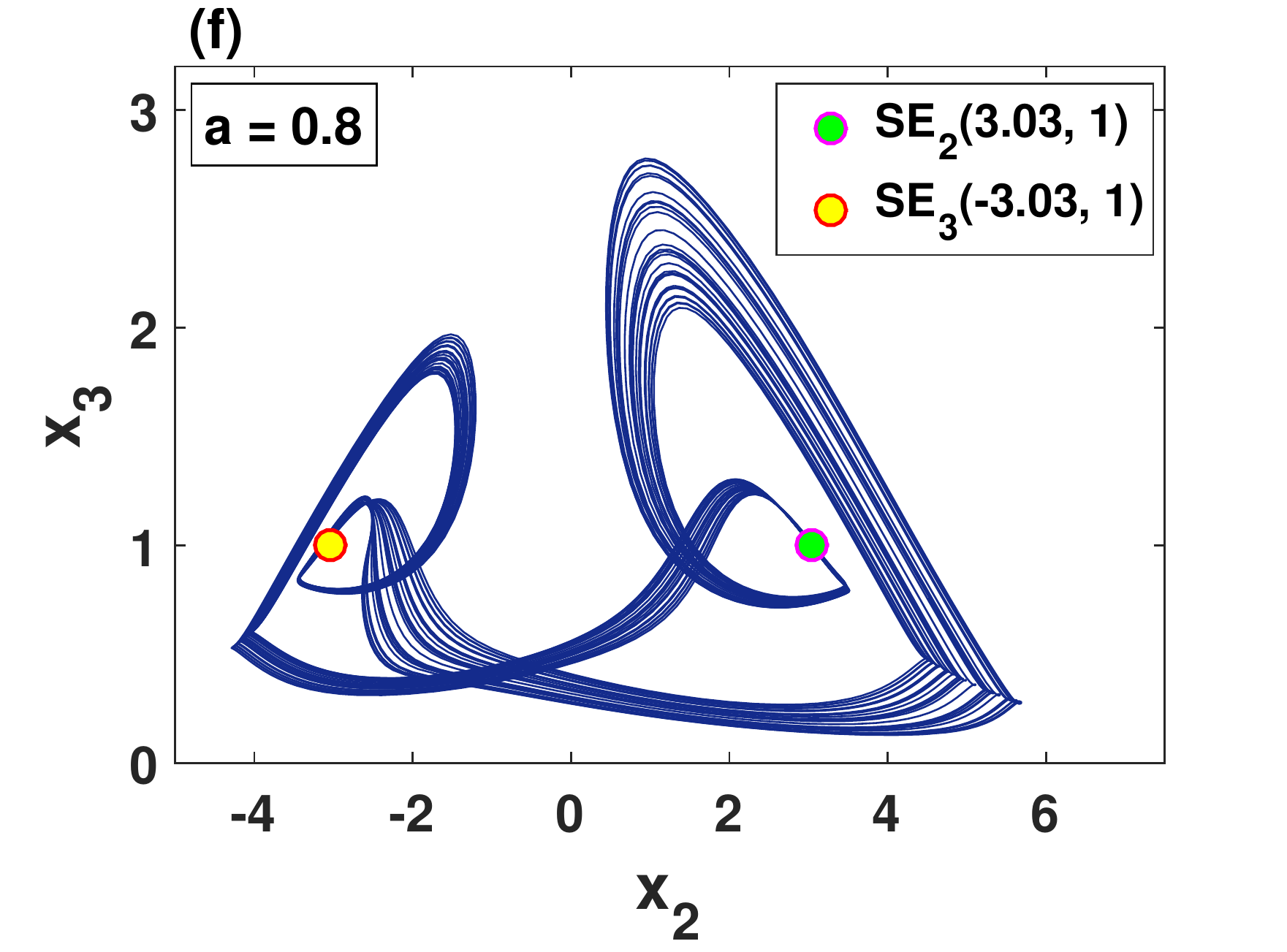}
	\end{subfigure}
	
	\begin{subfigure}[h]{0.31\textwidth}
		\includegraphics[width=6.1cm, height=3.83cm]{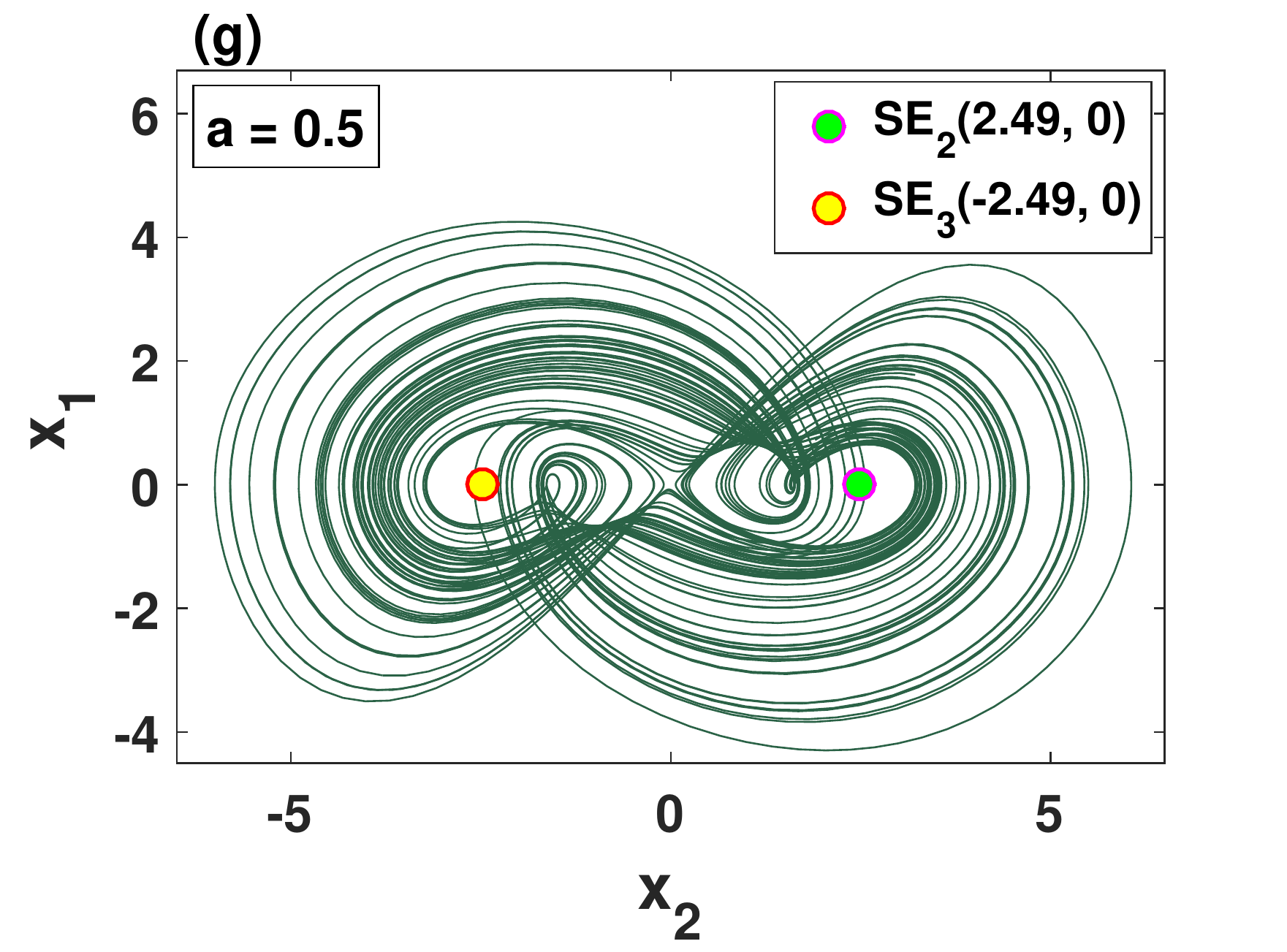}
	\end{subfigure}
	\centering
	\begin{subfigure}[h]{0.31\textwidth}
		\includegraphics[width=6.1cm, height=3.83cm]{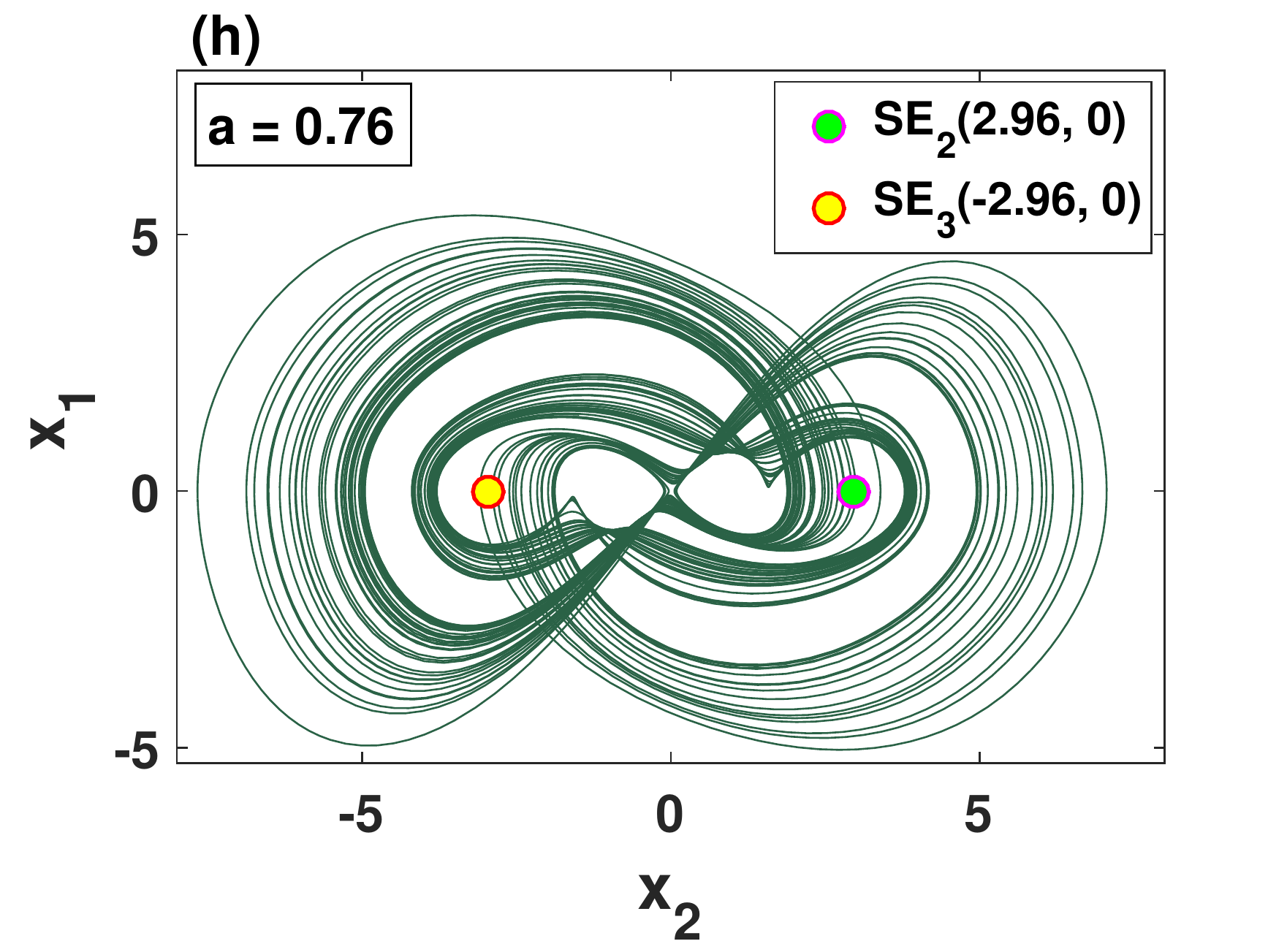}
	\end{subfigure}
	\centering
	\begin{subfigure}[h]{0.31\textwidth}
		\includegraphics[width=6.1cm, height=3.83cm]{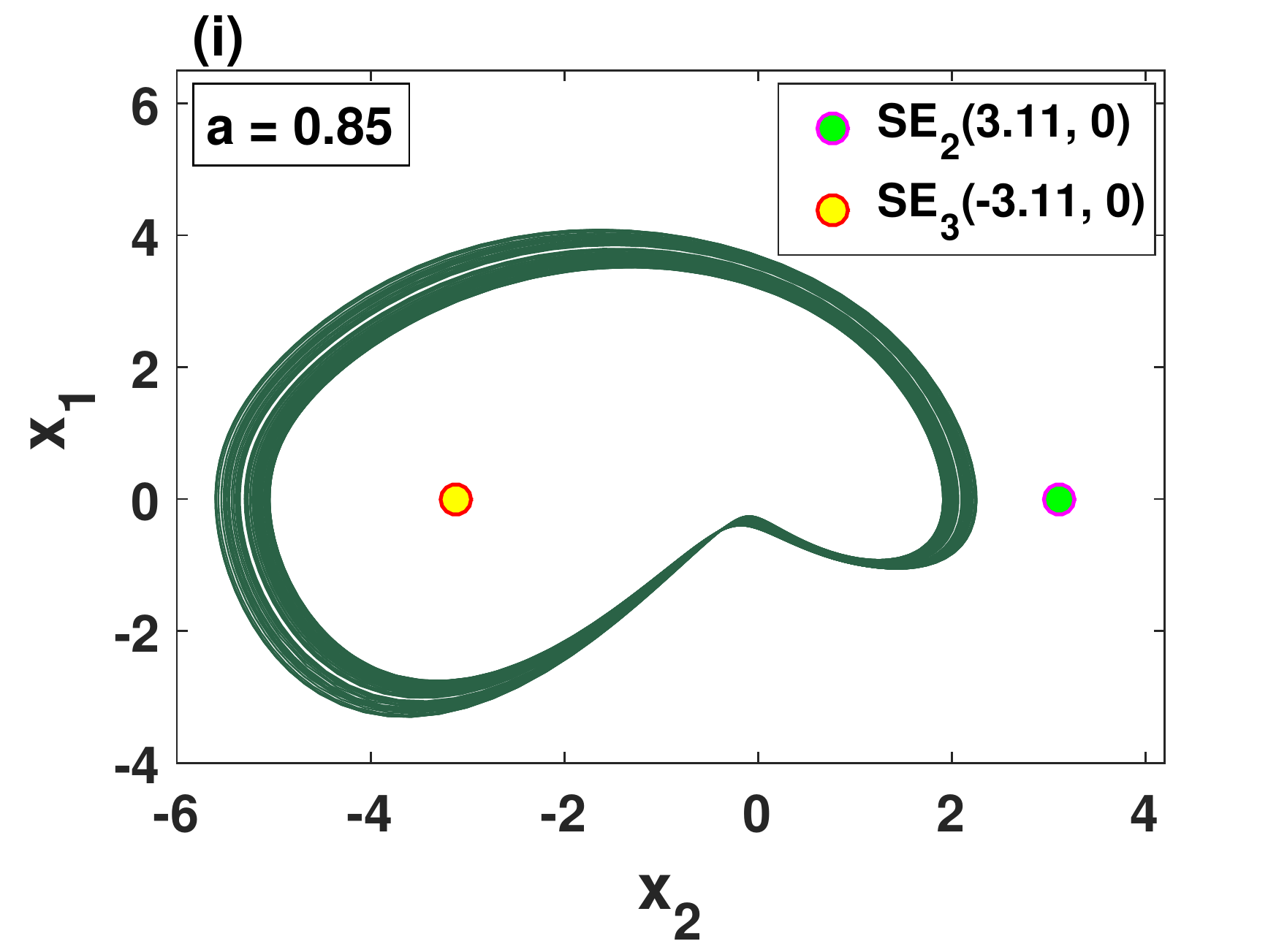}
	\end{subfigure}
	
	\begin{subfigure}[h]{0.31\textwidth}
		\includegraphics[width=6.1cm, height=3.83cm]{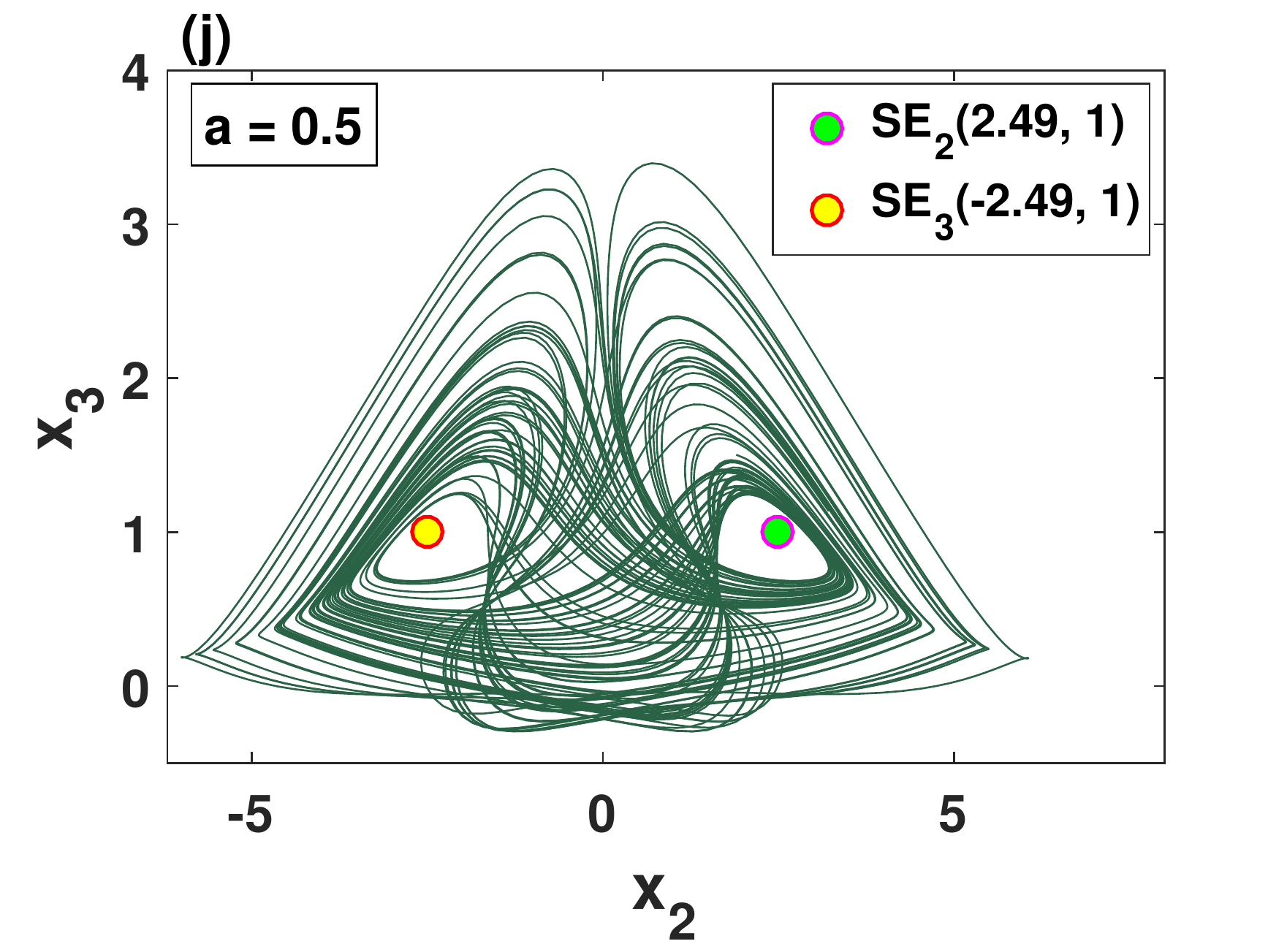}
	\end{subfigure}
	\centering
	\begin{subfigure}[h]{0.31\textwidth}
		\includegraphics[width=6.1cm, height=3.83cm]{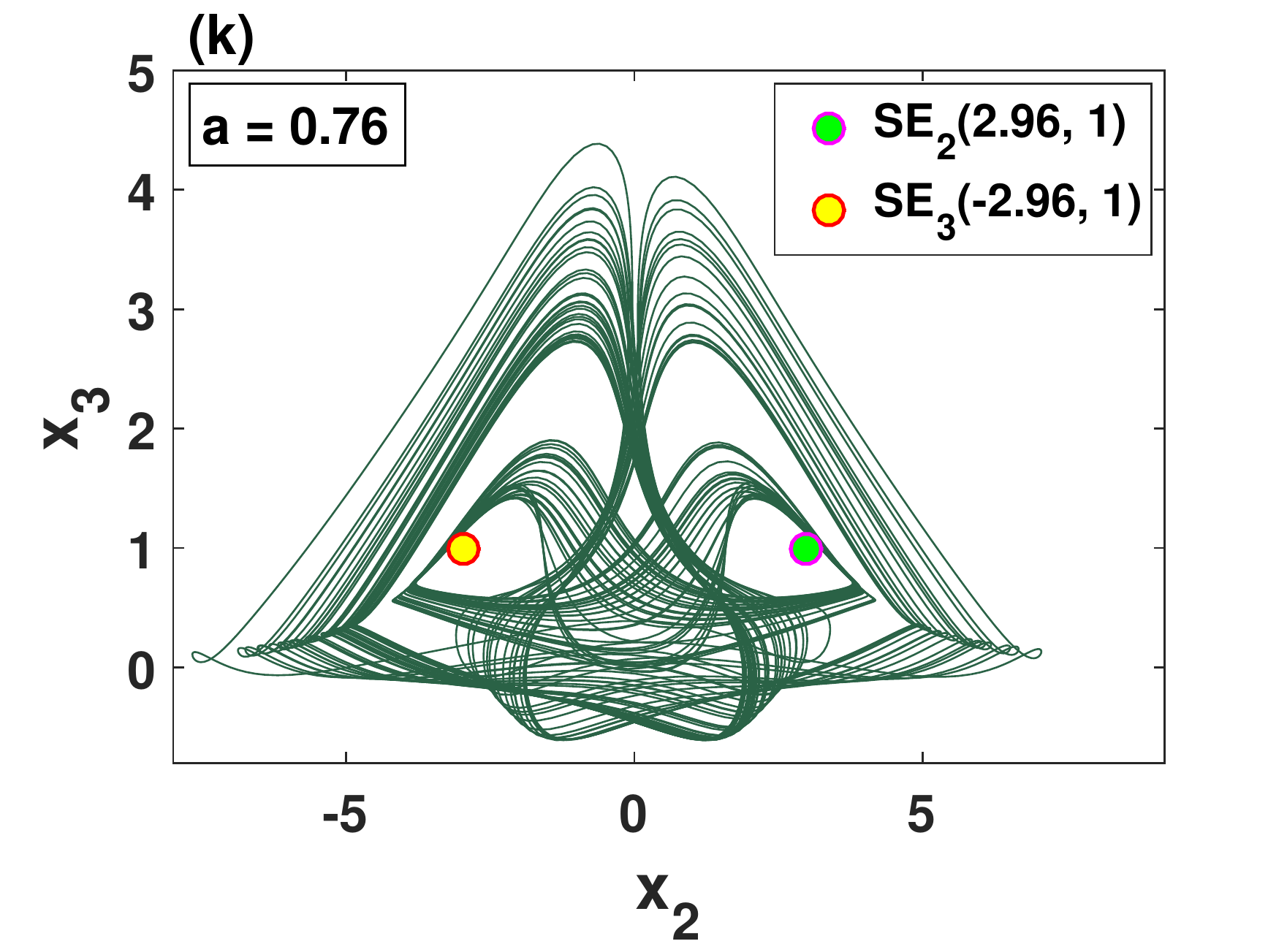}
	\end{subfigure}
	\centering
	\begin{subfigure}[h]{0.31\textwidth}
		\includegraphics[width=6.1cm, height=3.83cm]{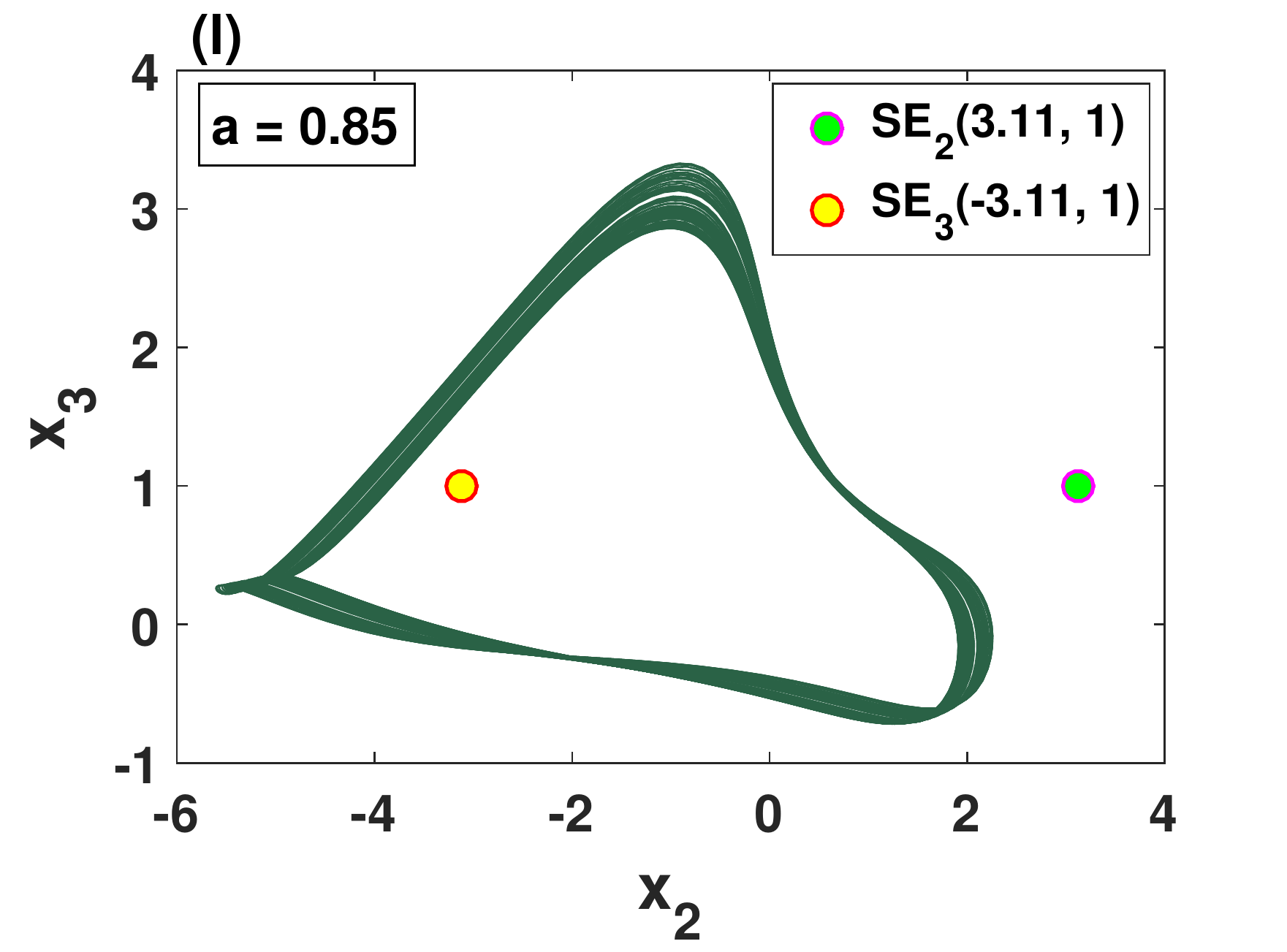}
	\end{subfigure}
	\caption{Process of degenerating the butterfly attractor, around the shifted symmetric equilibria (yellow and green), when the parameters $ \delta=0.5 $, $ \eta=0.1 $, $ h=2.2 $ and $ a $ varying: (a)-(f) with the controller $ G_{1}(x_{1}) $ and for the initial conditions $ (-2, 0.01, -1) $; (g)-(l) with the controller $ G_{2}(x_{1}) $ and for the initial conditions $ (-2, 0.099, -1) $.}
	\label{fig:phase2}
\end{figure*}

\section{The plasma perturbation model}
\label{section:section2}
Before going to the characteristic properties of the plasma perturbation model, we describe shortly its basic equations following Ref. \cite{constantinescu2011low} to derive the three coupled first-order ordinary differential equations. The starting point is the low-dimensional system of plasma perturbation, which is given by 
\begin{flalign}
\label{1}
\begin{cases}
\begin{aligned}
\frac{d^{2}}{dt_{n}^{2}}\xi_{n}&=(p_{n}^{\prime}-1) \cdot \xi_{n}-\delta \cdot \frac{d}{dt_{n}}\xi_{n},\\ \\
\frac{d}{dt_{n}}p_{n}^{\prime}&=\eta \cdot (h-p_{n}^{\prime}-\frac{\chi_{anom}}{\chi_{0}} \cdot \xi_{n}^{2} \cdot p_{n}^{\prime}),
\end{aligned}
\phantom{\hspace{1cm}}
\end{cases}
\end{flalign}
where $ \xi_{n} $ represents the normalized magnetic field, $ p_{n}^{\prime} $ is the normalized plasma pressure gradient, and $ \chi $ is the thermal diffusivity.

By introducing the variables $ x_{2}=\sqrt{\frac{\chi_{anom}}{\chi_{0}}} \cdot \xi_{n} $ and $ x_{3}=p_{n}^{\prime} $, the 3D nonlinear dynamical system is written in the following form
\begin{flalign}
\label{2}
\begin{cases}
\begin{aligned}
\frac{dx_{1}}{dt}&=x_{2}(x_{3}-1)-\delta x_{1},\\
\frac{dx_{2}}{dt}&=x_{1},\\
\frac{dx_{3}}{dt}&=\eta(h-x_{3}-x_{2}^2 x_{3}),
\end{aligned}
\phantom{\hspace{2.9cm}}
\end{cases}
\end{flalign}
where $ x_{1} $, $ x_{2} $, $ x_{3} $ are state variables. Here, $ \delta$, $h$, $\eta $ are considered as independent parameters, in which $ \delta $ denotes the relaxation/dissipation of perturbation, $ h $ is the normalized power input into the system, and $ \eta $ represents the characteristic relation between the two heat
diffusion coefficients. The system~(\ref{2}) describes the dynamics of the plasma pressure gradient and the amplitude of the magnetic field displacement.

To better illustrate the dynamics of the system~(\ref{2}), we now give its basic properties. Firstly, for the system (\ref{2}) to become dissipative, the state space contraction given by $ (\delta+\eta+x_{2}^{2})$ must be positive. Secondly, the system~(\ref{2}) is rotationally symmetric under the transformation $ (x_{1}, x_{2}, x_{3}) \longrightarrow (-x_{1}, -x_{2}, x_{3}) $. Finally, the system has three real equilibria $ E_{1}(0,\ 0,\ h) $, $ E_{2}(0,\ \sqrt{h-1},\ 1) $, and $ E_{3}(0,-\ \sqrt{h-1},\ 1) $.

Figure~\ref{fig:lya1} (a)-(c) shows the Lyapunov exponents of the system~(\ref{2})  versus parameters $ h $, $ \eta $, and $ \delta $ varying, respectively. Obviously, the system~(\ref{2}) exhibits periodic, quasi-periodic, and chaotic behaviors. When the parameters are selected as $ h = 2.2 $ , $ \eta= 0.1 $ and $ \delta= 0.5 $ with the initial conditions $ (1,0.5, 1) $, the system~(\ref{2}) shows a double-wing
chaotic attractor that resembles a butterfly, as illustrated in Figure~\ref{fig:phase1}. The corresponding Lyapunov exponents are obtained as $ LE_{1}=0.0736 $, $ LE_{2}=0 $ and $ LE_{3}=-0.8068 $, and the Lyapunov dimension is $ 2.0912 $.

\begin{figure*}[t]
	\centering
	\begin{subfigure}[h]{0.32\textwidth}
		\includegraphics[width=5.8cm, height=5.5cm]{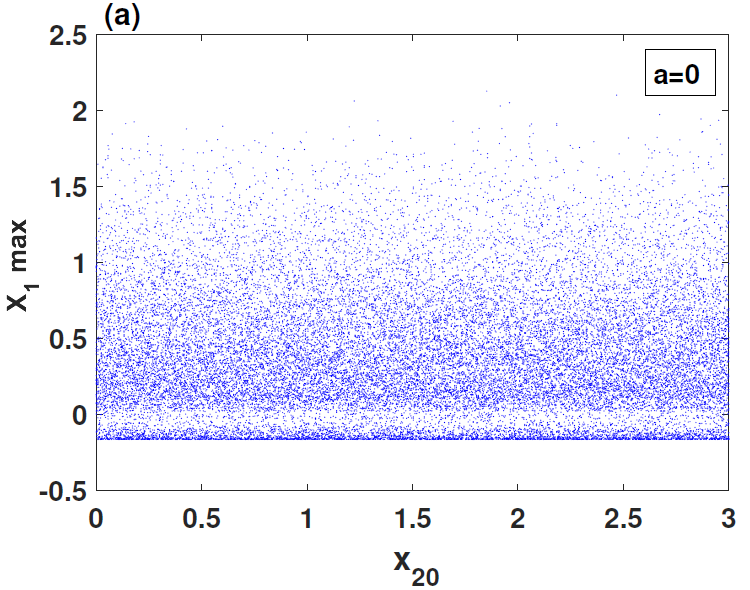}
	\end{subfigure} 
	\begin{subfigure}[h]{0.32\textwidth}
		\includegraphics[width=5.8cm, height=5.5cm]{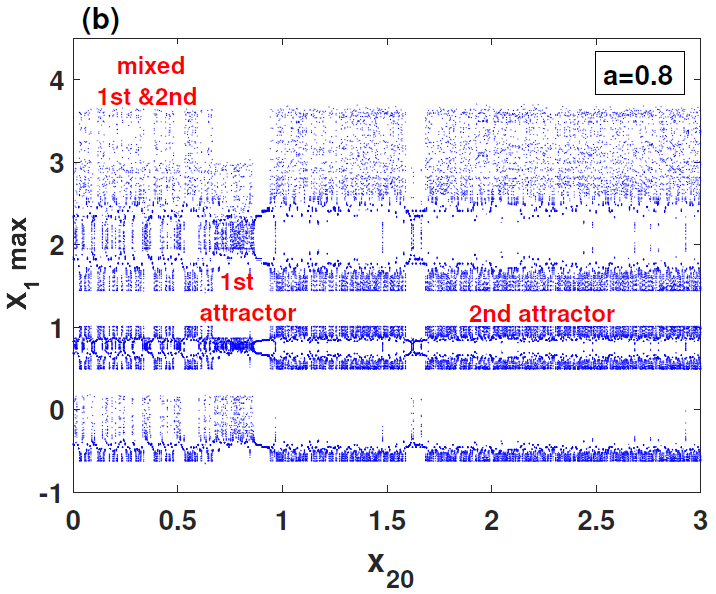}
	\end{subfigure} 
	\begin{subfigure}[h]{0.32\textwidth}
		\includegraphics[width=5.8cm, height=5.5cm]{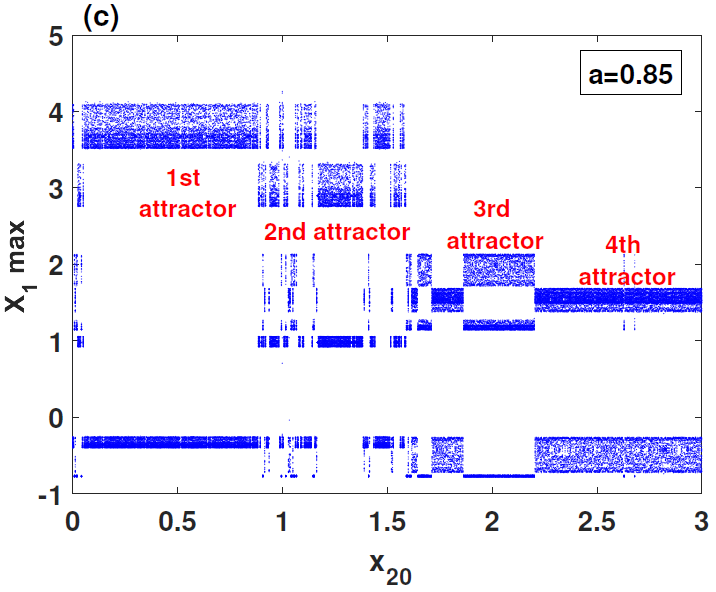}
	\end{subfigure} 
	\caption{Bifurcation diagrams of the system~(\ref{3}) versus initial conditions  $ (-2, x_{20}, -1) $ varying for the parameters $ \delta=0.5 $, $ \eta=0.1 $, $ h=2.2 $: (a) single chaotic attractor when $ a=0 $; (b) two different chaotic attractors coexist for the controller $ G_{1}(x_{1}) $ and when $ a=0.8 $; (c) four different chaotic attractors coexist for the controller $ G_{2}(x_{1}) $ and when $ a=0.85 $.}
	\centering
	\label{fig:bifurcation}
\end{figure*}

\section{Nonlinear controllers}
\label{section:section3}
The system~(\ref{2}) can be written as 
\begin{flalign*}
\begin{cases}
\begin{aligned}
\frac{dx_{1}}{dt}&=f_{1}(x_{1}, x_{2}, x_{3},\delta), \\
\frac{dx_{2}}{dt}&=f_{2}(x_{1}), \\
\frac{dx_{3}}{dt}&=f_{3}(x_{2}, x_{3}, h, \eta)
\end{aligned}
\phantom{\hspace{3.7cm}}
\end{cases}
\end{flalign*}
Firstly, since the variable $ x_{1} $ appears only once in the function $ f_{2}$, hence choosing a nonlinear controller $ G $ with respect to a variable $ x_{1} $ would keep the system with same number of equilibria. Secondly, it is reasonable to assume that the butterfly attractor can be degenerated if the symmetric equilibria of the system are shifted further apart from each other. Therefore, the butterfly attractor becomes one or more pair of coexisting attractors which can be selected by choosing appropriate initial conditions. In order to do that, the nonlinear controller $ G_{i}(x_{1}) $ must be greater than zero at the equilibrium point $ x_{1}=0 $. Finally, all equilibria of the system are symmetric with respect to the $ x_{3} $-axis, thus adding a nonlinear controller arbitrary may break the polarity balance of the symmetric property of the system. Consequently, introducing an even-function controller term to the right-hand side of $ f_{3} $ can preserve the symmetric property of the system. This idea leads to choose $ G_{1}(x_{1})= e^{-x_{1}^{2}}$ or $ G_{2}(x_{1})=cos(x_{1})$  as controllers, where both are nonlinear even-functions, and $ G_{1}(0)= G_{2}(0)=1$.

Thus, the nonlinear controller $ G_{i}(x_{1}) (i=1, 2) $ is added to the third equation of system~(\ref{2}), yielding
\begin{flalign}
\label{3}
\begin{cases}
\begin{aligned}
\frac{dx_{1}}{dt}&=x_{2}(x_{3}-1)-\delta x_{1},\\
\frac{dx_{2}}{dt}&=x_{1},\\
\frac{dx_{3}}{dt}&=\eta(h-x_{3}-x_{2}^2 x_{3})+ a\ G_{i}(x_{1}),
\end{aligned}
\phantom{\hspace{1.2cm}}
\end{cases}
\end{flalign}
where $ a $ is an amplitude controller parameter, and $ G_{i}(x_{1}) $ is chosen to be either $ G_{1}(x_{1})=e^{-x_{1}^{2}}$ or $ G_{2}(x_{1})= cos(x_{1})$.

In order to calculate the equilibria and stability of the system~(\ref{3}), we can choose the parameters $ \delta, h, \eta, a >0  $, then the equilibria of system~(\ref{3}) can be obtained by solving the following equations:
\begin{flalign}
\begin{cases}
\begin{aligned}
&x_{2}(x_{3}-1)-\delta x_{1}=0,\nonumber\\
&x_{1}=0,\nonumber\\
&\eta(h-x_{3}-x_{2}^2 x_{3})+ a\ G_{i}(x_{1})=0.\nonumber
\end{aligned}
\phantom{\hspace{2.3cm}}
\end{cases}
\end{flalign}
Obviously, when $ h+\frac{a}{\eta}\leq1 $, the system (\ref{3}) has only one shifted equilibrium point having the following form:
\begin{flalign*}
\begin{aligned}
&SE_{1}(0,\ 0,\ h+\dfrac{a}{\eta}),\nonumber
\end{aligned}
\phantom{\hspace{4.8cm}}
\end{flalign*}
whereas, the system (\ref{3}) has three real shifted equilibria, when $ h+\frac{a}{\eta}>1 $, as follows:
\begin{flalign*}
\begin{cases}
\begin{aligned}
&SE_{1}(0,\ 0,\ h+\dfrac{a}{\eta}),\nonumber\\
&SE_{2}(0,\ \sqrt{h+\dfrac{a}{\eta}-1},\ 1),\\
&SE_{3}(0,\ -\sqrt{h+\dfrac{a}{\eta}-1},\ 1).
\end{aligned}
\phantom{\hspace{3.2cm}}
\end{cases}
\end{flalign*}
Consequently, the system~(\ref{3}) has the same number of the equilibria of system~(\ref{2}) with the shifting of $ \dfrac{a}{\eta} $ in the $ x_{3} $-axis of $ SE_{1} $, and  $ x_{2} $-axis of $ SE_{2,3} $, as illustrated in Figure~\ref{fig:fixed}.

\begin{figure*}[t]
	\centering
	\begin{subfigure}[h]{0.32\textwidth}
		\includegraphics[width=6cm, height=5.5cm]{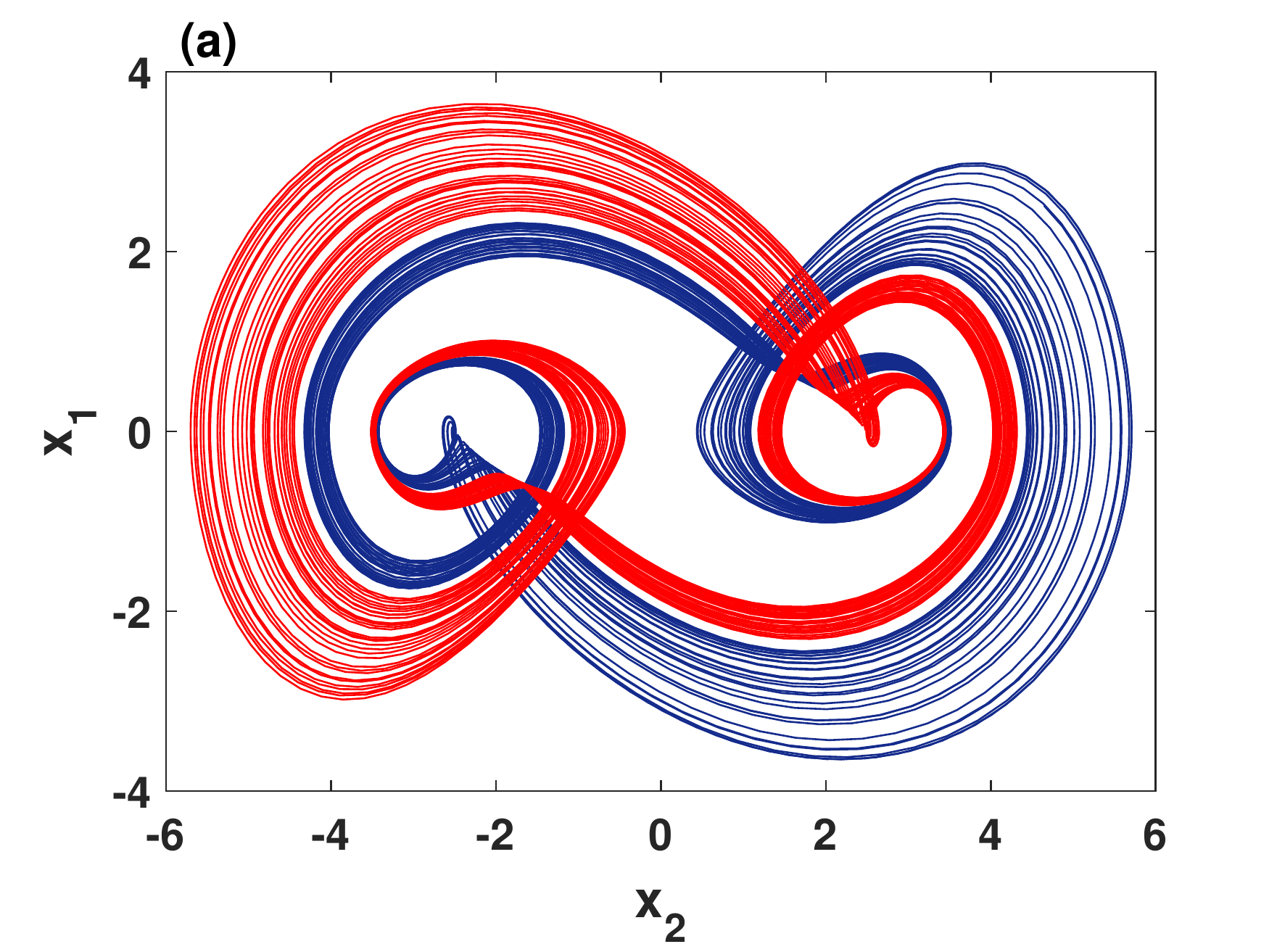}
	\end{subfigure}
	\begin{subfigure}[h]{0.32\textwidth}
		\includegraphics[width=6cm, height=5.5cm]{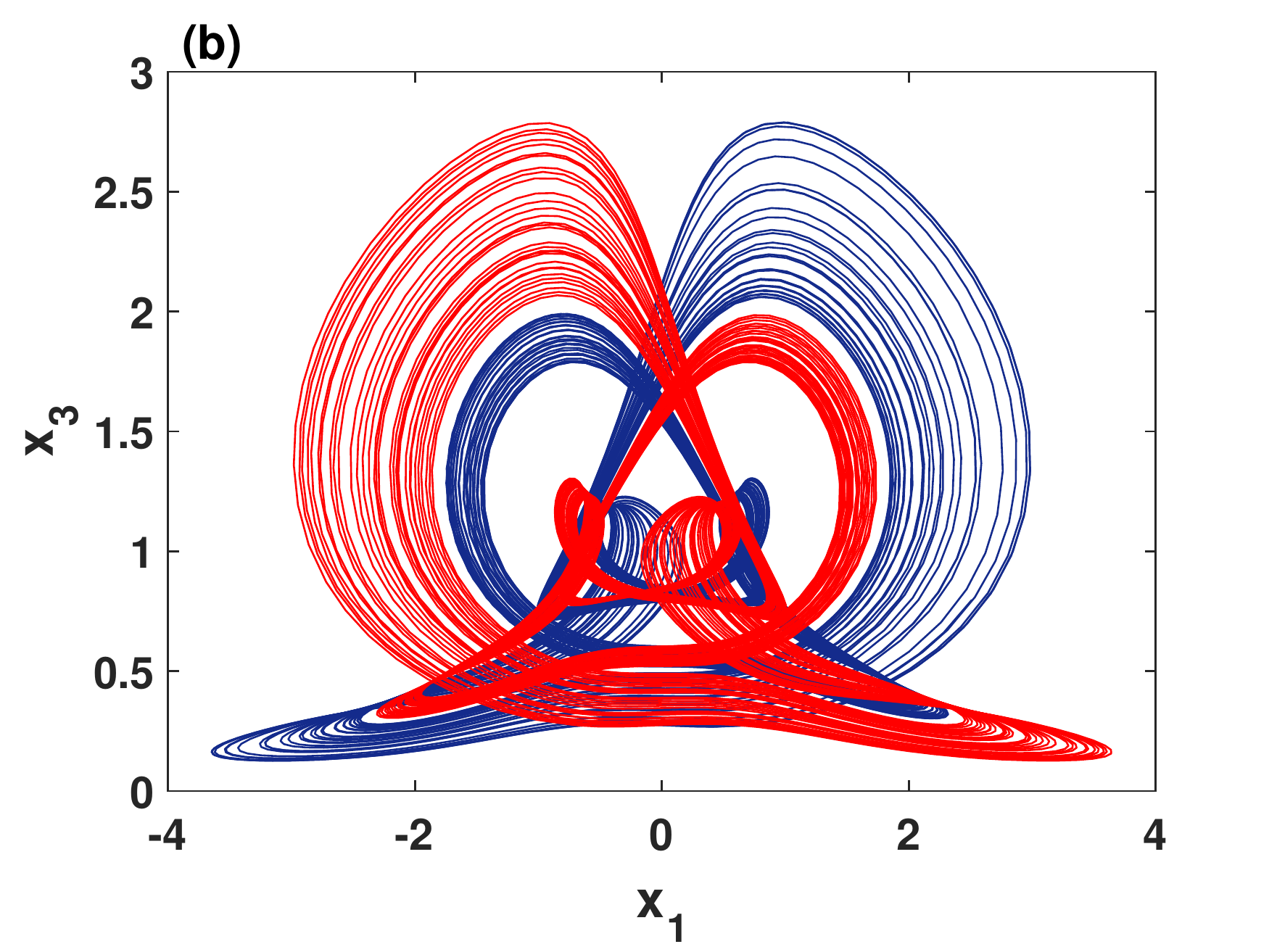}
	\end{subfigure}
	\begin{subfigure}[h]{0.32\textwidth}
		\includegraphics[width=6cm, height=5.5cm]{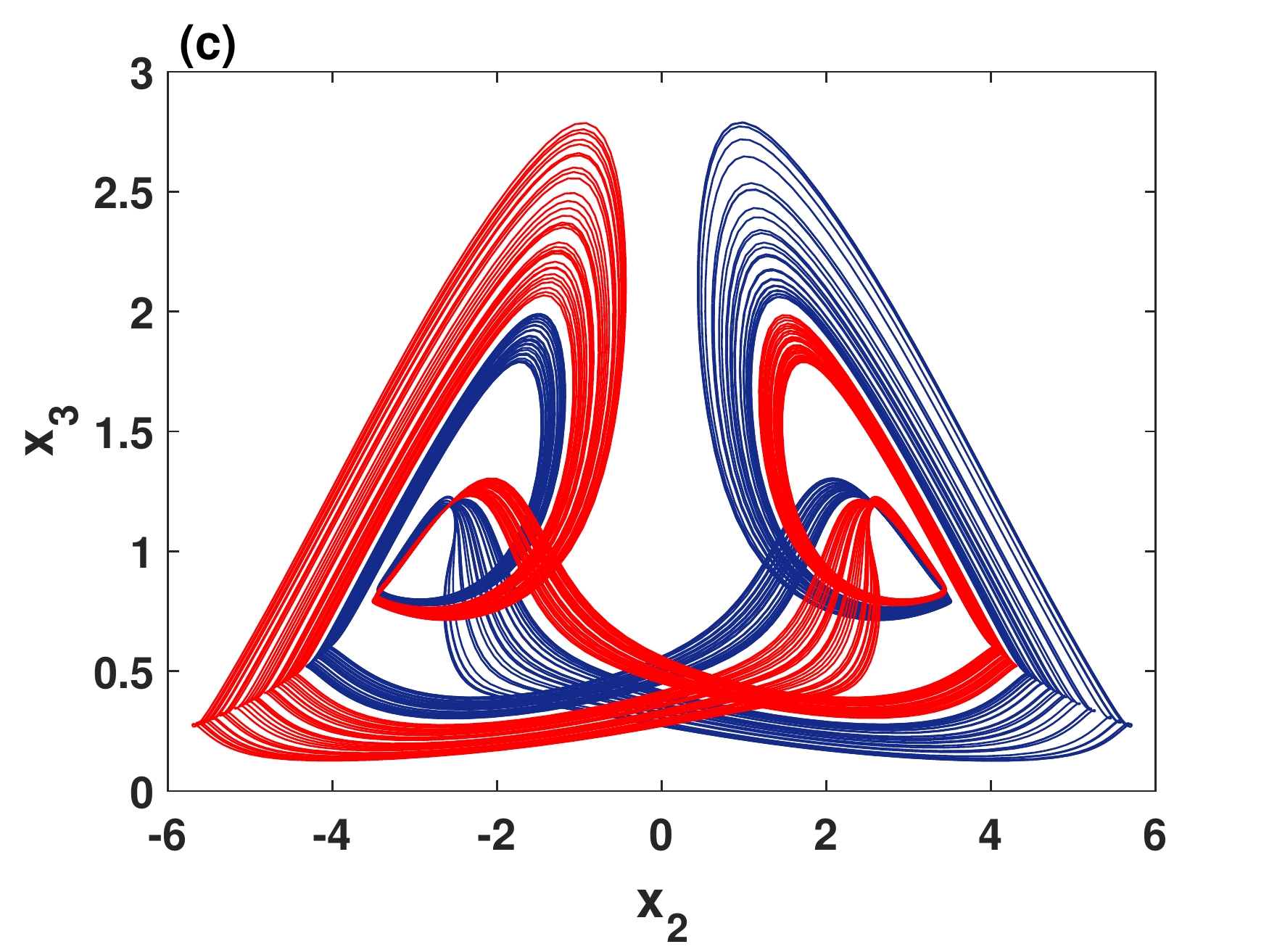}
	\end{subfigure}
	
	\centering
	\begin{subfigure}[h]{0.32\textwidth}
		\includegraphics[width=6cm, height=5.5cm]{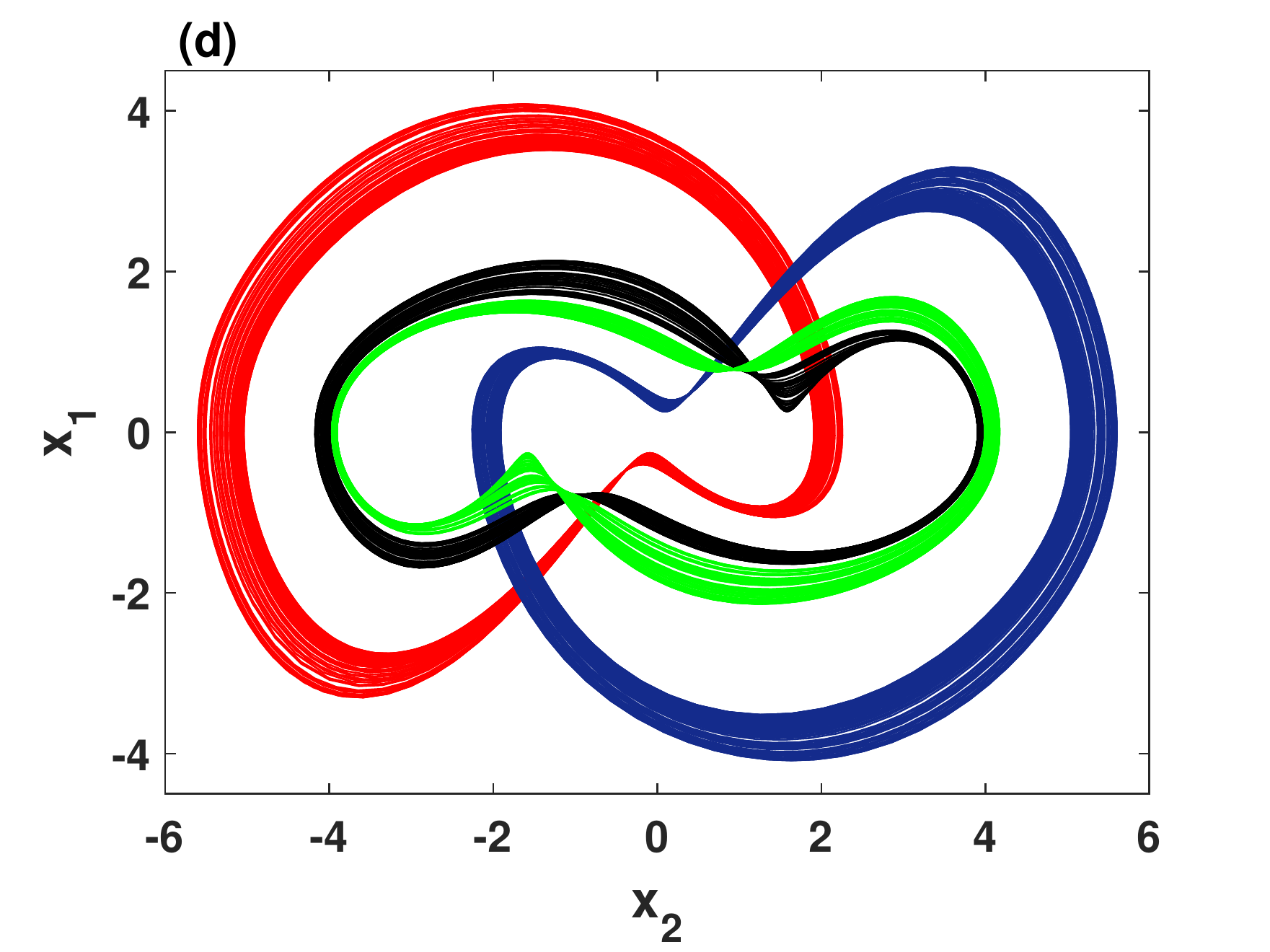}
	\end{subfigure}
	\begin{subfigure}[h]{0.32\textwidth}
		\includegraphics[width=6cm, height=5.5cm]{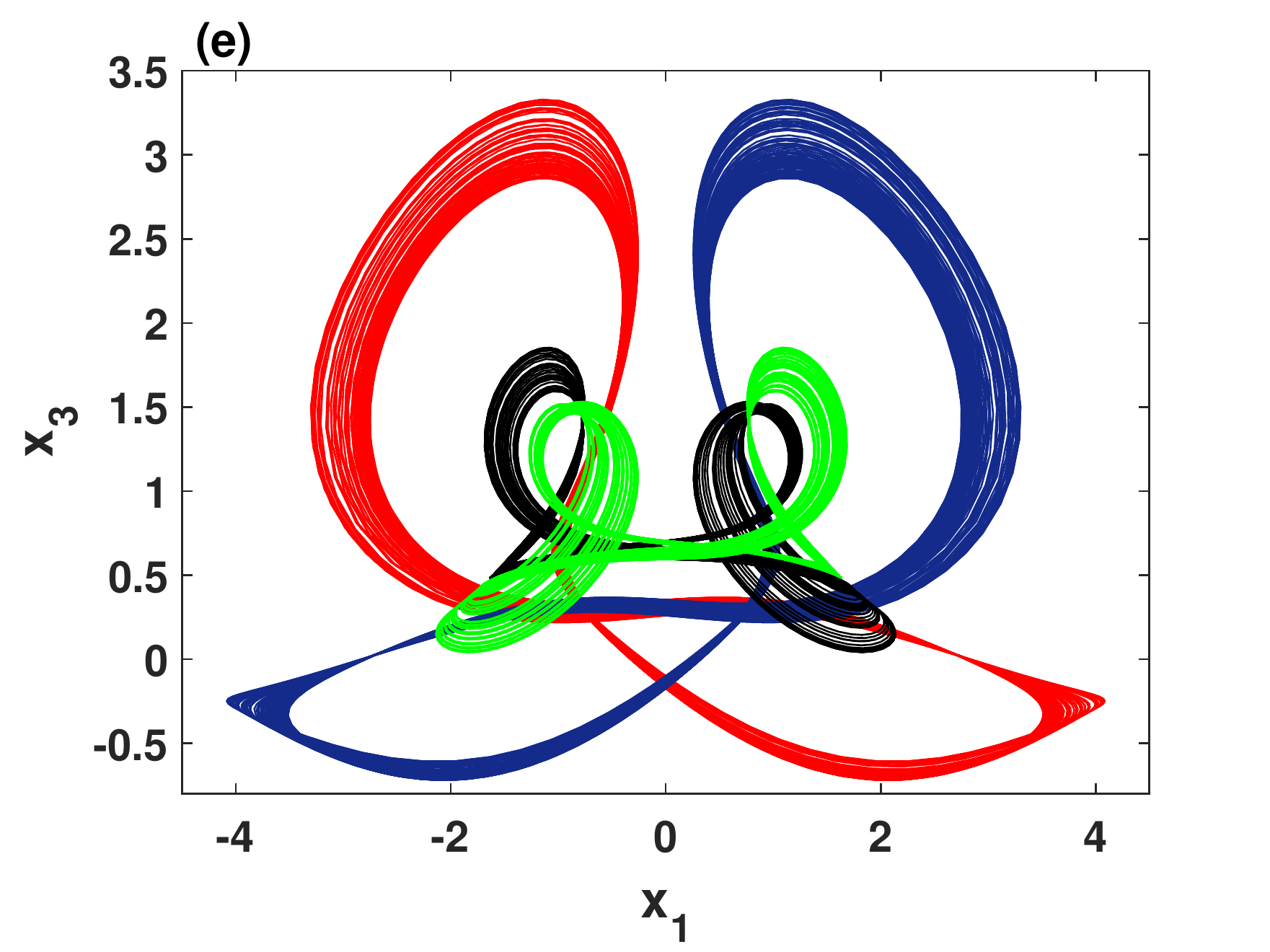}
	\end{subfigure}
	\begin{subfigure}[h]{0.32\textwidth}
		\includegraphics[width=6cm, height=5.5cm]{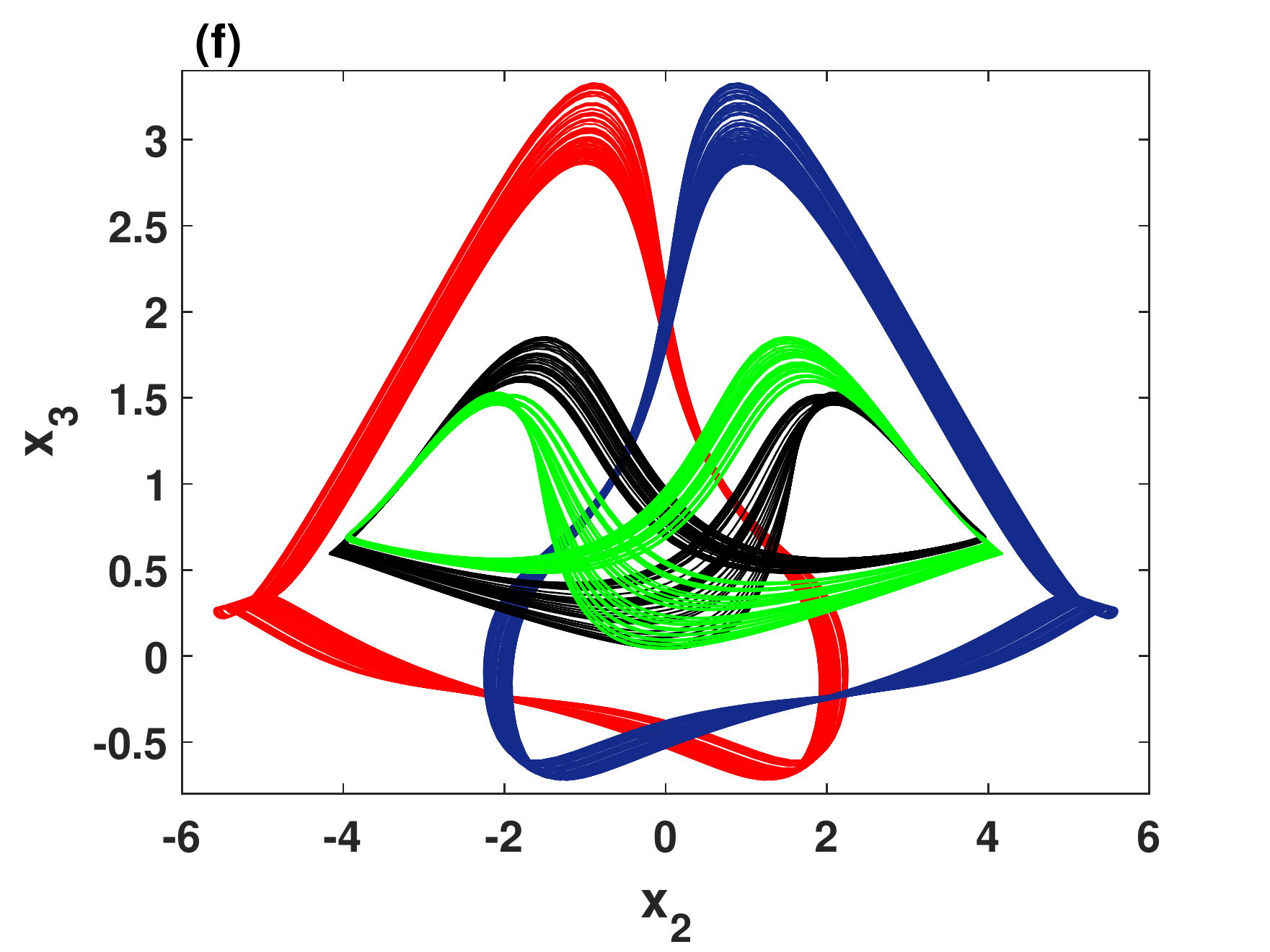}
	\end{subfigure}
	\caption{The coexisting chaotic attractors of the degenerated butterfly attractor when the parameters $ \delta=0.5 $, $ \eta=0.1 $, $ h=2.2 $: (a)-(c) symmetric pair of chaotic attractors coexist for the controller $ G_{1}(x_{1}) $ when $ a=0.8 $ and with the initial conditions $ (-2, 0.7, -1) $ (blue) and $ (-2, 2, -1) $ (red); (d)-(f) two symmetric pair of chaotic attractors coexist for the controller $ G_{2}(x_{1}) $  when $ a=0.85 $ and with the initial conditions $ (-2, 0.5, -1) $ (red), $ (-2, 1.2, -1) $ (blue), $ (-2, 2, -1)  $ (black), $ (-2, 3, -1) $ (green).}
	\centering
	\label{fig:phase3}
\end{figure*}

\begin{figure}[t]
	\centering
	\begin{subfigure}[b]{0.45\textwidth}
		\includegraphics[width=8cm, height=5cm]{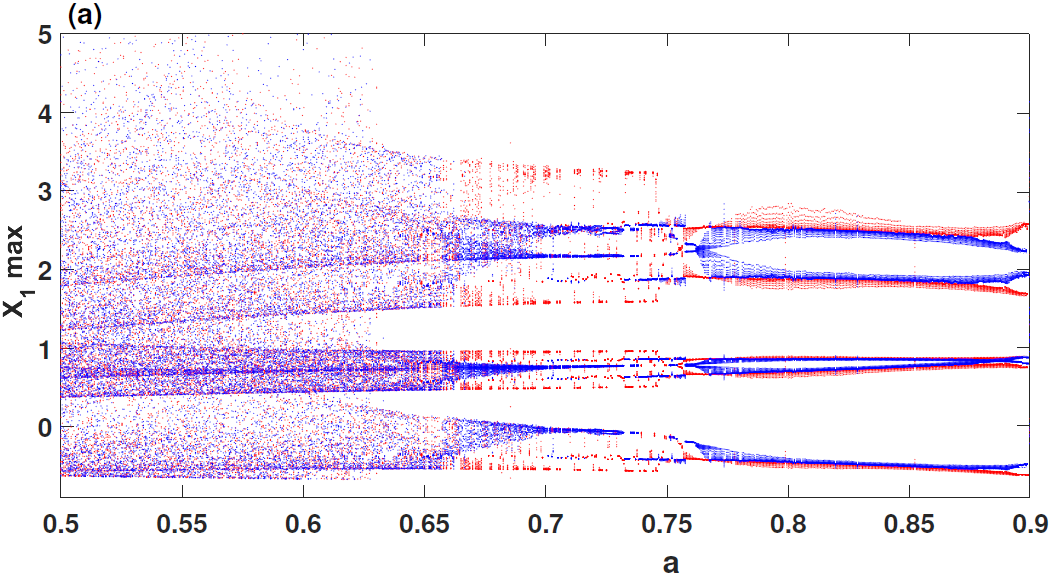}
	\end{subfigure}
	
	\begin{subfigure}[b]{0.45\textwidth}
		\includegraphics[width=8cm, height=5cm]{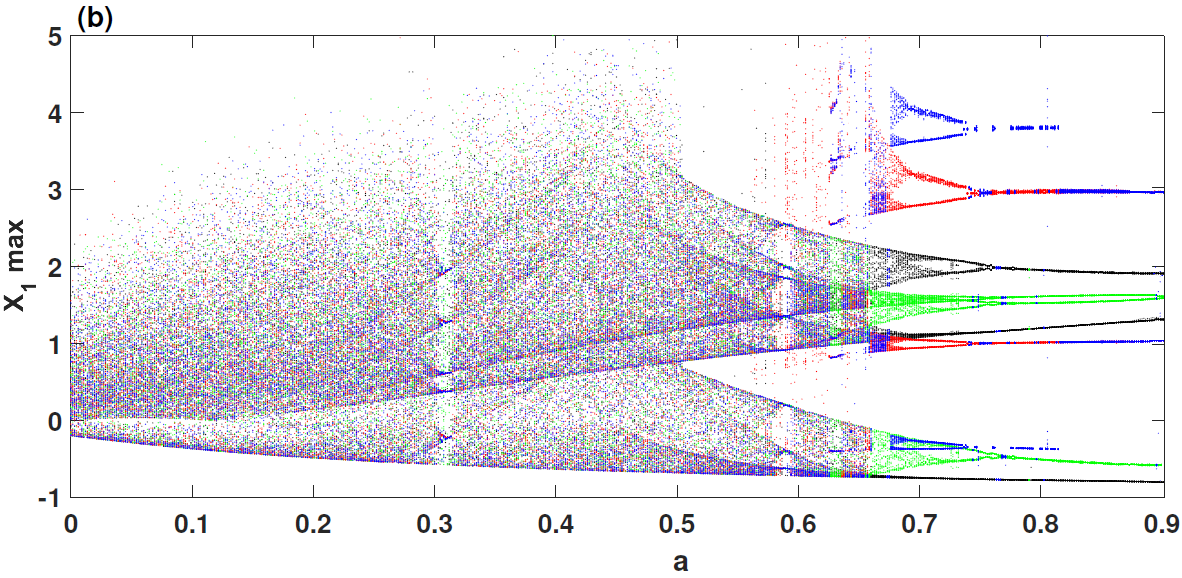}
	\end{subfigure}
	\caption{Bifurcation diagram to illustrate the degeneration of butterfly attractor into one and two symmetric pair of periodic attractors with the parameters $ \delta=0.5 $, $ \eta=0.09 $, $ h=2.5 $ and $ a $ varying: (a) for the controller $ G_{1}(x_{1}) $ with the initial conditions $ (2, 1.6, -2) $ (red) and $ (2, 1.6, 2) $ (blue); (b) for the controller $ G_{2}(x_{1}) $ with the initial conditions $ (2, 4, 2)$ (red), $ (-2, 4, -2)$ (green), $ (-2, 1.5, -2)$ (black), $ (-2, -1.5, -2)$ (blue).}
	\label{fig:bifurcation1}
\end{figure}

\begin{figure}[!b]
	\centering
	\begin{subfigure}[h]{0.23\textwidth}
		\includegraphics[width=4.5cm, height=5.5cm]{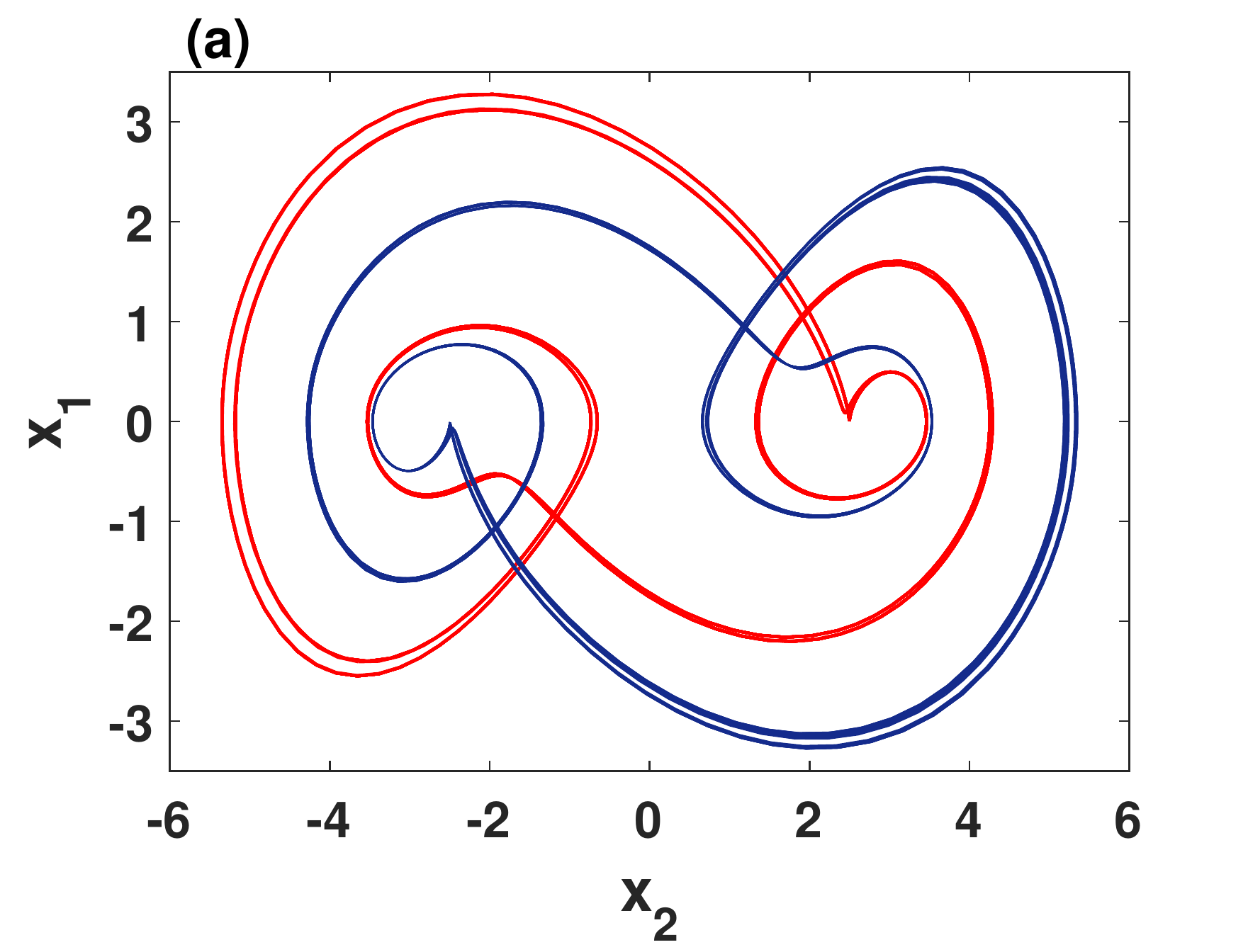}
	\end{subfigure}
	\begin{subfigure}[h]{0.23\textwidth}
		\includegraphics[width=4.5cm, height=5.5cm]{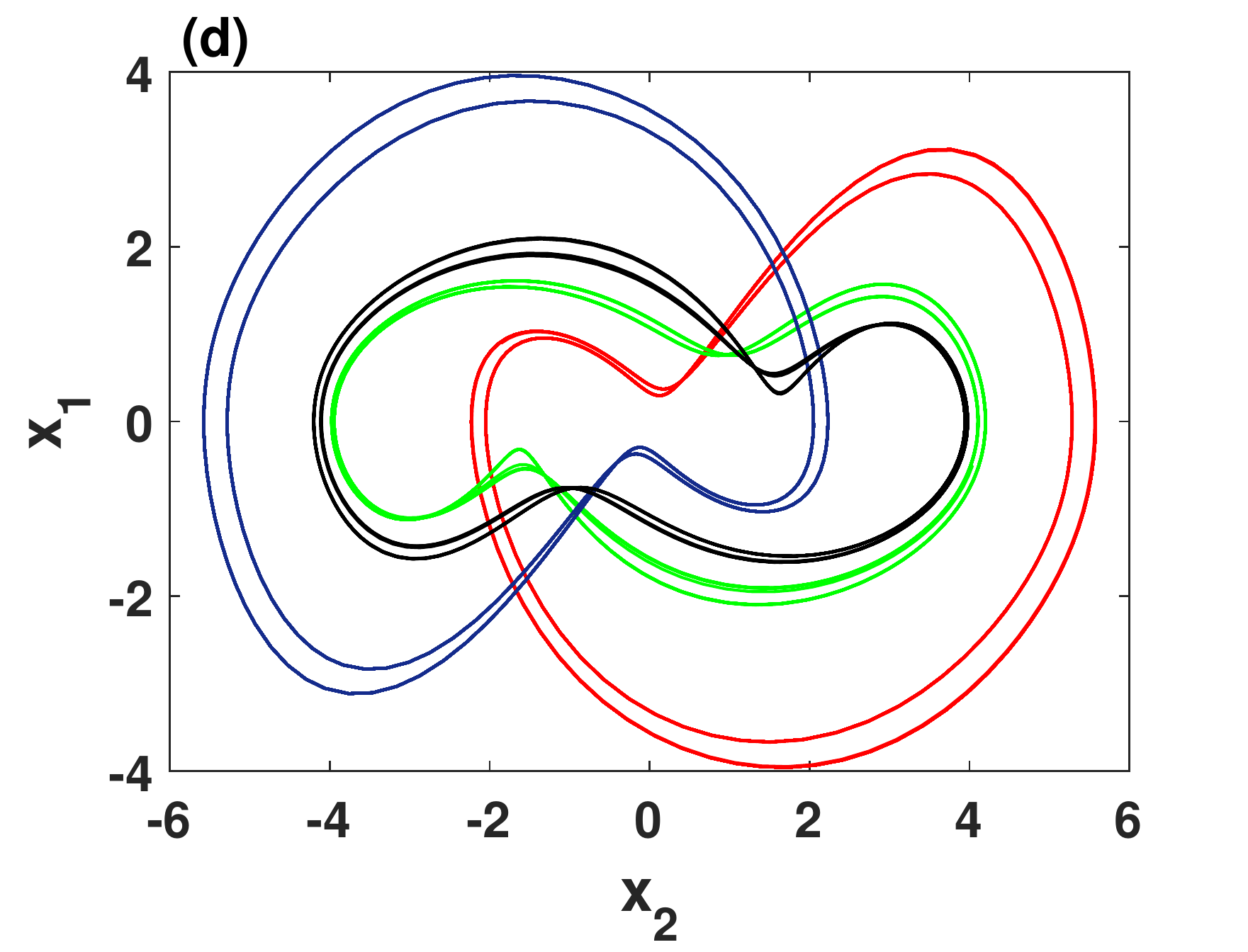}
	\end{subfigure}
	
	\centering
	\begin{subfigure}[h]{0.23\textwidth}
		\includegraphics[width=4.5cm, height=5.5cm]{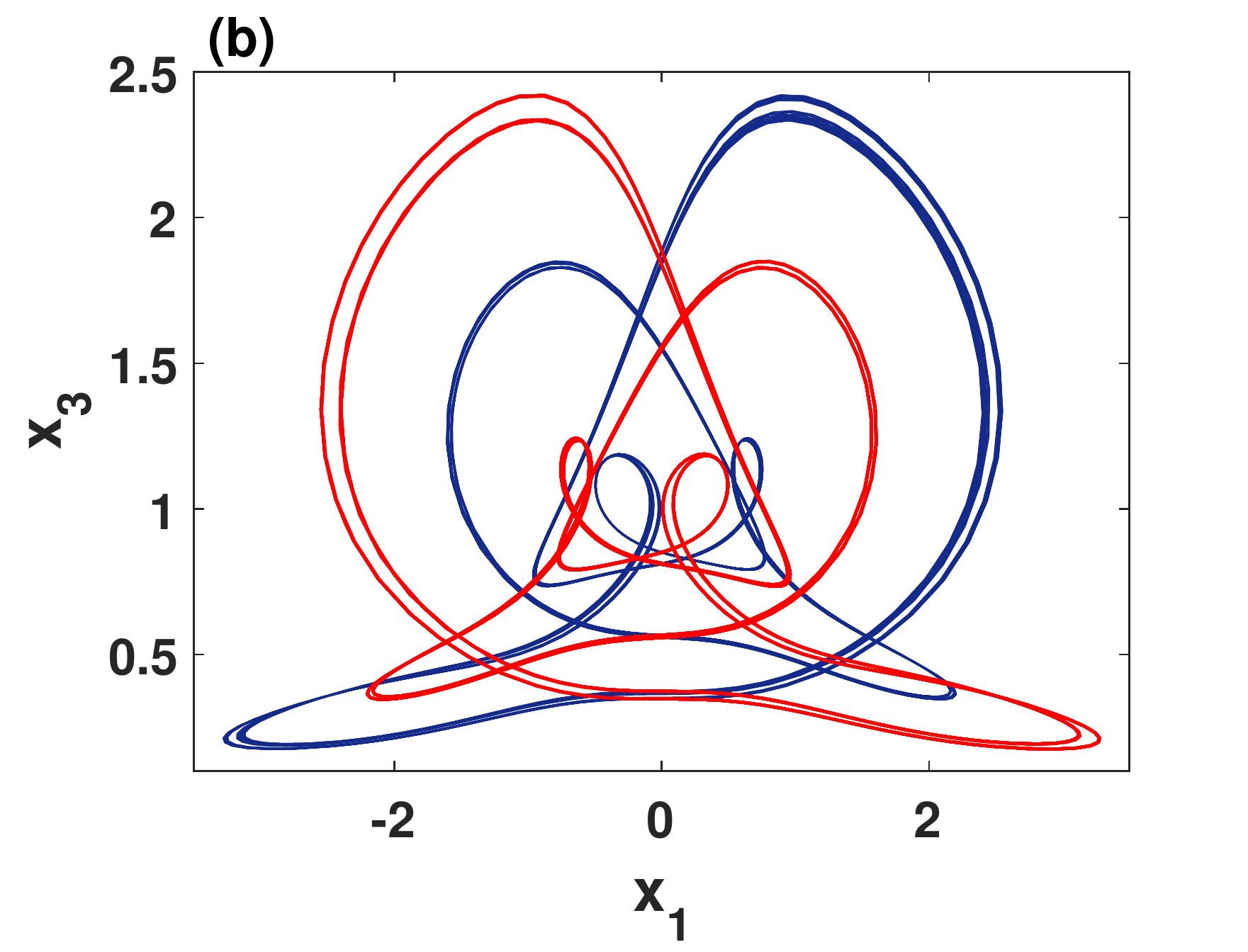}
	\end{subfigure}
	\begin{subfigure}[h]{0.23\textwidth}
		\includegraphics[width=4.5cm, height=5.5cm]{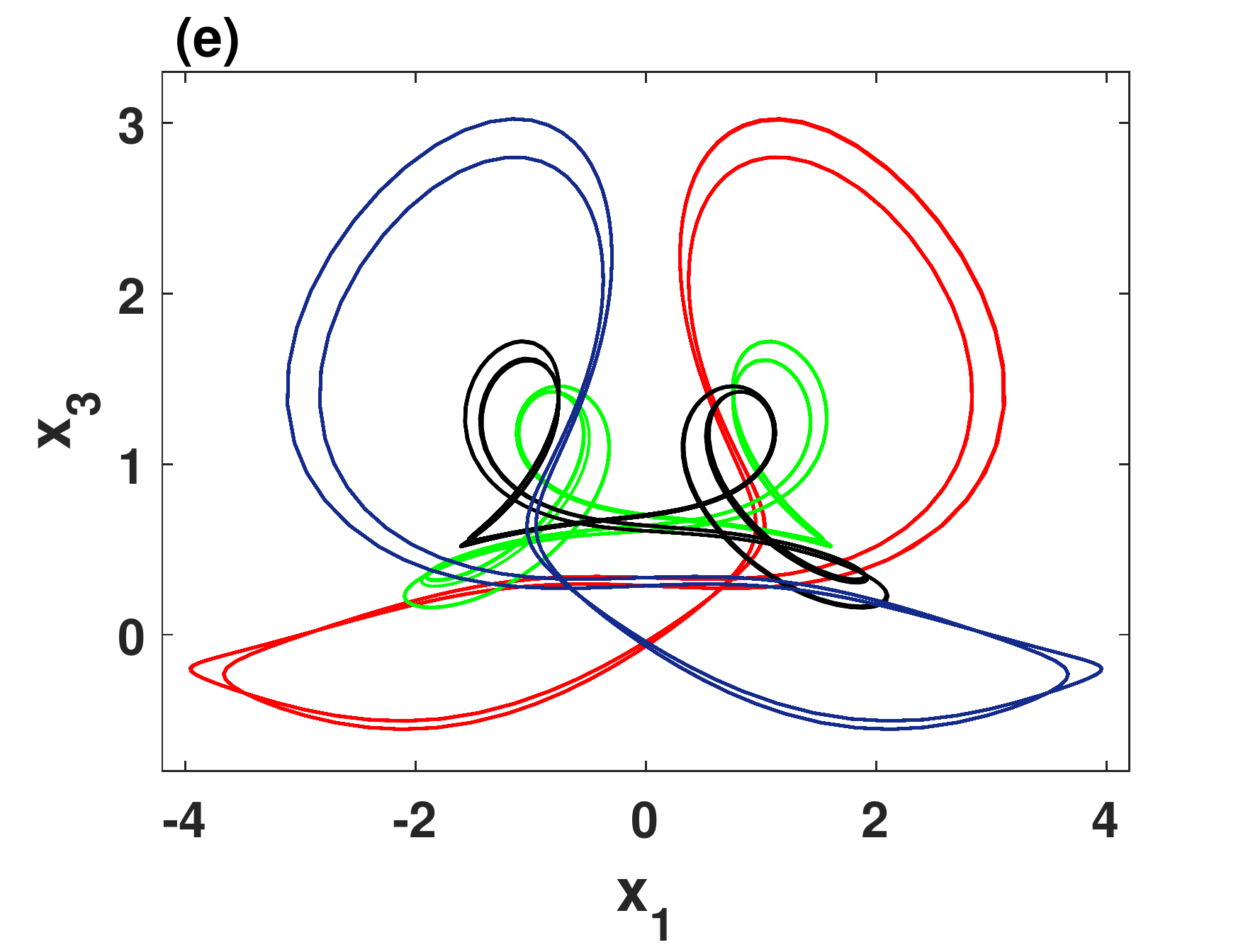}
	\end{subfigure}
	
	\centering
	\begin{subfigure}[h]{0.23\textwidth}
		\includegraphics[width=4.5cm, height=5.5cm]{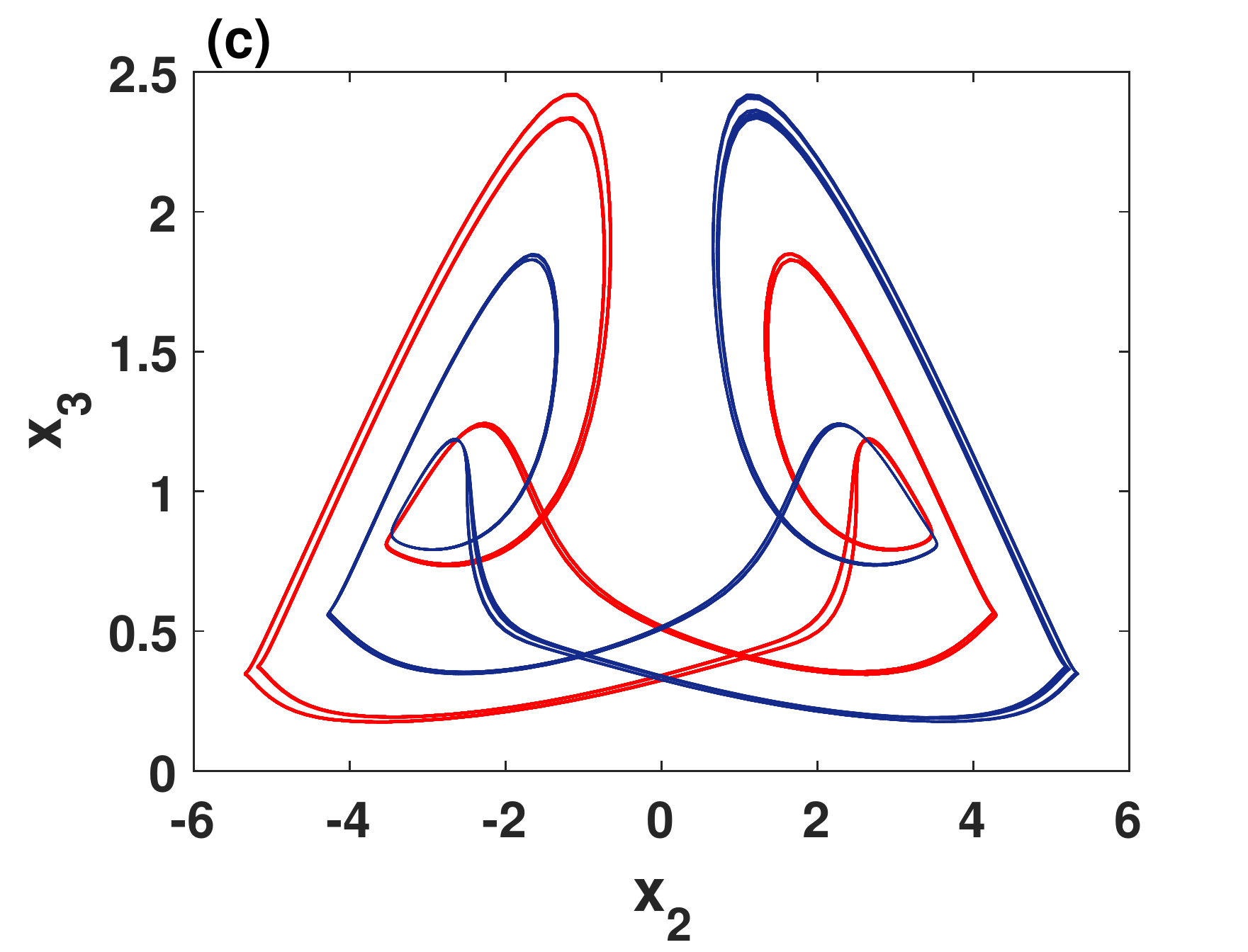}
	\end{subfigure}
	\begin{subfigure}[h]{0.23\textwidth}
		\includegraphics[width=4.5cm, height=5.5cm]{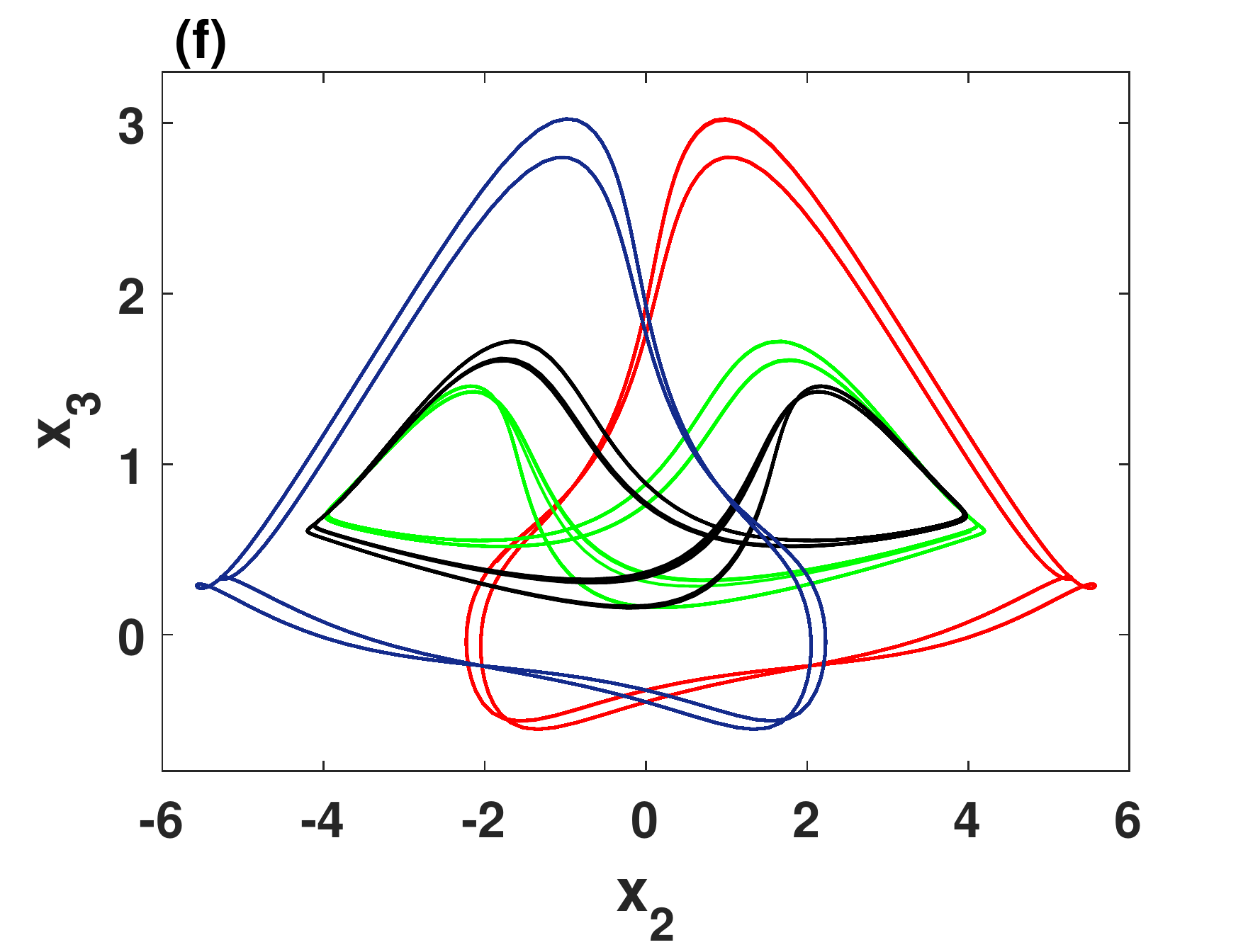}
	\end{subfigure} 
	\caption{Coexisting periodic attractors when the parameters $ \delta=0.5 $, $ \eta=0.09 $, $ h=2.5 $; (a)-(c) symmetric pair of period-2 limit cycles coexist for the controller $ G_{1}(x_{1}) $ when $ a=0.7 $ and with the initial conditions $ (2, 1.6, -2) $ (red) and $ (2, 1.6, 2) $ (blue); (d)-(e) two symmetric pair of period-2 limit cycles coexist for the controller $ G_{2}(x_{1}) $ when $ a=0.72 $ and with the initial conditions $ (2, 4, 2)$ (red), $ (-2, 4, -2)$ (green), $ (-2, 1.5, -2)$ (black), $ (-2, -1.5, -2)$ (blue).}
	\centering
	\label{fig:phase4}
\end{figure}

Meanwhile, it is easy to find that the system~(\ref{2}) is stable at $ E_{1} $ when $ h<1 $, and it is stable at $ E_{2,3} $ when $ \frac{2\eta h-2\eta}{ \delta\eta h\cdot(\eta h+\delta)}<1 $.

\begin{figure*}[t]
	\hspace{0.7cm}
	\begin{subfigure}[h]{0.4\textwidth}
		\includegraphics[width=7.4cm, height=4.5cm]{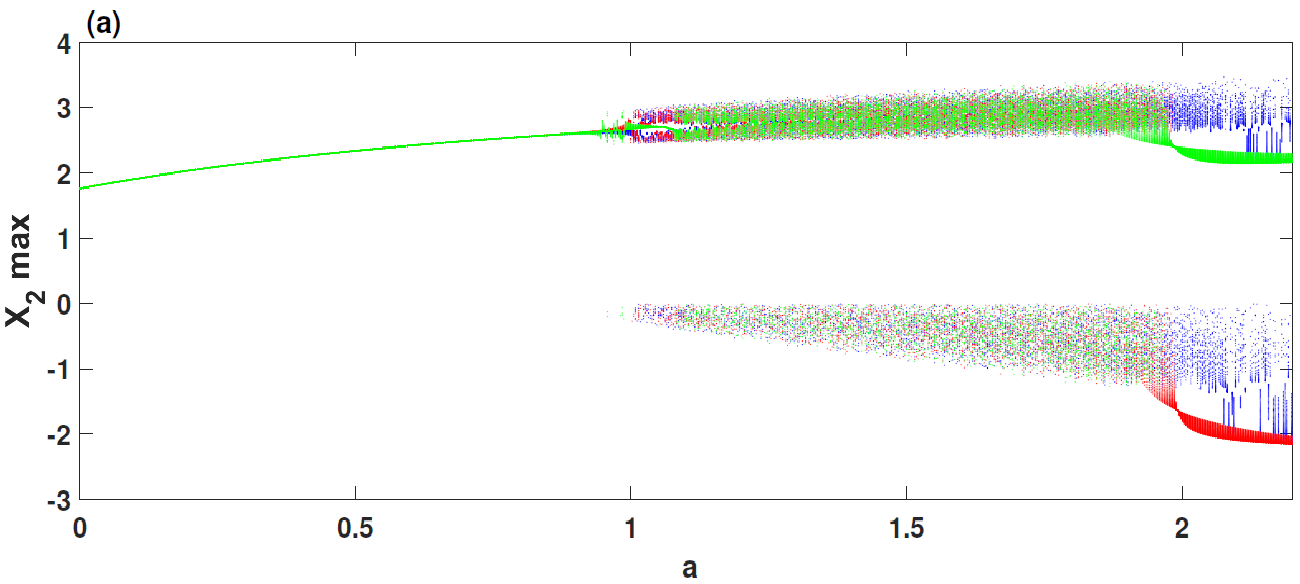}
	\end{subfigure} 
	\begin{subfigure}[h]{0.4\textwidth}
		\includegraphics[width=9cm, height=4.5cm]{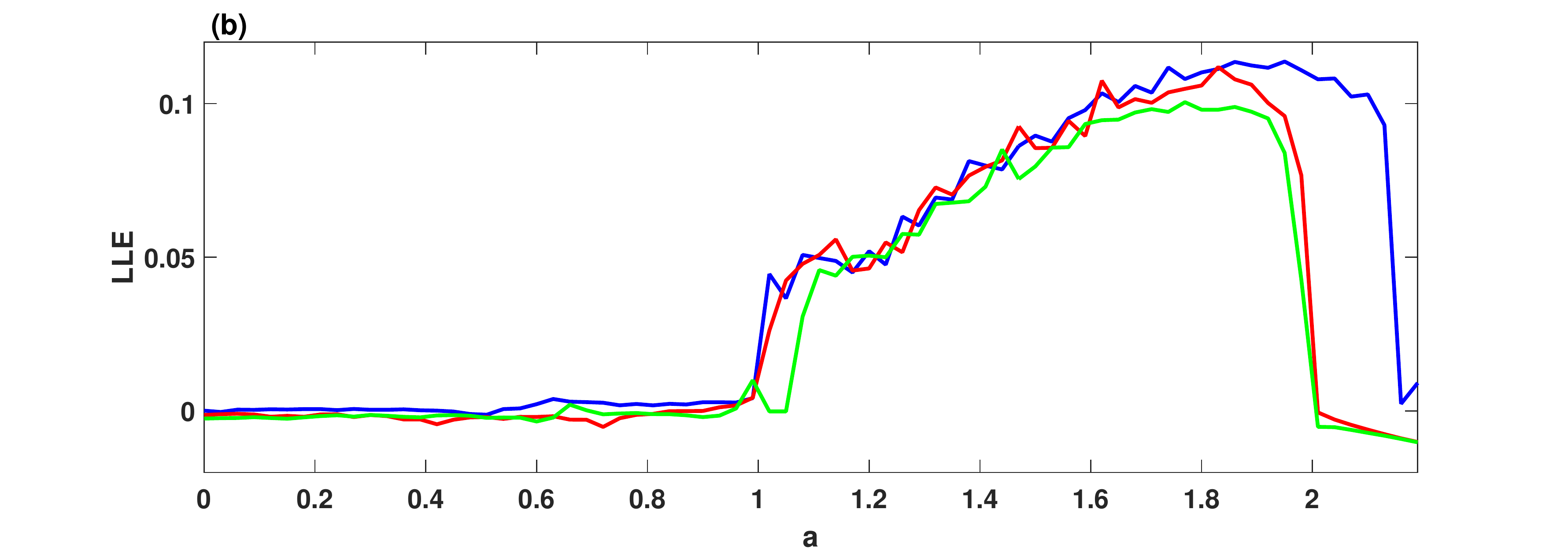}
	\end{subfigure}
	\caption{Dynamics of the system~(\ref{3}) versus parameter  $ a $  varying when $ \delta=0.45 $, $ \eta=0.6 $, $ h=2 $ for the controller $ G_{2}(x_{1}) $ with the initial conditions $ (1, 1, 3) $ (blue), $ (1, 4.4, 3) $ (red) and $ (1, -3.9, 3) $ (green): (a) the bifurcation diagram; (b) the largest Lyapunov exponents (LLE).}
	\centering
	\label{fig:bifurcation4}
\end{figure*}

\begin{figure*}[t]
	\hspace{2cm}
	\begin{subfigure}[h]{0.35\textwidth}
		\includegraphics[width=7cm, height=4.5cm]{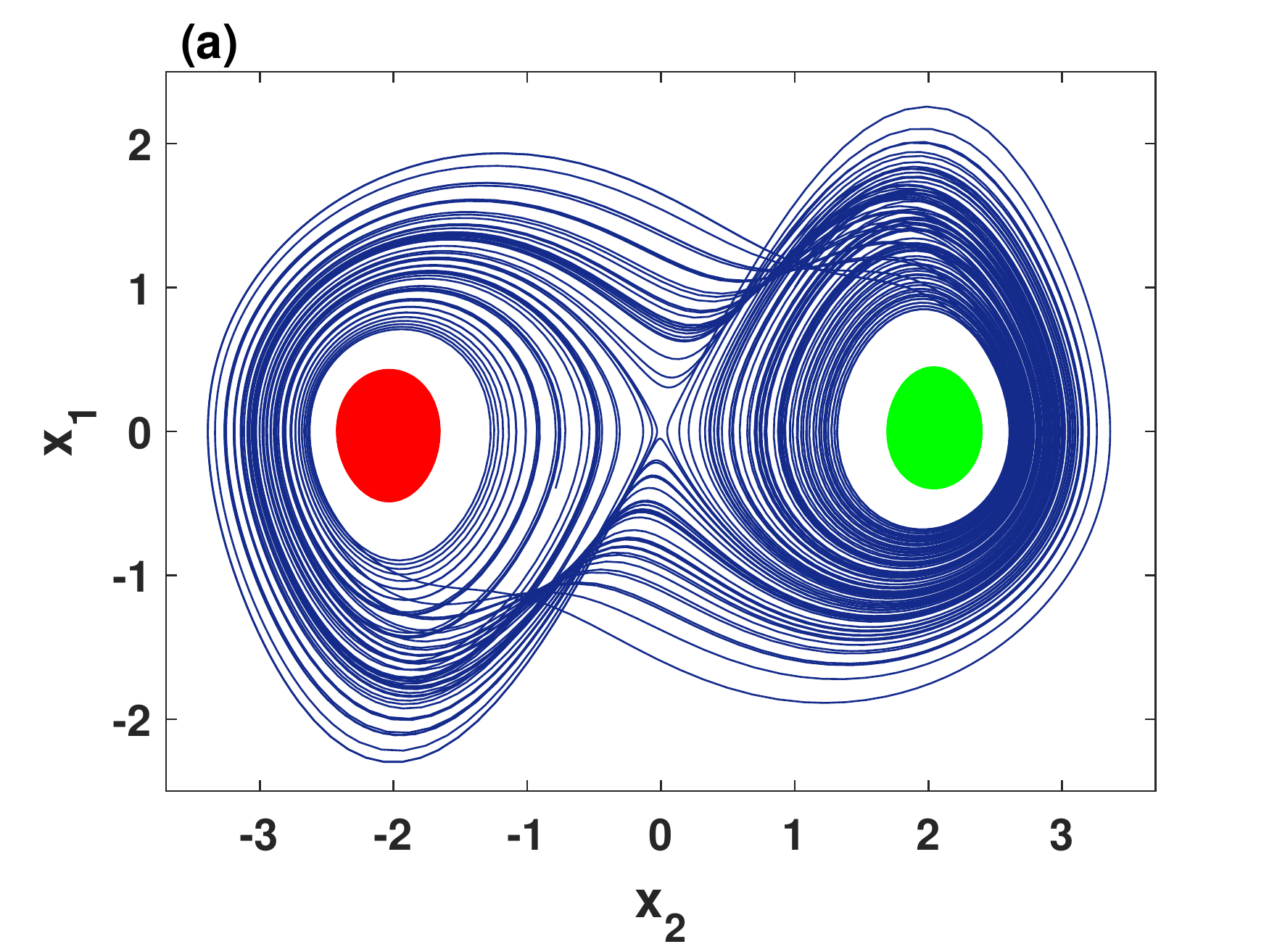}
	\end{subfigure}
	\begin{subfigure}[h]{0.35\textwidth}
		\includegraphics[width=7cm, height=4.5cm]{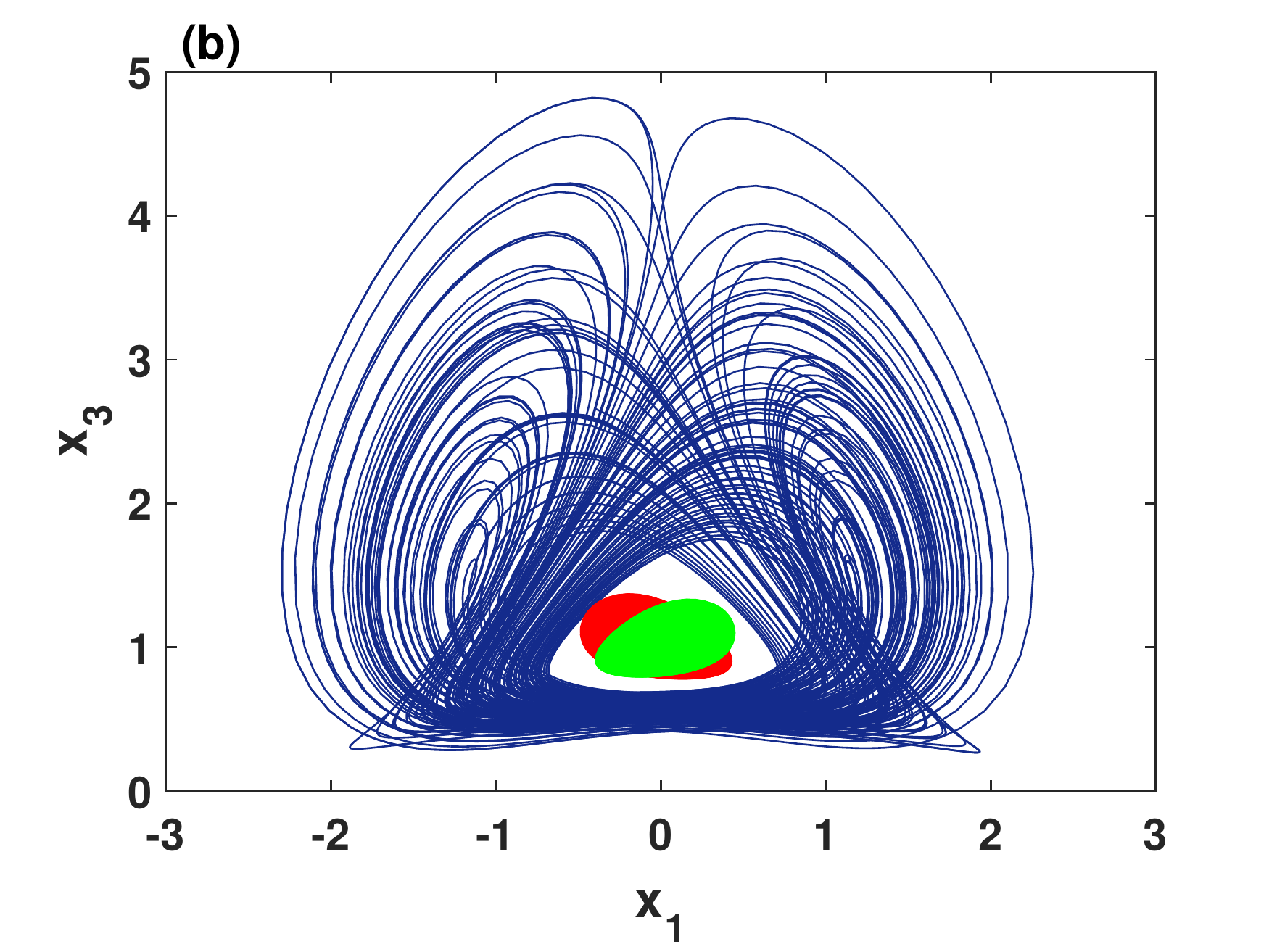}
	\end{subfigure}
	
	\hspace{2cm}
	\begin{subfigure}[h]{0.35\textwidth}
		\includegraphics[width=7cm, height=4.5cm]{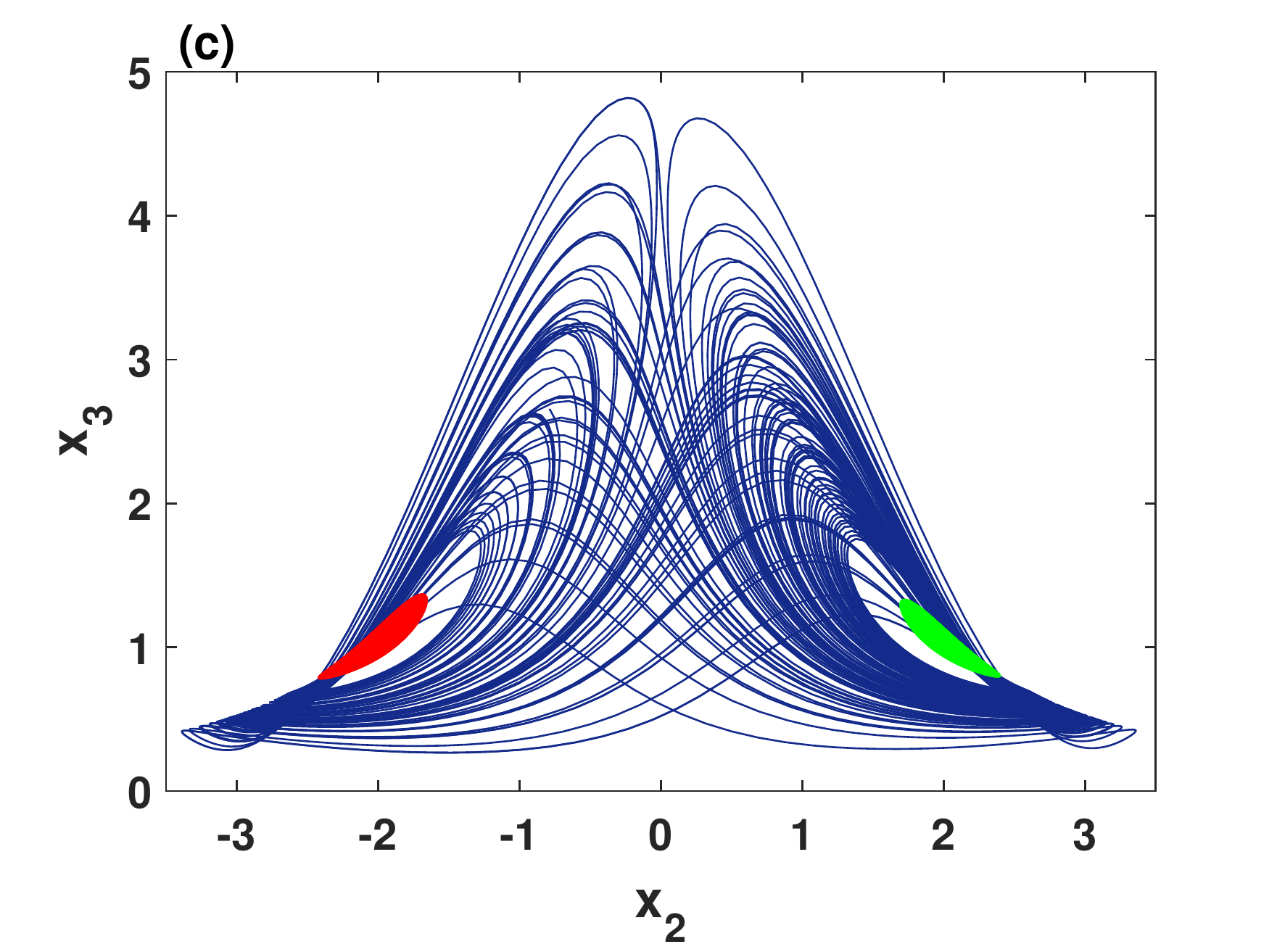}
	\end{subfigure}
	\begin{subfigure}[h]{0.35\textwidth}
		\includegraphics[width=7cm, height=4.5cm]{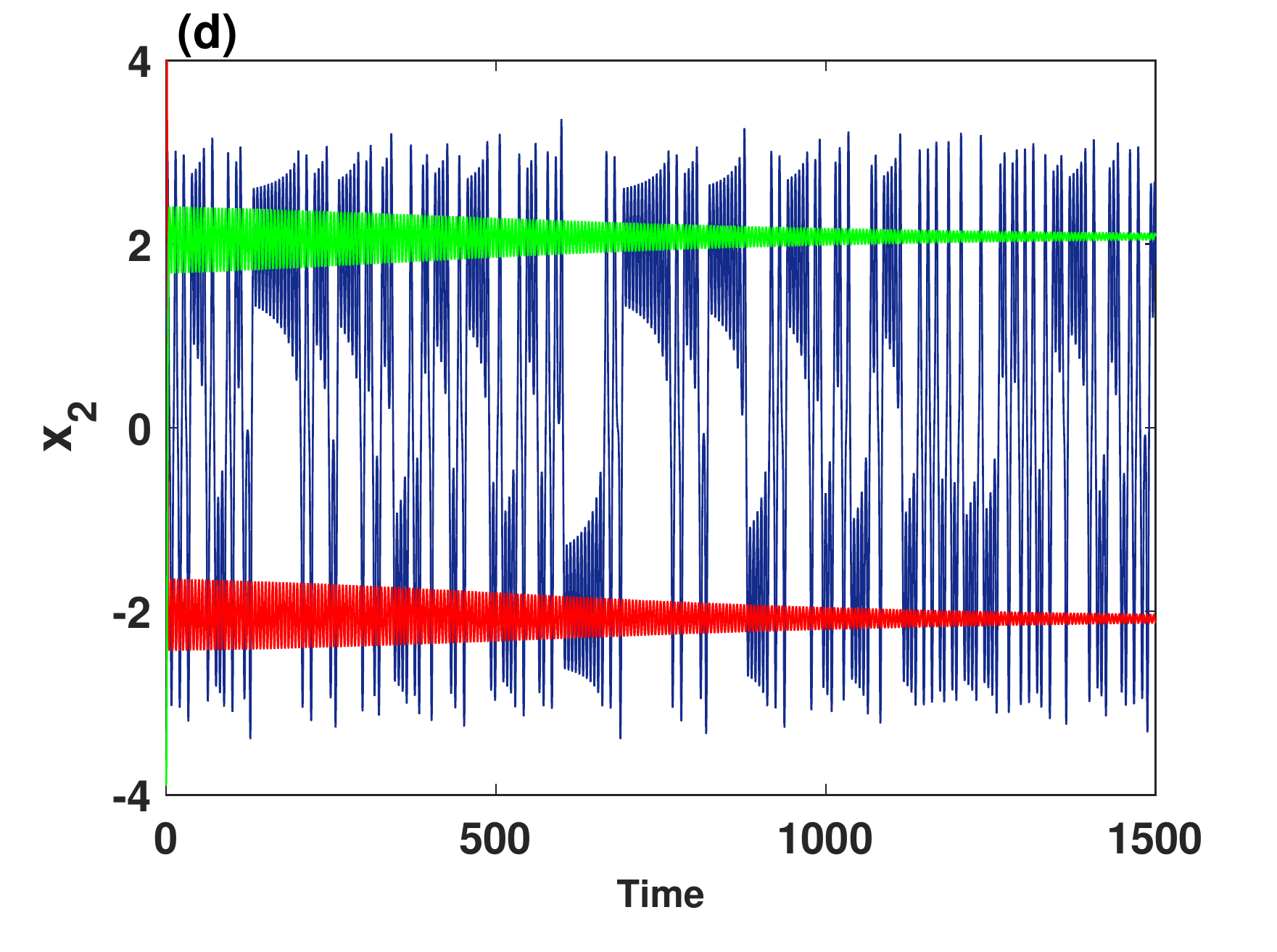}
	\end{subfigure}
	\caption{Three coexisting attractors of the system~(\ref{3}) for the controller $ G_{2}(x_{1}) $ with the parameters $ \delta= 0.45 $, $ \eta= 0.6 $, $ h=2 $, $ a=2 $: (a)-(c) different perspectives on the coexistence of symmetric chaotic attractor and two point attractors for the initial conditions $ (1, 1, 3) $ (blue), $ (1, 4.4, 3) $ (red) and $ (1, -3.9, 3) $ (green); (d) the corresponding time series of the state variables $ x_{2} $.}
	\centering
	\label{fig:phase6}
\end{figure*}

Linearizing system~(\ref{3}) at an arbitrary equilibrium point $ (x_{1}^{\ast}, x_{2}^{\ast}, x_{3}^{\ast}) $ is obtained by the following Jacobian matrix:
\begin{flalign*}
\begin{aligned}
J(x_{1}^{\ast}, x_{2}^{\ast}, x_{3}^{\ast})=
\begin{pmatrix}
-\delta &x_{3}-1&x_{2} \\
1 &0&0 \\
\frac{\partial f_{3}}{\partial x_{1}} & -2\eta x_{2} x_{3} & -\eta(1+x_{2}^{2})
\end{pmatrix}.
\end{aligned}
\phantom{\hspace{0.5cm}}
\end{flalign*}
where $ \frac{\partial f_{3}}{\partial x_{1}}= a\frac{\partial G_{i}(x_{1})}{\partial x_{1}}=-a\ sin(x_{1})$ or $ -2ax_{1}e^{-x_{1}^{2}} $. The eigenvalues of system~\ref{3} at the equilibria $ SE_{1} $ and $ SE_{2,3} $ can be obtained by solving Eq.~(\ref{4}) and Eq.~(\ref{5}), respectively.
\begin{flalign}
\label{4}
&\det(\lambda I-J_{SE_{1}})\\
\label{5}
&\det(\lambda I-J_{SE_{2,3}})
\phantom{\hspace{5cm}}
\end{flalign}
which yields
\begin{flalign}
\label{6}
\phantom{\hspace{0.6cm}}
f_{1}(\lambda)=&(\lambda+\eta)(\lambda^{2}+\delta \lambda+1-h-\frac{a}{\eta})=0,\\
\label{7}
f_{2}(\lambda)=&\lambda^{3}+(\eta h+\delta+a)\lambda^{2}+\delta(\eta h+a)\lambda \nonumber\\
&+2\eta(h+\frac{a}{\eta}-1)=0&&
\end{flalign}
It is obvious that Eq.~(\ref{6}) has always eigenvalue at $ SE_{1} $ with negative real part which is $ \lambda_{1}=-\eta$, whereas the real parts of the other eigenvalues are not always negative. By Routh-Hurwitz criterion, the eigenvalues $ \lambda_{2,3} $ at $ SE_{1} $ have negative real parts when $ (h+\frac{a}{\eta})<1 $. Furthermore, according to Routh-Hurwitz criterion, the real parts of the eigenvalues at $ SE_{2,3} $ of  Eq.~(\ref{7}) are negative if and only if 
\begin{flalign*}
\begin{cases}
\begin{aligned}
&(\eta h+\delta+a)>0,\\
&2\eta(h+\frac{a}{\eta}-1)>0,\\
&(\eta h+\delta+a)\cdot \delta(\eta h+a)>2\eta(h+\frac{a}{\eta}-1).
\end{aligned}
\phantom{\hspace{0.9cm}}
\end{cases}
\end{flalign*}
Consequently, we obtain the following proposition.
\begin{prop}\label{pro:1} The following statements are true for $ \delta, h, \eta, a >0  $.
	\begin{enumerate}
		\item If $ (h+\frac{a}{\eta})<1 $, then $ SE_{1} $ is a stable equilibrium.
		\item If $ \frac{2\eta h+2 a-2\eta}{ \delta(\eta h+a)\cdot(\eta h+\delta+a)}<1 $, then $ SE_{2,3} $ are stable equilibria. 
	\end{enumerate}
\end{prop}

\begin{figure*}[t]
	\begin{minipage}[b]{0.48\textwidth}
		\hspace{-0.5cm}
		\begin{subfigure}[b]{0.49\linewidth}
			\includegraphics[width=6.3cm, height=2.7cm]{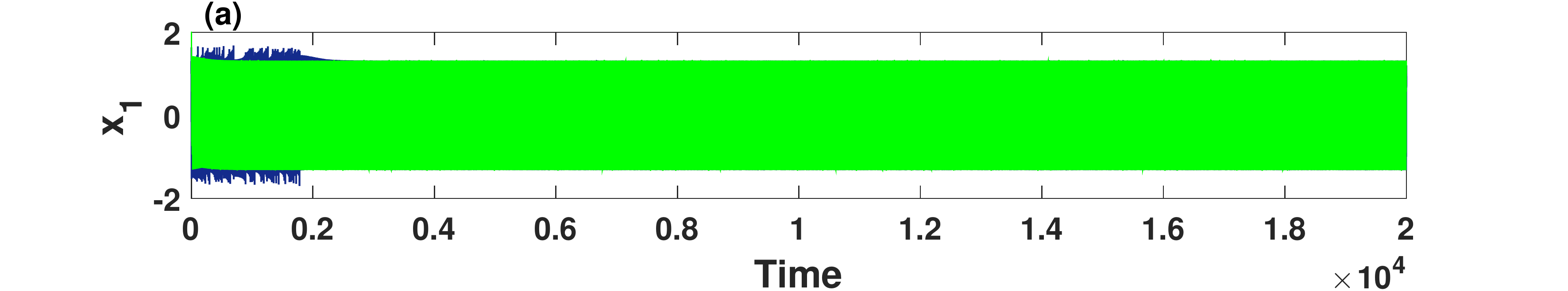}
		\end{subfigure}
		
		\hspace{-0.5cm}
		\begin{subfigure}[b]{0.49\linewidth}
			\includegraphics[width=6.3cm, height=2.7cm]{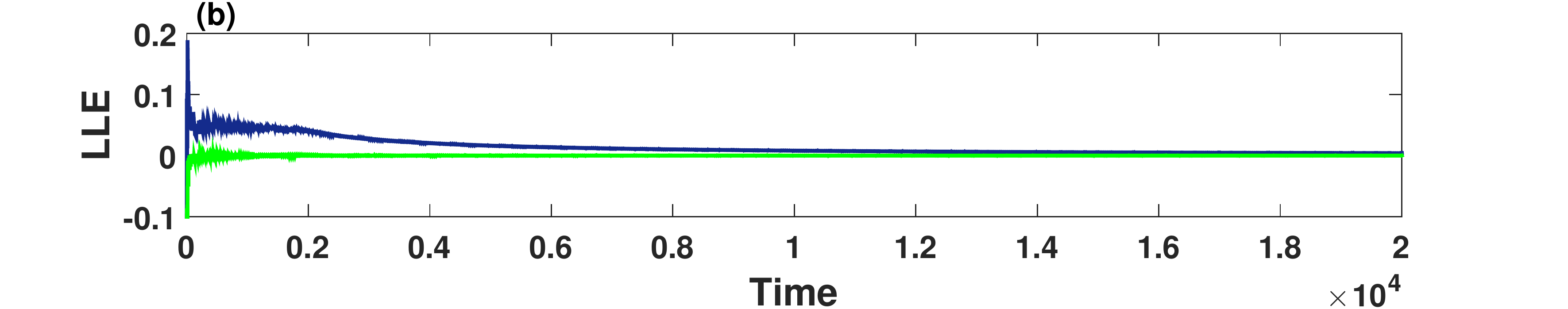}
		\end{subfigure}
	\end{minipage}
	\hfill
	\begin{subfigure}[b]{0.1\textwidth}
		\hspace{-7cm}
		\vspace{-0.2cm}
		\includegraphics[width=4.3cm, height=5.8cm]{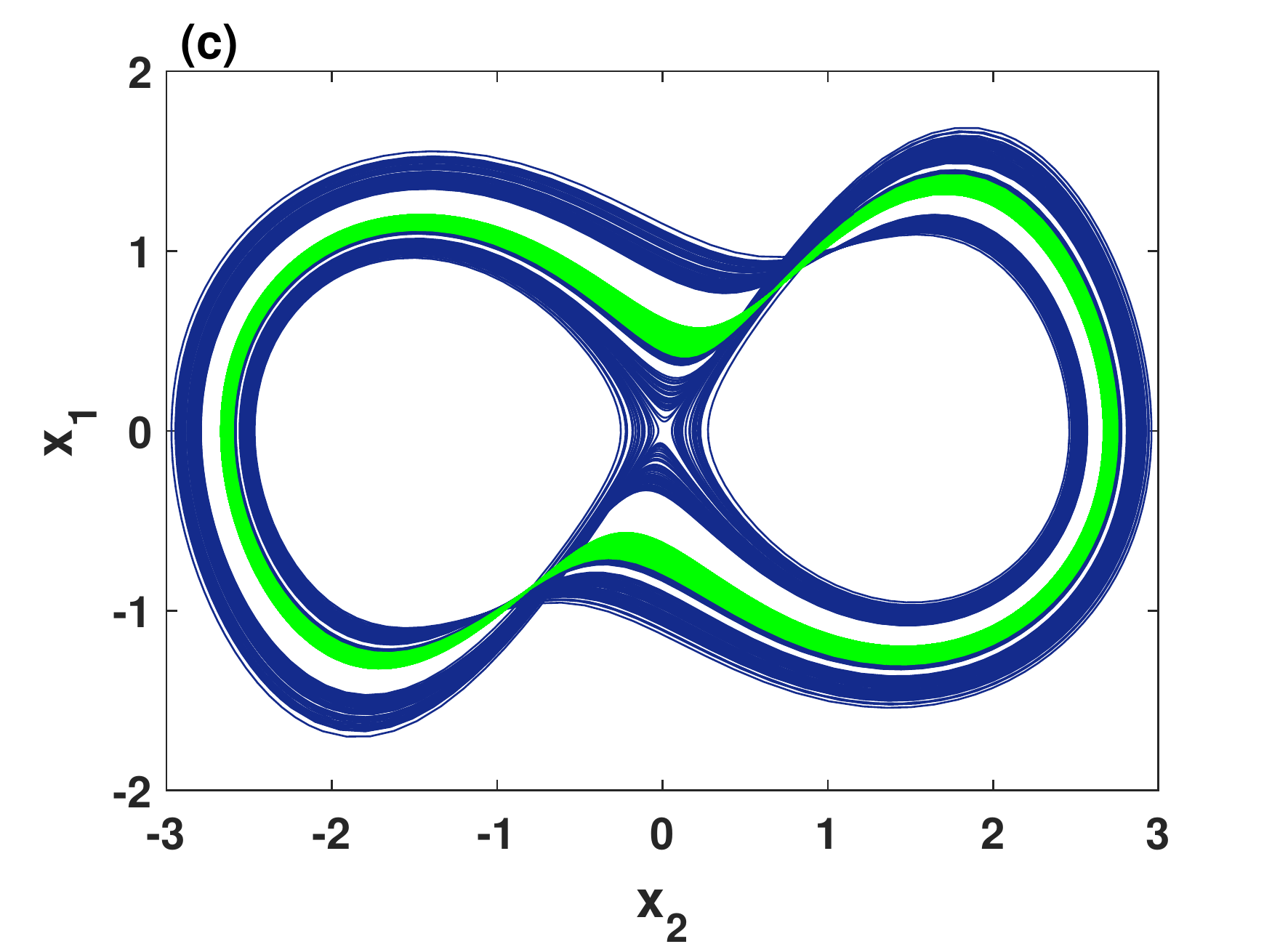}
	\end{subfigure}
	\begin{subfigure}[b]{0.1\textwidth}
		\hspace{-4.9cm}
		\vspace{-0.2cm}
		\includegraphics[width=4.3cm, height=5.8cm]{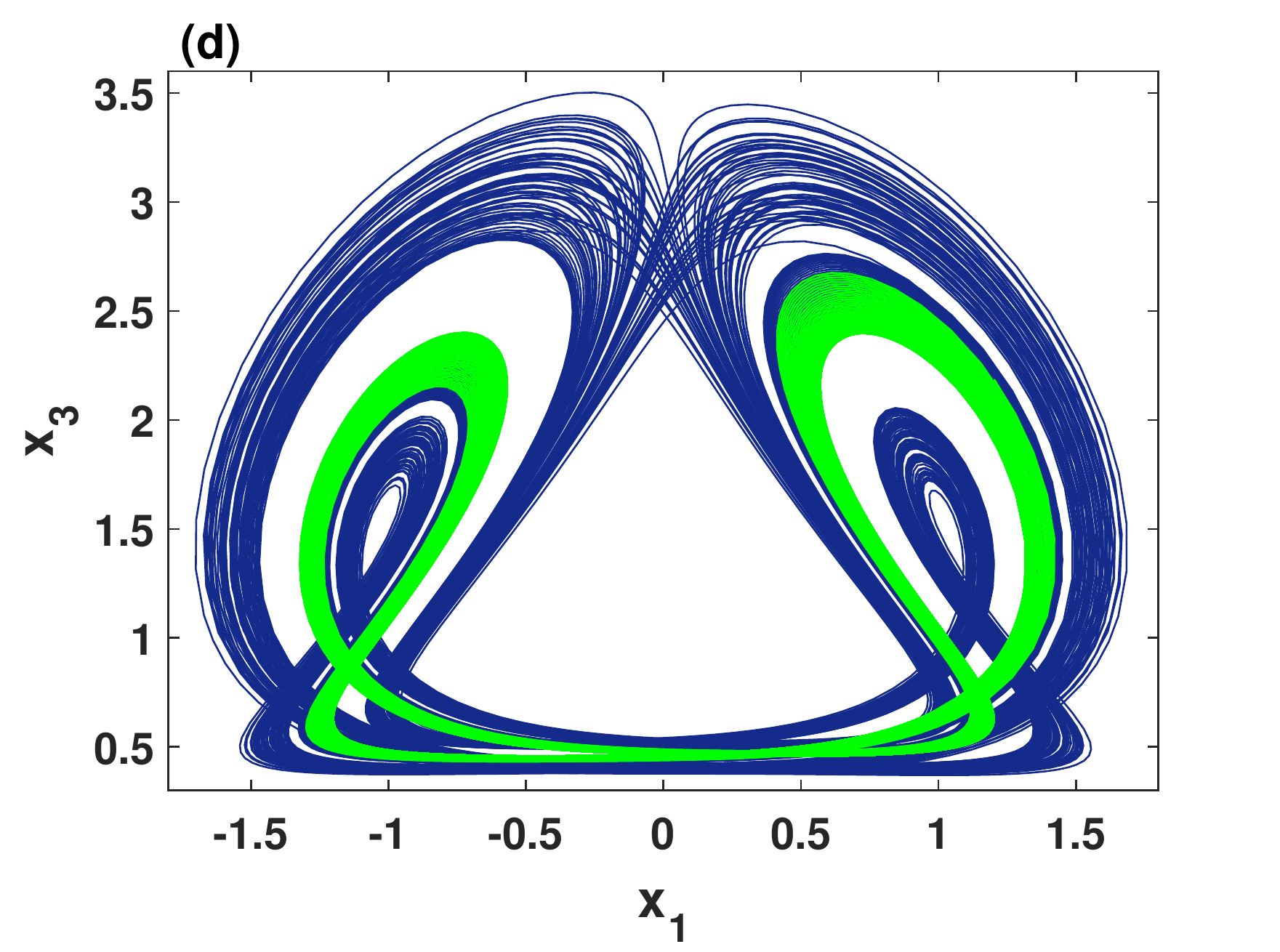}
	\end{subfigure}
	\begin{subfigure}[b]{0.1\textwidth}
		\hspace{-2.8cm}
		\vspace{-0.2cm}
		\includegraphics[width=4.3cm, height=5.8cm]{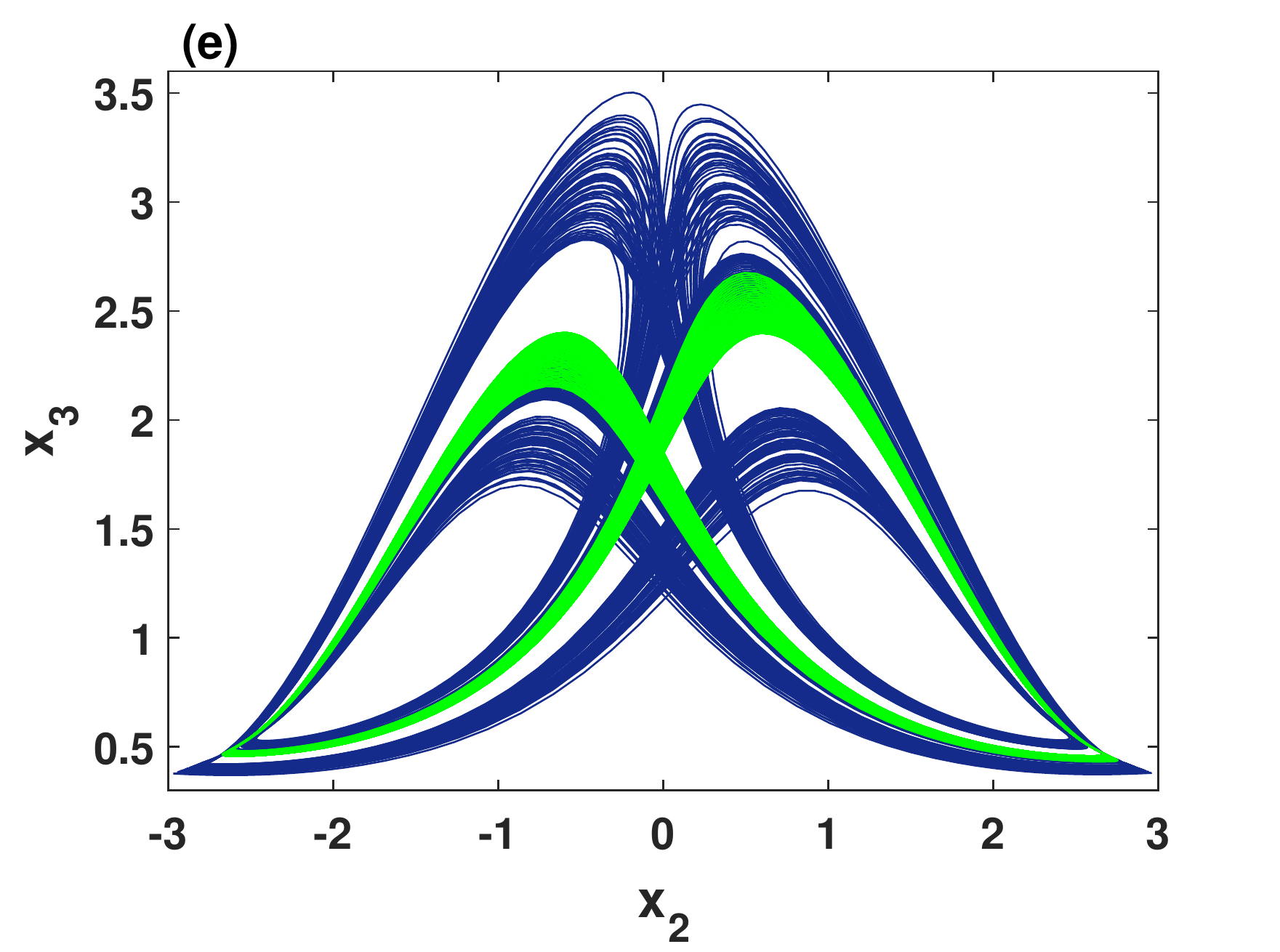}
	\end{subfigure}
	\vspace{0.2cm}
	\caption{The coexistence of transient transition chaos with completely quasi-periodic behavior of the system~(\ref{3}) when $ \delta=0.45 $, $ \eta=0.6 $, $ h=2 $, $ a=1.02 $, and for the initial conditions $ (1, 1, 3) $ (blue) and $ (1, -3.9, 3) $ (green): (a) the time-domain waveform; (b) the corresponding Largest Lyapunov Exponents (LLE); (c)-(e) the phase portraits.}
	\label{fig:phase7}
\end{figure*}

\begin{table}[!b]
	\setlength{\extrarowheight}{1pt}	
	\captionof{table}{ The corresponding Lyapunov exponents of the coexisting attractors that shown in Figure~(\ref{fig:phase3}).}
	\label{tb:tabel}
	\begin{tabular}{p{1.5cm} c c c} 
		\hline
		
		\hline
		
		\hline
		Controller & Parameter &Initial values& Lyapunov exponents \\
		\hline
		
		\hline
		
		\multirow{2}{*}{$ G_{1}(x_{1}) $} & \multirow{2}{*}{$ a=0.8 $} & red& $ (0.0179, 0, -1.3074) $\\ 
		&  &blue& $ (0.0229, 0, -1.3184) $\\[0.4cm] 
		
		\multirow{4}{*}{$ G_{2}(x_{1}) $} & \multirow{4}{*}{$ a=0.85 $}& red& $ (0.0329, 0, -1.2426) $\\ 
		&  &blue& $ (0.0388, 0, -1.2492) $\\
		&  &black& $ (0.0322, 0, -1.2743) $\\
		&  & green & $ (0.0282, 0, -1.2688) $\\
		\hline
		
		\hline
		
		\hline
	\end{tabular}
\end{table}

\section{Degenerating butterfly attractor}
\label{section:section4}
Using the controller $ G_{i}(x_{1}) $, this section illustrates how the shifting of equilibria can change the spacing of chaotic attractors, and then degenerate the butterfly attractor of system~(\ref{2}) into several independent chaotic attractors.  In addition, under certain values of the shifted equilibria, the butterfly attractor is also degenerated into various periodic attractors. The interesting things is that the butterfly wings can be recombined by coexisting symmetric pair of chaotic attractors, coexisting two symmetric pair of chaotic attractors, coexisting symmetric pair of limit cycles, and coexisting two symmetric pair of limit cycles.

\subsection{Degenerating the butterfly attractor into one or two symmetric pair of chaotic attractors}
In order to study the effect of shifting the equilibria of system~(\ref{2}) using the controller $ G_{i}(x_{i}) $, we fix the parameters $ \delta, \eta, h $ and initial conditions of the system~(\ref{3}), and then gradually increase the value of $ a $. For the controller $ G_{1}(x_{1}) $ and when $ \delta=0.5 $, $ \eta=0.1 $, $ h=2.2 $ with the initial conditions $ (-2, 0.01, -1) $, three values of $ a $ are selected as $ 0.5 $, $ 0.76 $ and $ 0.8 $, and the corresponding phase portraits are depicted in Figure~\ref{fig:phase2} (a)-(f). It can be observed that unlike the attractors in system~(\ref{2}) (see Figure~\ref{fig:phase1}), the spacing of attractors change when the shifted symmetric equilibria ($ SE $) become apart from each other. Figure~\ref{fig:phase2} (c) and (f) show that when $ SE_{2,3} $ become $ (0, \pm 3.03, 1) $, the butterfly attractor is broken into a single-wing chaotic attractor with Lyapunov exponents (LE) $ (0.0179, 0, -1.3097) $. Moreover, by choosing $ \delta=0.5 $, $ \eta=0.1 $, $ h=2.2 $ with the controller $ G_{2}(x_{1}) $, the parameter $ a $ is selected to be  $ 0.5 $, $ 0.76 $ and $ 0.85 $, the corresponding phase portraits are depicted in Figure~\ref{fig:phase2} (g)-(l). As can be seen in Figure~\ref{fig:phase2} (i) and (l), the butterfly attractor is broken into a compressed single-wing chaotic attractor when $ a=0.85 $. At $ a=0.85 $, the LE of the single-wing chaotic attractor is $ (0.0368, 0, -1.2445) $, the eigenvalues of $ SE_{1} $ are $ -0.1 $, $ -3.37 $, $ 2.87 $, and the eigenvalues of $ SE_{2,3} $ are $ -1.84 $, $ 0.13\pm 1.01\ i $. Therefore, the equilibrium $ SE_{1} $ is a saddle-node, and the equilibria $ SE_{2,3} $ are saddle-focus. 

To observe how many coexisting attractors can be produced after the broken of butterfly attractor into single-wing attractor, we consider the bifurcation diagrams of the initial conditions $ (-2, x_{20}, -1) $ varying with $ a $ equal to $ 0 $, $ 0.8 $ and $ 0.85 $, as illustrated in Figure~\ref{fig:bifurcation} (a)-(c), respectively. For the initial value of $ x_{20} $ varies in $ [0, 3] $, the bifurcations show that there is only one chaotic attractor when $ a=0 $, but there are two different chaotic attractors for the controller $ G_{1}(x_{1}) $ with $ a=0.8 $ and four different chaotic attractors for the controller $ G_{2}(x_{1}) $ with $ a=0.85 $, as shown Figure~\ref{fig:bifurcation}. To display these two and four independent chaotic attractors, the phase portraits are plotted in Figure~\ref{fig:phase3} for different sets of initial conditions. For $ a=0.8 $ with the controller $ G_{1}(x_{1}) $, symmetric pair of chaotic attractors coexist when the orbits of system~(\ref{3}) are initiated with $ (-2, 0.7, -1) $ (blue) and $ (-2, 2, -1) $ (red), as shown in Figure~\ref{fig:phase3} (a)-(c). Whereas, two symmetric pair of chaotic attractors coexist for $ a=0.85 $ with the controller $ G_{2}(x_{1}) $ when the orbits of system~(\ref{3}) are initiated with $ (-2, 0.5, -1) $ (red), $ (-2, 1.2, -1) $ (blue), $ (-2, 2, -1) $ (black), and $ (-2, 3, -1) $ (green), as shown in Figure~\ref{fig:phase3} (d)-(f). The LE values of these coexisting attractors, which appeared in Figure~\ref{fig:phase3}, are  demonstrated in Table~\ref{tb:tabel}.

Through the above analysis, we can conclude that under certain values of the shifted equilibria, the butterfly attractor of the plasma perturbation system~(\ref{2}) is degenerated into one or two symmetric pair of strange attractors. Meanwhile, the butterfly attractor can be recombined by choosing appropriate initial conditions, in which each set of initial conditions can produce one component of the butterfly attractor.

\subsection{Degenerating the butterfly attractor into one or two symmetric pairs of limit cycles}
Stretching the equilibria of system~(\ref{3}) apart enough from each other gives rise to degenerate the butterfly attractor into periodic attractors. When the parameters $ \delta=0.5 $, $ \eta=0.09 $, $ h=2.5 $ and  $ a $ is gradually increased, the dynamic evolution of the system~(\ref{3}) is given by the bifurcation diagrams under two sets of initial conditions with $ G_{1}(x_{1}) $, and four sets of initial conditions with $ G_{2}(x_{1}) $, as shown in Figure~\ref{fig:bifurcation1} (a) and (b), respectively. Figure~\ref{fig:bifurcation1} (a) shows that the system~(\ref{3}) experiences single chaotic attractor when $ a=0.65 $ with $ SE_{2,3}=(0, \pm2.95, 1) $, coexisting symmetric single-wing chaotic attractors when $ a=0.68 $  with $ SE_{2,3}=(0, \pm3, 1) $, coexisting symmetric pair of period-2 limit cycles when $ a=0.7 $ with $ SE_{2,3}=(0, \pm3.04, 1) $, and then coexisting two quasi-periodic attractors when $ a=0.8 $ with $ SE_{2,3}=(0, \pm3.22, 1) $. On the other hand, Figure~\ref{fig:bifurcation1} (b) shows that the system~(\ref{3}) experiences single chaotic attractor when $ a=0.5 $ with $ SE_{2,3}=(0, \pm2.65, 1) $, coexisting two symmetric pair of chaotic attractors when $ a=0.67 $  with $ SE_{2,3}=(0, \pm2.99, 1) $, coexisting two symmetric pair of period-2 limit cycles when $ a=0.72 $ with $ SE_{2,3}=(0, \pm3.08, 1) $, and then coexisting two symmetric pair of period-1 limit cycles when $ a=0.8 $ with $ SE_{2,3}=(0, \pm3.22, 1) $. To further demonstrate the degenerating of butterfly attractor into one or two symmetric pair of period-2 limit cycles, we display the phase portraits for $ G_{1}(x_{1}) $ with two sets of initial conditions, and for $ G_{2}(x_{1}) $ with four sets of initial conditions, as shown in Figure~\ref{fig:phase4} (a)-(c) and (d)-(f), respectively. Obviously, Figure~\ref{fig:phase4} show that the butterfly wings are recombined by coexisting one or two symmetric pair of period-2 limit cycles.

Consequently, it can be concluded that depending on the distance between the symmetric equilibria $ SE_{2,3} $, the behavior of system respectively changes from single chaotic attractor to the coexistence of one or two symmetric pair of chaotic attractors, and then to the coexistence of several periodic attractors.

\section{Multistability behaviors with complex transient transition behaviors}
\label{section:section5}
This section investigates the existence of multistability behaviors in the system~(\ref{3}) with the nonlinear controller $ G_{2}(x_{1}) $, including coexisting two point attractors with butterfly attractor, coexisting transient transition chaos with quasi-periodic behavior, and coexisting symmetric Hopf bifurcations.

\subsection{Butterfly attractor and two point attractors}
Suppose that the parameters $ \delta=0.45 $, $ \eta=0.6 $ and $ h=2 $, then by Proposition~\ref{pro:1}, the equilibrium $ SE_{1} $ is unstable for all positive value of the amplitude controller parameter $ a $. Meanwhile, the equilibria $ SE_{2,3} $ of system~(\ref{3}) are unstable when $ a\in[0, 1.947] $ and asymptotically stable when $ a> 1.947$. To display the interesting phenomenon of multistability behaviors for these set of parameters, the bifurcation diagram and the corresponding Largest Lyapunov
exponents (LLE) are depicted when  $ a $ is gradually increased from $ 0 $ to $ 2.2 $ for three different sets of initial conditions, as shown in Figure~\ref{fig:bifurcation4}. The orbit of blue color begins with the initial conditions $ (1, 1, 3) $, the orbit colored in red is initiated with $ (1, 4.4, 3) $, and the orbit colored in green is initiated with $ (1, -3.9, 3) $. According to Figure~\ref{fig:bifurcation4}, it can be observed that there exists a periodic behavior in the system~(\ref{3}) when $ a=2 $ for the initial conditions $ (1, -3.9, 3) $ (green) and $ (1, 4.4, 3) $ (red). Meanwhile, the system~(\ref{3}) shows a chaotic behavior with the initial conditions $ (1, 1, 3) $ (blue). The corresponding eigenvalues of the Jacobian matrix at the equilibria $ SE_{1} $ and $ SE_{2,3} $ for these parameters are given by
\begin{flalign*}
\begin{cases}
\begin{aligned}
&SE_{1}:-0.6,\ \ -2.3188, \ \ 1.8688  ,\\
&SE_{2,3}:-3.6462,\ \ -0.0019\pm 1.194\ i .
\end{aligned}
\phantom{\hspace{2cm}}
\end{cases}
\end{flalign*}
Since $ SE_{1} $ has one positive real eigenvalue and two negative real eigenvalues, it is therefore an unstable saddle point. At the equilibria $ SE_{2,3} $, the corresponding eigenvalues have one real eigenvalue and a pair of complex conjugate eigenvalues, with all negative real parts, hence $ SE_{2,3} $ are stable focus-nodes. Therefore, the phase portraits and the corresponding time-domain waveform are depicted in Figure~\ref{fig:phase6} to illustrate the coexistence of butterfly attractor with two point attractors. Obviously, the orbits that initiate with the initial conditions $ (1, -3.9, 3) $ (green) and $ (1, 4.4, 3) $ (red) converge on the symmetric stable focus-nodes $ SE_{2,3} $. Whereas, the orbit colored in blue initiates with $ (1, 1, 3) $, which emerges from the unstable saddle point $ SE_{1} $, exhibiting double-wing chaotic attractor that resembles a butterfly surrounded the symmetric equilibria $ SE_{2,3} $.

\begin{figure}[t]
	\hspace{-0.7cm}
		\includegraphics[width=10.5cm, height=3.5cm]{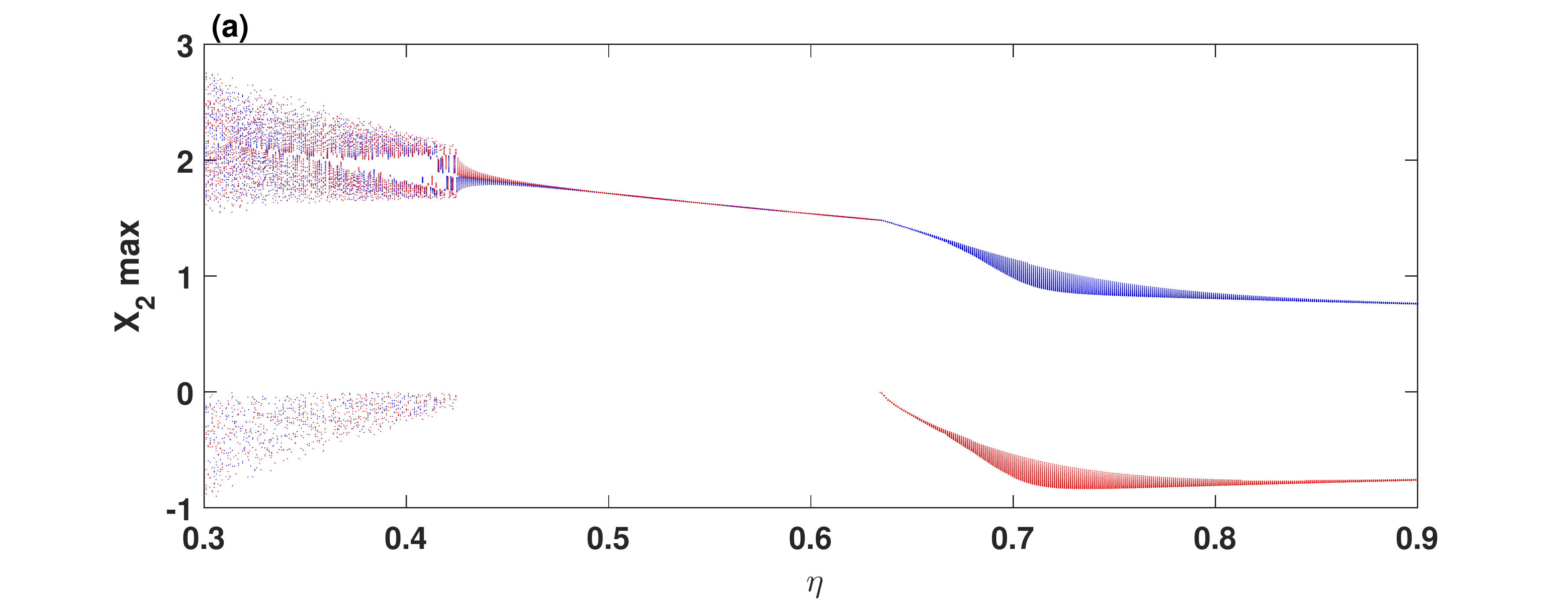}

     \begin{minipage}[b]{0.47\textwidth}
    	\hspace{-0.1cm}
    	\begin{subfigure}[b]{0.43\linewidth}
    		\includegraphics[width=4.8cm, height=5.5cm]{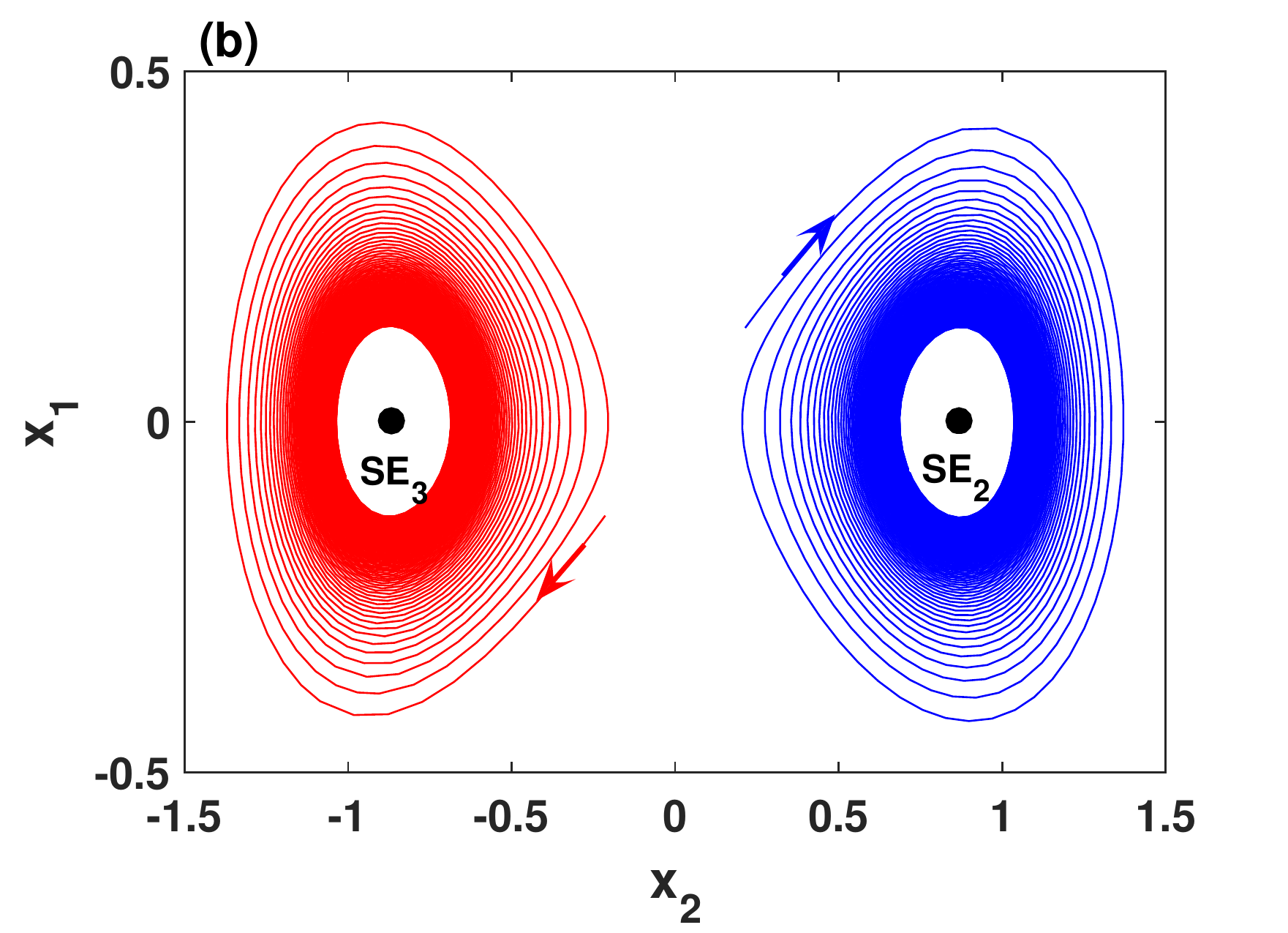}
    	\end{subfigure}
    	\hspace{0.55cm}
    	\begin{subfigure}[b]{0.43\linewidth}
    		\includegraphics[width=4.8cm, height=5.5cm]{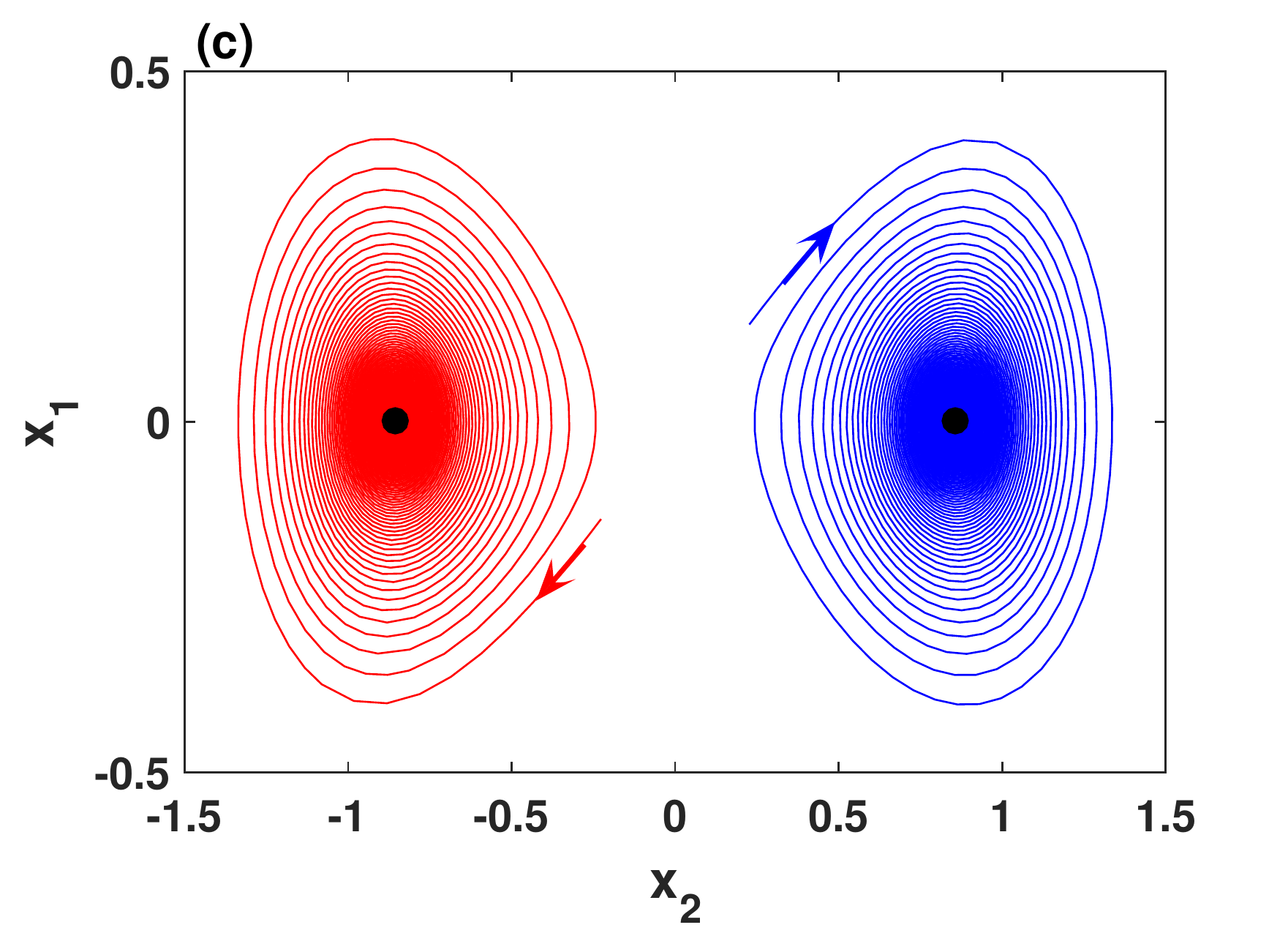}
    	\end{subfigure}
    \end{minipage}
	\caption{Hopf bifurcation in the system~(\ref{3}) when the parameters $ \delta=0.5 $, $ h=1 $, $ a=0.52 $ and $ \eta $ varying for the initial conditions $ (0, 1, 0) $ (blue) and $ (0, -1, 0) $ (red): (a) the bifurcation digram of the system when $ 0.3\leq\eta \leq 0.9 $; (b) the coexistence of attracted orbits to the stable symmetric limit cycles occurs when $ \eta_{0}>\eta=0.69$; (c) the coexistence of attracted orbits to the symmetric equilibria $ SE_{2,3} $ occurs when $ \eta_{0}<\eta=0.71 $.}
	\centering
	\label{fig:bifurcation5}
\end{figure}

\subsection{The coexistence of complex transient transition behavior of chaos with quasi-periodic}
According to the bifurcation diagram and LLE in the Figure~\ref{fig:bifurcation4}, it is easy to find that there exists a periodic behavior in the system~(\ref{3}) when $ a\in[1, 1.05] $ for the initial conditions $ (1, -3.9, 3) $ (green). Meanwhile, the system~(\ref{3}) seemingly shows a chaotic behavior for the initial conditions $ (1, 1, 3) $ (blue) and $ (1, 4.4, 3) $ (red). For the sake of verifying these complex coexisting behaviors, a suitable value of parameter $ a $ is chosen, as can be seen below. When $ a=1.02 $ and the initial conditions of the system~(\ref{3}) are selected as $ (1, 1, 3)$ (blue) and $ (1, -3.9, 3) $ (green), the phase portraits, the corresponding time-domain waveform and the LLE are depicted in Figure~\ref{fig:phase7}. The interesting phenomenon is observed that the system with the initial conditions $ (1, 1, 3)$ (blue) has a transition from a chaotic behavior to quasi-periodic with LE $ (0, 0, -2.6392) $ in the long time evolution, as shown in Figure~\ref{fig:phase7} (a) and (b). These two figures display that there is a positive Lyapunov exponent in a finite time interval at the beginning, and then the value of LLE gradually tends to be zero with the time evolution. Consequently, the system has a complex transient transition behaviors. In addition, when the initial conditions are taken as $ (1, -3.9, 3) $ (green), the system simultaneously exhibits a completely quasi-periodic behavior with the Lyapunov exponent $ (0, 0, -2.6420) $ in the long time evolution. Therefore, we are convinced that system~(\ref{3}) has very rich and complicated dynamics as well as the coexistence of transient transition with quasi-periodic behavior.

\subsection{The existence of symmetric Hopf bifurcations}
One may ask if the plasma perturbation model with the nonlinear controller $ G_{2}(x_{1}) $ has Hopf bifurcation. The answer is yes. This subsection will illustrate numerically the existence of symmetric Hopf bifurcations in the system~(\ref{3}). When we fix $ \delta=0.5 $, $h=1$, $ a=0.52 $, the bifurcation digram versus the parameter $ \eta $ varying is depicted in Figure~(\ref{fig:bifurcation5}) (a) for the initial conditions $ (0, 1, 0) $ (blue) and $ (0, -1, 0) $ (red). It can be seen that the system has coexisting two different periodic behaviors when $ \eta\geq0.64 $. However, according to Proposition~\ref{pro:1} (2), the system~(\ref{3}) is asymptotically stable at $ SE_{2,3} $ when $ \eta>0.6939 $.

Firstly, the nondegeneracy condition can be determined by calculating the roots of the characteristic equation at the equilibria $ SE_{2,3} $. For the bifurcation parameter $ \eta_{0}=0.6939 $, the system~(\ref{3}) at  $ SE_{2,3} $ has the eigenvalues with one pair of purely imaginary roots which are $ \lambda_{2,3}=\pm 0.7790 i $. That meas, the nondegeneracy condition is satisfied. Secondly, the differentiation characteristic equation~(\ref{7}) with respect to the parameter $ \eta $ is as follows:
\begin{flalign*}
\frac{d\lambda(\eta)}{d\eta}=\frac{2-2h-\delta h \lambda-h \lambda^{2}}{3\lambda^{2}+2(\eta h+\delta+a)\lambda+\delta(\eta h+a)},
\end{flalign*}
which leads to:
\begin{flalign*}
\frac{d\lambda(\eta)}{d\eta}\Big|_{\eta=\eta_{0}, \lambda=0.7790 i}=-0.2065 - 0.1334i
\end{flalign*}
since the differentiation characteristic equation has nonzero real part, then the transversality condition is also verified. Finally, we also found the direction of the Hopf bifurcation which is $ -1 $. Thus, the Hopf bifurcation of system~(\ref{3}) at the equilibria $ SE_{2,3} $ is nondegenerate and supercritical.

Furthermore, two numerical simulations are given in Figure~(\ref{fig:bifurcation5}) (b) and (c). For $ \eta_{0}>\eta=0.69$, the orbit of the system is attracted to the stable symmetric limit cycles with the initial conditions $ (0, 1, 0) $ (blue) and $ (0, -1, 0) $ (red), as shown in Figure~(\ref{fig:bifurcation5}) (b). By selecting $ \eta_{0}<\eta=0.71 $, the orbit of system with the initial conditions $ (0, 1, 0) $ (blue) is attracted to the stable equilibrium $ E_{2} $, meanwhile, the orbit with the initial conditions $ (0, -1, 0) $ (red) is attracted to the other stable equilibrium $ E_{3} $, as shown in Figure~(\ref{fig:bifurcation5}) (c).

\section{Conclusions}
\label{section:section6}
This work has studied the plasma perturbation model, which has a symmetric double-wing chaotic attractor that resembles a butterfly. Using a desirable chaotification technique, the equilibria of the system are shifted apart from each other. As a result, the spacing of chaotic attractors are changed, and then the butterfly wings are broken into one or two symmetric pair of coexisting chaotic attractors. In addition, the butterfly chaotic wings can be degenerated into several periodic orbits when the equilibria of the system are stretched apart enough from each other. Other multistability behaviors have been observed in the plasma system with nonlinear controller, in which the butterfly chaotic attractor coexists with two point attractors, and the transient transition chaos also coexists with quasi-periodic behavior. Finally, the coexistence of the symmetric Hopf bifurcations has revealed in the plasma system for particular values of the parameters.

\section*{References}

\bibliography{mybibfile}

\end{document}